\documentclass[twocolumn]{aastex63}
\pdfoutput=1

\usepackage{amsmath}
\graphicspath{{./}{figures/}}

\received{July 8th, 2020}
\revised{September 2nd, 2020}
\accepted{September 18th, 2020}%\today}
\submitjournal{ApJ}

\shorttitle{Unveiling the Mysteries of PWN \pwn\ }
\shortauthors{Hattori et al.}
\newcommand{\nustar}{\textit{NuSTAR}}
\newcommand{\hitomi}{\textit{Hitomi}}
\newcommand{\chandra}{\textit{Chandra}}
\newcommand{\psr}{J1833$-$1034}
\newcommand{\pwn}{G21.5$-$0.9}
\newcommand{\flux}{erg\,s$^{-1}$\,cm$^{-2}$ }
\newcommand{\ctwoline}{[\ion{C}{2}]~157.7~$\mu$m} 
\newcommand{\ooneline}{[\ion{O}{1}]~63.2~$\mu$m} 
\newcommand{\othreeline}{[\ion{O}{3}]~88.4~$\mu$m} 

\begin{document}

\title{The Nonstandard Properties of a ``Standard" PWN: Unveiling the Mysteries of PWN \pwn\  Using its IR and X-ray emission}

\correspondingauthor{Soichiro Hattori}
\email{soichiro@nyu.com}

\author{Soichiro Hattori}
\affil{NYU Abu Dhabi, PO Box 129188, Abu Dhabi, United Arab Emirates}
\affiliation{Center for Astro, Particle, and Planetary Physics (CAP$^3$), NYU Abu Dhabi, PO Box 129188, Abu Dhabi, United Arab Emirates}

\author[0000-0003-4136-7848]{Samayra~M.~Straal}
\affil{NYU Abu Dhabi, PO Box 129188, Abu Dhabi, United Arab Emirates}
\affiliation{Center for Astro, Particle, and Planetary Physics (CAP$^3$), NYU Abu Dhabi, PO Box 129188, Abu Dhabi, United Arab Emirates}

\author{Emily Zhang}
\affil{Columbia University, 2960 Broadway, New York, NY 10027, USA}

\author[0000-0001-7380-3144]{Tea Temim}
\affil{Space Telescope Science Institute, 3700 San Martin Drive, Baltimore, MD 21218, USA}

\author[0000-0003-4679-1058]{Joseph~D. Gelfand}
\affil{NYU Abu Dhabi, PO Box 129188, Abu Dhabi, United Arab Emirates}
\affiliation{Center for Astro, Particle, and Planetary Physics (CAP$^3$), NYU Abu Dhabi, PO Box 129188, Abu Dhabi, United Arab Emirates}
\affiliation{Center for Cosmology and Particle Physics (CCPP, Affiliate), New York University, 726 Broadway, Room 958, New York, NY 10003}

\author[0000-0002-6986-6756]{Patrick O. Slane}
\affil{Harvard-Smithsonian Center for Astrophysics, 60 Garden Street, Cambridge, MA 02138, USA}

\begin{abstract}

The evolution of a pulsar wind nebula (PWN) depends on properties of the progenitor star, supernova, and surrounding environment. As some of these quantities are difficult to measure, reproducing the observed dynamical properties and spectral energy distribution (SED) with an evolutionary model is often the best approach in estimating their values. \pwn, powered by the pulsar \psr, is a well observed PWN for which previous modeling efforts have struggled to reproduce the observed SED. In this study, we reanalyze archival infrared (IR; \textit{Herschel, Spitzer}) and X-ray (\textit{Chandra, NuSTAR, Hitomi}) observations. The similar morphology observed between IR line and continuum images of this source indicates that a significant portion of this emission is generated by surrounding dust and gas, and not synchrotron radiation from the PWN. Furthermore, we find the broadband X-ray spectrum of this source is best described by a series of power laws fit over distinct energy bands. For all X-ray detectors, we find significant softening and decreasing unabsorbed flux at higher energy bands. Our model for the evolution of a PWN is able to reproduce the properties of this source when the supernova ejecta has a low initial kinetic energy $E_{\mathrm{sn}} \approx 1.2 \times 10^{50}\,\mathrm{ergs}$ and the spectrum of particles injected into the PWN at the termination shock is softer at low energies. Lastly, our hydrodynamical modeling of the SNR can reproduce its morphology if there is a significant density increase of the ambient medium ${\sim} 1.8$~pc north of the explosion center.

\end{abstract}

\keywords{Supernova remnants (1667), Pulsars (1306), Infrared astronomy (786), X-ray astronomy (1810)} %http://astrothesaurus.org/concept-select/

\section{Introduction}
\label{sec:intro}

Stars with mass in the range $M \gtrsim 8~M_\odot$ are believed to end their lives in a core-collapse supernova event (\citealt{baade34}). In many cases, this produces a highly magnetized and rapidly rotating neutron star (i.e., pulsar). Pulsar's dissipate their rotational energy by powering a relativistic outflow of electrons and positrons, commonly referred to as the ``pulsar wind". The expanding magnetic bubble of particles, created by the interaction of the relativistic pulsar wind with the ambient medium, is the pulsar wind nebula (PWN). For young PWNe the ambient medium is the slow-moving supernova ejecta in the host supernova remnant (SNR), but for older PWNe it can also be interstellar medium (ISM) after the pulsar exits the SNR \citep{slane17}. The reader is directed to \citet{slane17}, \citet{gaensler06}, \citet{chevalier05}, \citet{arons2004}, and \citet{amato2020} for detailed explanations on PWNe. 

As the evolution of a PWN depends on the central neutron star, the composition of the pulsar wind, and its surrounding environment, modeling PWNe allows us to determine the physical characteristics of all components of the system--which are difficult, if not impossible, to determine by other means (e.g., \citealt{torres2017}). In addition, evolutionary models allow us to infer properties of the progenitor star and supernova.  Furthermore, such models also determine the spectrum of particles accelerated inside these sources, needed to determine the currently unknown physical mechanism \citep{sironi13} by which such objects produce some of the most energetic particles observed in the Universe.  The modeling is done by generating the dynamical properties and spectral energy distribution (SED) which can be fit against measurements gathered over the entire electromagnetic spectrum.

\begin{figure}[tbh]
    \centering
    \includegraphics[width=\columnwidth]{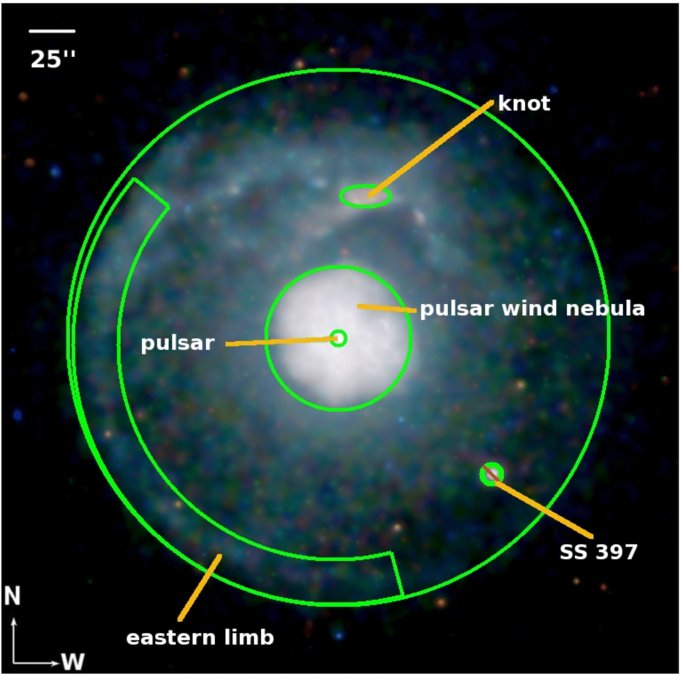}
    \caption{Image of \pwn\ with the individual components labeled. Reproduced from \citet{matheson10}.}
    \label{fig:g21_indiv_components}
\end{figure}

PWN \pwn\ is a bright X-ray emitting source well-suited for modeling as it has been observed by many telescopes spanning the electromagnetic spectrum. Initially detected in 1970 (\citealt{altenhoff1970, wilson1970}) in the radio band and later observed in the X-ray band in 1981 \citep{becker1981}, its central pulsar \psr\ was detected in 2006 \citep{camilo06} whose measured period $P$ and period-derivative $\dot{P}$ suggest a high spin-down luminosity 
and a low characteristic age. 
As a PWN with a circular morphology (Figure \ref{fig:g21_indiv_components}) associated with a young ($\lesssim 10^4$ years) pulsar, \pwn\ is appropriate for an analysis using ``one-zone" models (e.g., \citealt{reynolds84}, see \citealt{gelfand17} for a recent review). 

\begin{table*}[tbh]
    \caption{Previous broken power-law measurements of the X-ray spectrum of PWN \pwn}
    \label{table:prev_xray_obs}
    \centering
    \begin{tabular}{cccccccc}
    \hline
    \hline
    Observatory & Energy Range & $N_H$  & $\Gamma_1$ & $E_{\rm break}$ & $\Gamma_2$ & $F_{2-8}$ & $F_{15-50}$\\
    $\cdots$ & [keV] & [$10^{22}$~cm$^{-2}$] & $\cdots$ & [keV] & $\cdots$ &  [$10^{-11}~\frac{\rm erg}{\rm cm^2~s}$] & [$10^{-11}~\frac{\rm erg}{\rm cm^2~s}$]\\
    \hline
    {\it NuSTAR} & 3$-$45 & $\equiv2.99$ & $1.996_{-0.012}^{+0.013}$ & $9.1_{-1.4}^{+1.2}$ & $2.093_{-0.012}^{+0.013}$ & $5.27\pm0.08$ & $5.11\pm0.08$\\
    {\it Hitomi} & 0.8$-$80 & $3.22\pm0.03$ & $1.74\pm0.02$ & $7.1\pm0.3$ & $2.14\pm0.01$ & $4.80\pm0.02$ & $4.54\pm0.04$\\
    \hline
    \hline
    \end{tabular} \\ 
    \footnotesize {\bf Note}: The {\it NuSTAR} results are from \cite{nynka14} and the {\it Hitomi} results are from \cite{hitomi18}. $F_{2-8}$ indicates the $2-8$~keV flux and $F_{15-50}$ indicates the $15-50$ keV flux.  Errors are 90\% confidence intervals.\\ 
\end{table*}

However, previous attempts to model this system have not succeeded in simultaneously reproducing the radio and X-ray spectrum
\citep{tanaka11,torres14,hitomi18}. The modeling is further complicated by the discrepant measurement of the X-ray spectrum by three observatories, \chandra, \nustar, and \hitomi\ \citep{guest19, nynka14, hitomi18}. One such difference is shown in Table \ref{table:prev_xray_obs}, where the parameters for the broken power-law model between \nustar\ and \hitomi\ are in disagreement.  Furthermore, the infrared (IR) emission observed from this source \citep{gallant99}, often assumed to be dominated by synchrotron radiation from the PWN (e.g., \citealt{tanaka11,torres14,hitomi18}), may be contaminated by emission from surrounding gas and dust.  To address these concerns, we reanalyzed archival IR and X-ray observations of this source.  

This paper is structured as follows. In \S\ref{obs} we describe the IR and X-ray observations and detector-specific data reduction and analysis of PWN \pwn. In \S\ref{sec:xray-analysis} we describe our piecewise power-law fitting approach to analyze the X-ray spectra and present our results. In \S\ref{sec:discussion} we discuss the implications of these results in our modeling for this source. We summarize our findings in \S\ref{sec:summary}. 

\section{Observations and Data Analysis}
\label{obs}
In this section, we describe our analysis of archival IR ({\it Herschel, \it Spitzer} \S\ref{sec:ir}) and X-ray ({\it Chandra} \S\ref{sec:chandra}, {\it NuSTAR} \S\ref{sec:nustar}, and {\it Hitomi} \S\ref{sec:observations_hitomi}) observations of this source. 

\subsection{{\it Infrared Observations}}
\label{sec:ir}

\pwn\ was observed with the Photodetector Array Camera (PACS) Integral Field Unit \added{(IFU)} Spectrometer \citep{poglitsch10} aboard \added{the} \textit{Herschel} Space Observatory on 2013 April 07. The range spectroscopy mode was used to cover the \ooneline\ and 145.5~\micron, \othreeline, and \ctwoline\ emission lines. The total \replaced{field-of-view}{field of view} of one IFU pointing is $47\arcsec\ \times 47\arcsec$, consisting of 25 spaxels. In order to cover the entire PWN in \pwn\, we obtained a 2~$\times$~2 mosaic IFU mosaic of the source, as well as a single-pointing off-source background observation for each line. The IFU cubes were analyzed using HIPE version 15.0.1 \citep{ott2010}. The analysis included trimming of the spectral edges and a subtraction of the baseline continuum obtained by a 2-degree polynomial fit across the line-free spectral region.  While narrow background lines were detected in the baseline-subtracted and spatially-integrated spectrum of the off-source IFU pointing, both narrow and broad lines were detected in the IFU cubes centered on the PWN. The broad lines have a full-width-at-half-maximum (FWHM) of 850~${\rm km\:s}^{-1}$ for the \ctwoline~\micron\ and 1000~${\rm km\:s}^{-1}$ for the \ooneline~63.2~\micron\ line and likely arise from SN ejecta material. The corresponding ejecta velocities are then 425$\pm$75~${\rm km\:s}^{-1}$ and 500$\pm$20~${\rm km\:s}^{-1}$ for the \ctwoline\ and \ooneline, respectively. If the observed line emission arises predominantly from ejecta with a low tangential velocity, the expansion velocity measured from the lines would represent a lower limit on the true velocity, which could be up to a factor of two higher, giving an expansion velocity range between 350 and 1000~${\rm km\:s}^{-1}$. In a radiative shock, the emission that we observe likely arises from highest-density material at the contact discontinuity, in which case the observed velocity represent the expansion velocity of the PWN rather than the free expansion velocity of the ejecta. However, since the shock velocities that produce the IR lines are relatively low, the free-expansion velocity of the ejecta material at the PWN boundary is within a similar range.  

\begin{figure*}[tbh]
    \center
    \includegraphics[width=0.9\textwidth]{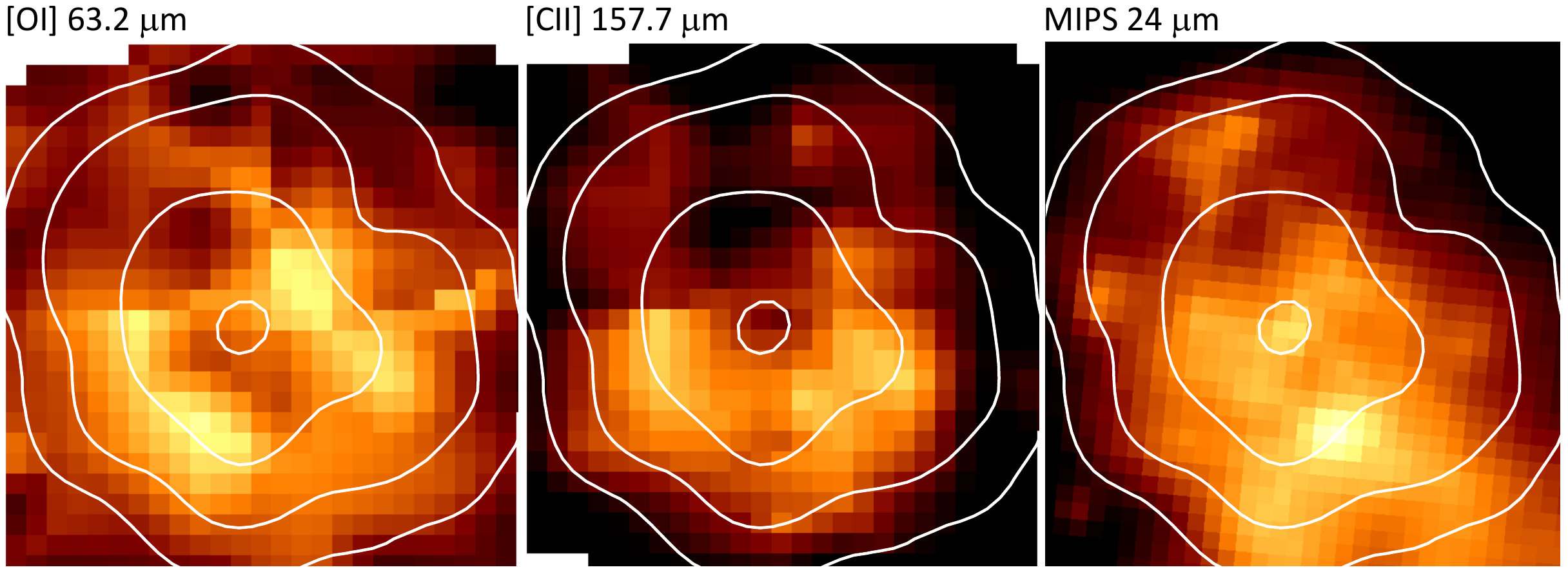}
    \caption{Emission line maps of the \ion{O}{1}~63.2~\micron\ and \ion{C}{2}~157.7~\micron\ broad ejecta lines in the PWN region are shown in the left and middle panels, respectively. The MIPS~24~\micron\ image of the same region is shown in the right panel. The X-ray contours from the PWN are shown in white.}
    \label{fig:irfig}
\end{figure*}

We produced emission line maps of the \ooneline\ and \ctwoline\ ejecta lines by integrating the spectra across the broad-line component, while excluding the narrow line that arises from the background emission. The maps are shown in Figure~\ref{fig:irfig} with the X-ray contours from the PWN overlaid in white.

\begin{figure*}[tbh]
    \includegraphics[width=1.0\textwidth]{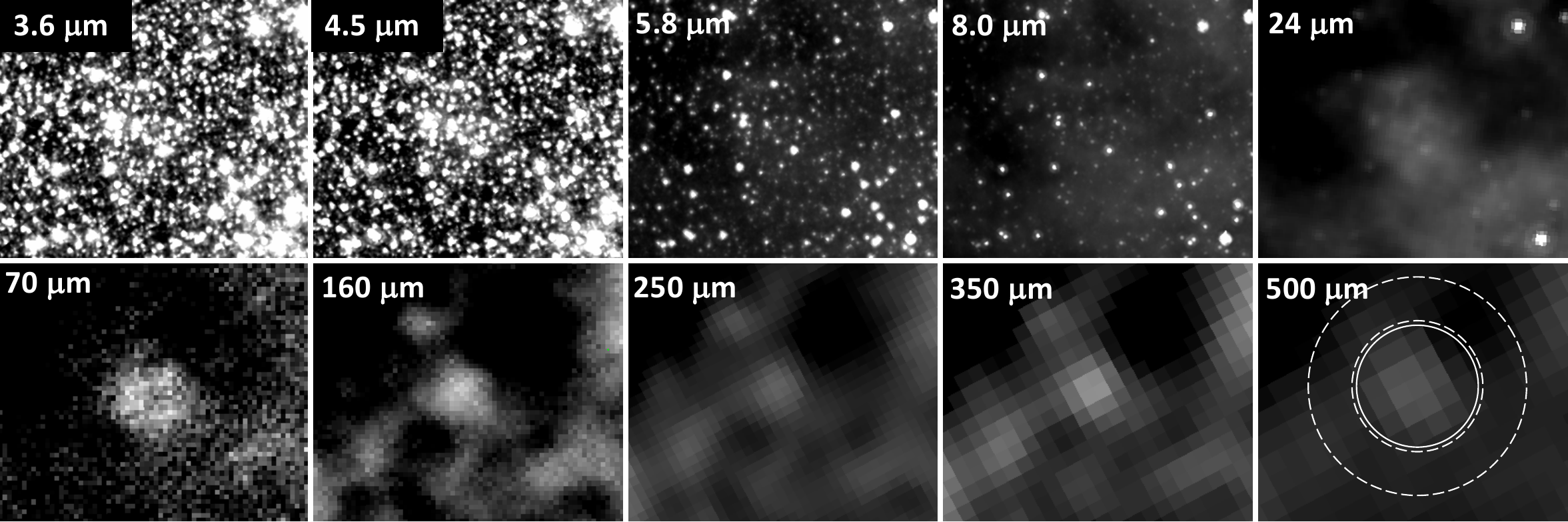}
    \caption{The \textit{Spitzer} 3.6, 4.5, 5.8, 8.0, and 24~\micron\ images and \textit{Herschel} 70, 160, 250, 350, and 500~\micron\ images of the PWN region in SNR \pwn. The last panel shows the extraction aperture and background annulus used to extract the total flux densities in the \textit{Spitzer} 24~\micron\ image and all the \textit{Herschel} images. The corresponding flux densities are listed in Table~\ref{tab:ir_fluxes}.}
    \label{fig:irfig2}
\end{figure*}

Total IR flux densities of the PWN region in \pwn\ were estimated from the images obtained with the Infrared Array Camera (IRAC) and Multiband Imaging Photometer (MIPS) aboard \textit{Spitzer} (PID 3647, PI: Slane), and the Photodetector Array Camera and Spectrometer (PACS) and Spectral and Photometric Imaging REceiver (SPIRE) instruments aboard \textit{Herschel} (Obs ID 1342218642), spanning a wavelength range between 3.6 and 500~\micron. For the MIPS, PACS, and SPIRE images, we extracted the total flux densities using an aperture centered on the PWN with a radius of 41.5\arcsec\ and a background annulus with inner and outer radii of 41.5\arcsec\ and 74.0\arcsec, respectively.  The IR images and the extraction aperture are shown in Figure~\ref{fig:irfig2} and total flux densities listed in Table~\ref{tab:ir_fluxes}.
\added{The IR background annulus slightly overlaps with the northern enhancement detected in X-ray (see Figures \ref{fig:g21_indiv_components} \&  \ref{fig:g21_chandra}). However, as little-to-no IR emission is detected from this feature, its inclusion in the background annulus should not significantly affect our analysis.}
The uncertainties on the flux densities in this case are dominated by the uncertainties of the local background emission. The IRAC images show very faint emission from the PWN, superposed on a dense stellar field. To make a rough estimate of the PWN emission in the IRAC bands, we estimated the surface brightness in a very small region free of stellar sources and assumed that this surface brightness is constant across the entire area of the PWN. The estimated background-subtracted surface brightness values at 3.6, 4.5, 5.8, and 8.0~\micron\ are 0.33, 0.37, 1.6, and 4.0~MJy/sr. The total flux densities were estimated by multiplying by a PWN area of $8.7\times10^{-8}$ steradians. 

\begin{table*}[tbh]
    \caption{Observed (\S\ref{sec:ir}) and Predicted synchrotron (\S\ref{sec:pwnmodel}) IR flux density of the PWN}
    \label{tab:ir_fluxes}
    \makebox[0.90\linewidth]{%
    \centering
    \begin{tabular}{lcccccc}
    \hline
    \hline
         Instrument & Wavelength $\lambda$ & Total Flux Density $S_\nu^{\rm tot}$ & \multicolumn{2}{c}{PWN Flux density $S_{\nu}^{\rm pwn}$}  & \multicolumn{2}{c}{Residual Flux Density $S_{\nu}^{\rm resid}$} \\
         & & & Variable $p$ & $p\equiv1.85690$ & Variable $p$ & $p \equiv 1.85690$ \\
         & (\micron) & (Jy) & (Jy) & (Jy) & (Jy) & (Jy) \\
         \hline
         SPIRE & 500  & 1.0$\pm$0.2  & 0.64 & 0.67 & 0.36 & 0.33 \\
         SPIRE & 350  & 1.2$\pm$0.2  & 0.48 & 0.50 & 0.72 & 0.70 \\
         SPIRE & 250  & 1.6$\pm$0.2  & 0.35 & 0.37 & 1.25 & 1.23 \\
         PACS & 160 & 3.7$\pm$1.0 & 0.21 & 0.22 & 3.49 & 3.48 \\
         PACS & 70  & 3.4$\pm$1.5 & 0.11 & 0.11 & 3.29 & 3.29 \\
%         MIPS & 70  & 1.9$\pm$0.4 & 0.11 & \\
         MIPS & 24  & 0.22$\pm$0.03 & 0.037 & 0.039 & 0.183 & 0.181 \\
         IRAC & 8.0  & $\sim$0.33 & 0.014 & 0.015 & 0.316 & 0.315 \\
         IRAC & 5.8  & $\sim$0.13 & 0.011 & 0.011 & 0.119 & 0.119 \\
         IRAC & 4.5 & $\sim$0.031 & 0.009 & 0.009 & 0.022 & 0.022 \\
         IRAC & 3.6  & $\sim$0.027 & 0.007 & 0.008 & 0.020 & 0.019 \\
         \hline
         \hline
    \end{tabular}
    }
\end{table*}

\subsection{Chandra Observations}
\label{sec:chandra}

\begin{figure*}[tbh]
    \includegraphics[width=0.49\textwidth]{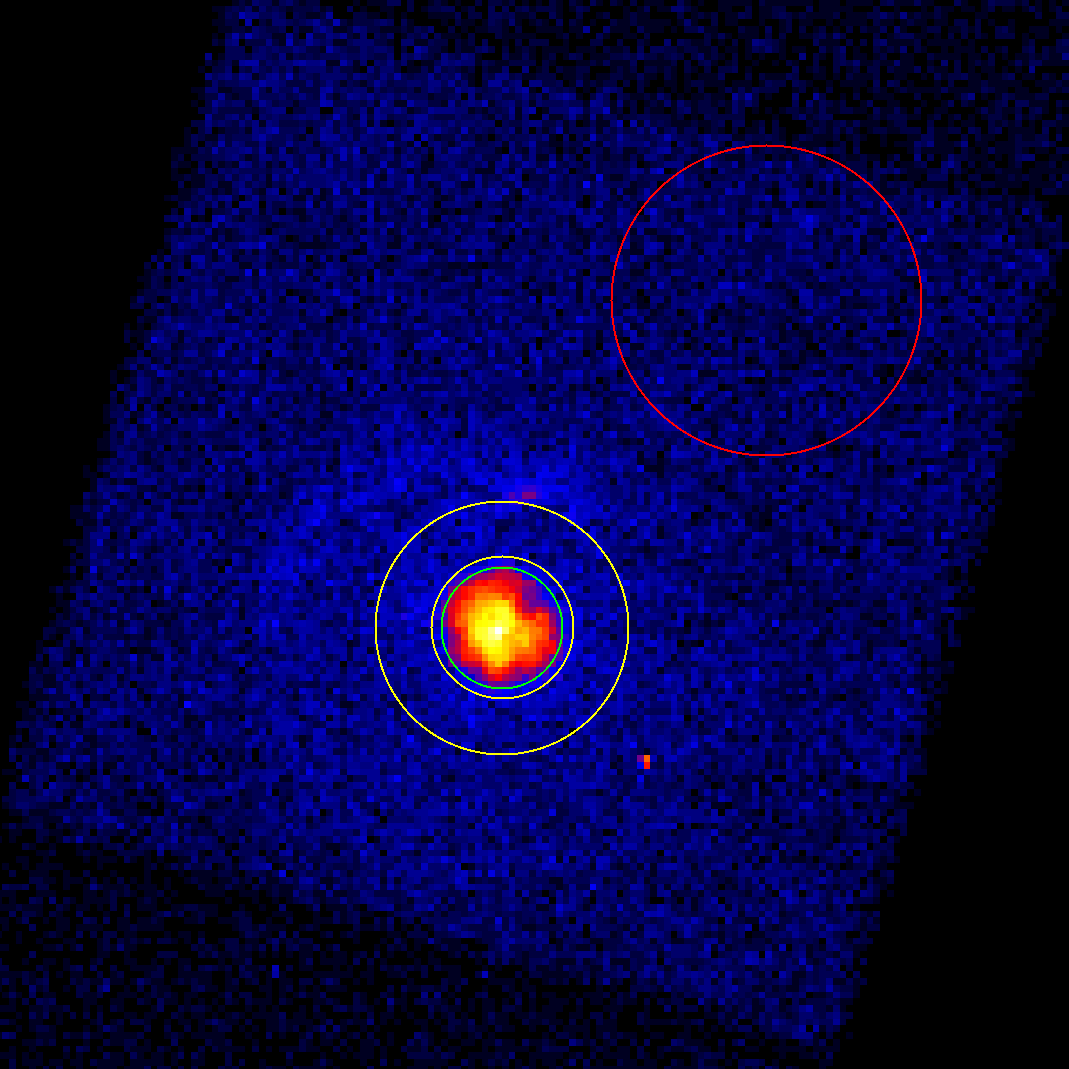}
    \includegraphics[width=0.49\textwidth]{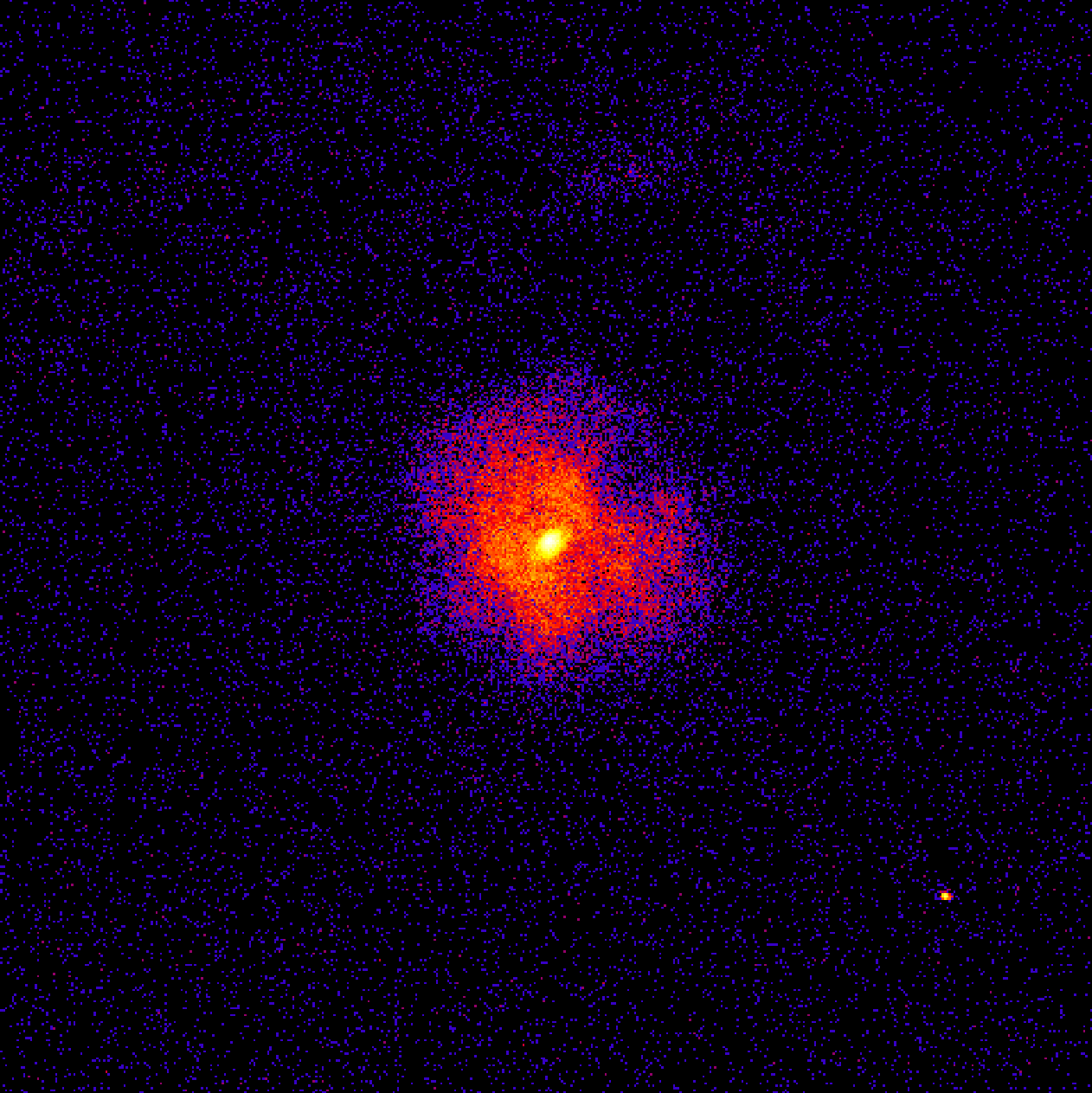}
    \caption{\chandra\ images of \pwn. Images are taken from ObsID 1433 (Table \ref{tab:chandra_obs}). {\it Left}: Image showing the \chandra\ source region (green), the \chandra\ background region (red), and the IR background annulus (yellow). {\it Right}: An enlarged view of the PWN.}
    \label{fig:g21_chandra}
\end{figure*}

\chandra\ has regularly observed \pwn\ with both its Advanced CCD Imaging Spectrometer (ACIS) and High Resolution Camera (HRC). For this study we reanalyzed ACIS-S observations where \pwn\ fell on the S3 chip, the back-illuminated chip where the best imaging and energy resolution is obtained. A single ACIS chip has a field of view of $8.3\arcmin \times 8.3\arcmin$ with an imaging resolution of $\sim 1\arcsec$ over the energy range 0.2--10 keV.

Software used for this analysis include CIAO version 4.10 \citep{Fruscione2006} and its accompanying Sherpa version \citep{freeman2001, doe2007}, as well as SAOImage DS9 version 7.6 \citep{joye2003, sao2000}.

After querying and downloading 16 ACIS-S observations from the \chandra\ Data Archive where \pwn\ fell on the S3 chip, error-causing subarray files (ObsIDs 1554, 3693, 10646, 14263, 16420) were deleted.

After analyzing each observation independently, we found that the results for ObsIDs 1230 and 159 deviated significantly from those for all other ObsIDs. Upon investigation, this discrepancy was attributed to these observations being taken with focal plane temperatures of -$100^\circ$C, as compared with -$110^\circ$C or -$120^\circ$C focal plane temperatures for all the other observations. The accuracy of the temperature-dependent gain correction decreases for temperatures below -$112^\circ$C, so ObsIDs 1230 and 159 were deemed too warm to provide reliable spectral results.

Spectra were extracted for the stack of all 11 remaining observations, summarized in Table \ref{tab:chandra_obs}, over the energy band 0.5-8 keV using the CIAO tool specextract. These spectra were extracted using a $35\arcsec$ radius circular region covering the entire central PWN as shown in Figure \ref{fig:g21_chandra}.

\begin{table}[tbh]
\caption{Details of {\it Chandra} observations }
\label{tab:chandra_obs}
\centering
\begin{tabular}{cccc}
\hline
\hline
ObsID & Start Time & Data Mode & Exposure [s] \\
\hline
1433 & 1999-11-15 22:31:18 & FAINT & 14970 \\
1717 & 2000-05-23 09:24:15 & FAINT & 7540 \\
1770 & 2000-07-05 03:42:36 & FAINT & 7220 \\
1838 & 2000-09-02 01:09:11 & FAINT & 7850 \\
2873 & 2002-09-14 01:09:17 & FAINT & 9830 \\
3700 & 2003-11-09 12:20:43 & VFAINT & 9540 \\
5159 & 2004-10-27 13:32:57 & VFAINT & 9830 \\
5166 & 2004-03-14 22:12:41 & VFAINT & 10020 \\
6071 & 2005-02-26 09:08:53 & VFAINT & 9640 \\
6741 & 2006-02-22 02:57:52 & VFAINT & 9830 \\
8372 & 2007-05-25 12:06:03 & VFAINT & 10010 \\
\hline
\hline
\end{tabular}
\end{table}
%for all observations, instrument is ACIS-S & type is CAL

\subsection{\it NuSTAR Observations and Data Reduction}
\label{sec:nustar}
The Nuclear Spectroscopic Telescope Array ({\it NuSTAR}) is a high-energy (3-79\,keV) X-ray space observatory consisting of two co-aligned telescopes with detectors placed at each of their focal plane modules (referred to as FPMA and FPMB) \citep{nustar}. Each {\it NuSTAR} telescope has a field of view of $12\arcmin \times 12\arcmin$ with a full-width half maximum (FWHM) of $18\arcsec$ and a half-power diameter (HPD) of $58\arcsec$. The FWHM spectral resolution is 400\,eV at 10\,keV.                                                
\nustar\ observed \pwn\ on nine separate occasions for a total of $\sim$383~ks on {\it each} of its two on-board FPM detectors (see Table \ref{tab:nustar_obs}). Of the nine observations, two of them (ObsID 10002014001, 40001016001) were taken in the STELLAR spacecraft mode, making them unsuitable for science observations due to the spacecraft roll maneuver of $\sim$1~deg/day as mentioned in the \href{https://heasarc.gsfc.nasa.gov/W3Browse/nustar/numaster.html}{{\it NuSTAR} Master Catalog}\footnote[1]{\url{https://heasarc.gsfc.nasa.gov/W3Browse/nustar/numaster.html}}. As such, we did not analyze these observations. In addition, due to the short effective exposure time of ObsID 10002014006, we did not analyze this observation. Of the remaining six observations, the previous \citep{nynka14} paper analyzed four observations (ObsID 10002014003, 10002014004, 40001016002, 40001016003) totalling $\sim$190~ks, but did not analyze two observations (ObsID 10002014002, 10002014005) which would add an additional $\sim$178~ks. We analyzed six observations, including the two previously unanalyzed observations, for a total exposure time of $\sim$368~ks on each FPM. As each observation was done by both detectors on {\it NuSTAR}, we analyzed a total of twelve data-sets.

For all twelve data-sets we followed the standard pipeline processing (HEASoft v6.24 \citep{heasarc2014} and NuSTARDAS v1.80) as explained in the \href{https://heasarc.gsfc.nasa.gov/docs/nustar/analysis/nustar_swguide.pdf}{{\it NuSTAR} Data Analysis Software Guide}\footnote[2]{\url{https://heasarc.gsfc.nasa.gov/docs/nustar/analysis/nustar_swguide.pdf}} prior to spectral fitting. We ran the processing script {\sf nupipeline} (v0.4.6) with the default options to produce the cleaned and calibrated event files, referred to as ``Level 2 Data Products" in the guide. The default pipeline option does not perform any South Atlantic Anomaly (SAA) filtering (done via the command {\sf nucalcsaa}). While the SAA filter may be required for fainter sources, \pwn\ is a relatively bright source ($>$1 count per second) and therefore there is likely no need for SAA filtering as mentioned in the official {\it NuSTAR website} (\href{https://www.nustar.caltech.edu/page/background}{Background Filtering}\footnote[3]{\url{https://www.nustar.caltech.edu/page/background}}). 

Once the cleaned and calibrated event files were created, we generated the associated redistribution/response matrix (RMF) and ancillary response (ARF) files by running {\sf nuproducts} (v0.3.3) with the option {\sf extended=yes}. The {\sf extended=yes} option is necessary to generate the ARF appropriate for an extended source. 

\begin{figure}
    \centering
    \includegraphics[width=0.95\columnwidth]{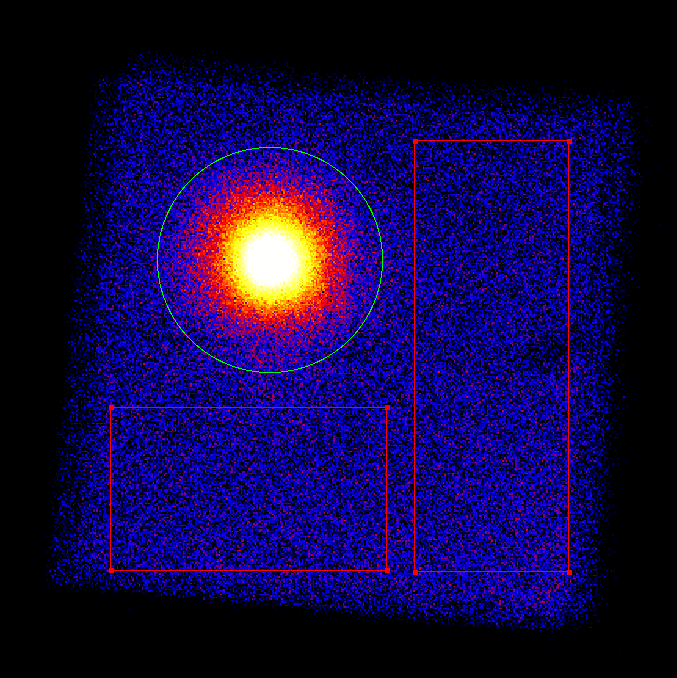}
    \caption{A representative {\it NuSTAR} image of \pwn\ with the source and background region used for the spectral analysis described in \S\ref{sec:nustar}. The green circle is the source region and the two red rectangles are the background regions. While \nustar\ has two on-board detectors, the images between the two detectors for this source at the given observations were nearly identical.} 
    \label{fig:nustar}
\end{figure}

The source spectra was extracted using a $177\arcsec$ radius circular region centered on the PWN and the background spectra were extracted using two rectangular regions away from the source {Figure \ref{fig:nustar}}. This conventional background extraction method may induce small uncertainties/fluctuations in the background as the {\it NuSTAR} background is known to be non-uniform across it's field of view {\citep{nustarnuskybgd}}. However, since \pwn\ is roughly 10 times brighter than the background in most of the energy range we do spectral fitting for these background uncertainties should be negligible. We also see no stray light emission from nearby bright X-ray sources in the field of view that may contribute to the background during any of the observations. We confirmed that stray light is not an issue during the observations of this source using the {\it NuSTAR} Science Operation Center's \href{http://www.srl.caltech.edu/NuSTAR_Public/NuSTAROperationSite/CheckConstraint.php}{{\it NuSTAR} constraint check page}\footnote[4]{\url{http://www.srl.caltech.edu/NuSTAR_Public/NuSTAROperationSite/CheckConstraint.php}}.

\begin{deluxetable*}{ccccrr}
%\tablenum{1}
\tablecaption{NuSTAR Observations of \pwn\ \label{tab:nustar_obs}}
\tablehead{
\colhead{ObsID} & \colhead{Start Time} & \colhead{Obs Type} & \colhead{Spacecraft Mode} & \colhead{Exposure A [s]} & \colhead{Exposure B [s]}
%& \colhead{\citet{nynka14}} 
}
\startdata
10002014001 & 2012-07-27 14:36:07 & CAL & STELLAR & 12990 & 13003 \\ %& No \\
10002014002 & 2012-07-28 01:01:07 & CAL & INERTIAL & 44456 & 44447 \\ %& No \\
10002014003 & 2012-07-29 01:21:07 & CAL & INERTIAL & 44723 & 44722 \\ % & Yes \\
10002014004 & 2012-07-30 01:33:37 & CAL & INERTIAL & 28023 & 28011 \\ % & Yes \\
10002014005 & 2012-07-31 19:38:33 & CAL & INERTIAL & 133782 & 133760 \\ % & No \\
10002014006 & 2012-08-03 20:51:07 & CAL & INERTIAL & 1944 & 1956 \\ % & No \\
40001016001 & 2013-02-26 05:31:07 & SNR & STELLAR & 50 & 50 \\ % & No  \\
40001016002 & 2013-02-26 05:56:07 & SNR & INERTIAL & 29704 & 29679 \\ % & Yes \\
40001016003 & 2013-02-26 22:11:07 & SNR & INERTIAL & 87721 & 87646 \\ % & Yes 
\enddata
%\tablecomments{}
\end{deluxetable*}

%% Hitomi events incl source and background regions
\begin{figure*}
    \centering
    \gridline{\fig{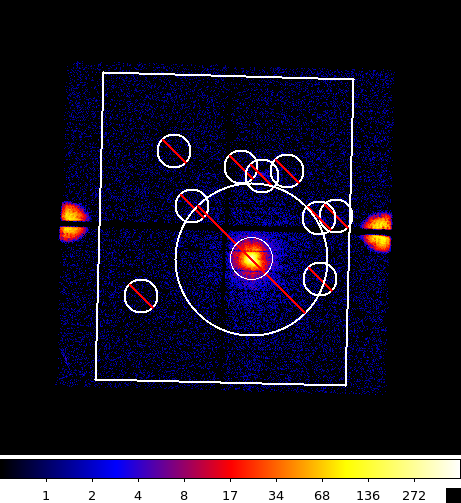}{0.31\textwidth}{\hitomi\ SXI }
        \fig{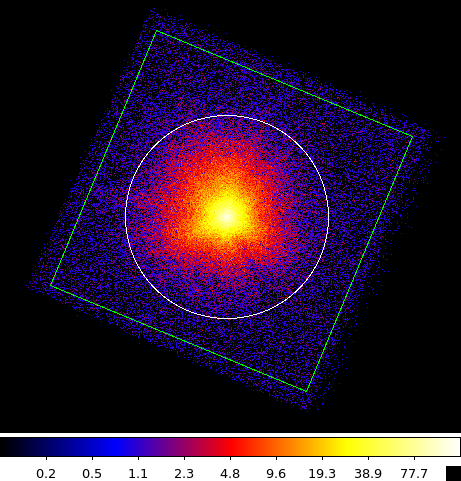}{0.31\textwidth}{\hitomi\ HXI 1 }
        \fig{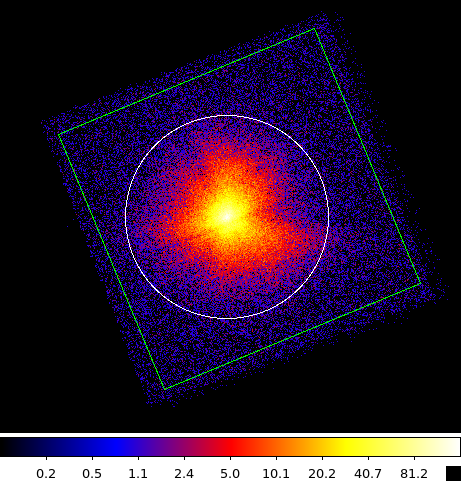}{0.31\textwidth}{\hitomi\ HXI 2 }
    }
    \caption{\hitomi\ detector images of \pwn. From left to right, the SXI and both HXI detectors are shown. In the SXI image the source region is shown as the innermost circle. The background region is shown as the white square where the source region, a larger annulus around the source, and previously known X-ray point sources in the FoV are excluded. In the HXI images the source region is shown as the white circle and the background region by the green rectangle where the source region is excluded. Source and background regions are taken from the \hitomi\ step by step analysis guide version 6.1. }\label{fig:hitomi_img}
\end{figure*}

%%%%%%%%% HITOMI %%%%%%%%%%
\subsection{\hitomi\ Observations and Data Reduction} %SMS
\label{sec:observations_hitomi}
During its mission's lifetime the \hitomi\ X-ray observatory \citep{takahashi2016} observed PWN \pwn\ as part of its commissioning and verification phase under the observing ID's 100050010 - 100050040 between 2016 March 19-23. % How many seconds: Eff exposure, SXI:51 ks, HXI: 99ks
Data was recorded to all four instruments, the Soft X-ray Imager (SXI), the Soft X-ray Spectrometer (SXS), the Hard X-ray Imager (HXI), and the Soft Gamma-ray Detector (SGD).  However, during this observing run the effective area of the SXS was reduced and the two SGD detectors were either in their turn-on phase or no data was being recorded \citep{hitomi18}.   Therefore, observations of these instruments are not incorporated in our analysis.  We report on the re-processing and re-analysis of the data obtained with the SXI and both HXI detectors in the $0.8-80$\,keV energy range. 

The data reduction was performed following the \hitomi\ step-by-step analysis guide version 6.1, using the \hitomi\ software version 6, as incorporated in version 6.26.1 of the HEAsoft tools\footnote[5]{\url{https://heasarc.gsfc.nasa.gov/docs/hitomi/analysis/}}.
Updated calibration tools were applied using the \hitomi\ CALDB version 10, released on 15 February 2018. 
With an angular resolution of the HXI detectors of $<1.7$\arcmin\ and the SXI detector of $<1.3$\arcmin\ \citep{takahashi2014}, PWN \pwn\ is not spatially resolved and hence forward analyzed as a point source.  
The HXI detectors were treated as independent instruments and the data analyzed separately. The event files of the HXI1 and HXI2 detector were merged prior to source and background selection. Source and (off-source) background regions, as provided by the analysis guide, were inspected and applied to the data.  Even though the \hitomi\ analysis guide notions that the off-source background spectrum may still include some source emission, affecting the derived flux, no non-X-ray Background (NXB) spectrum is available, leaving an off-source background extraction as sole solution.
This background region comprises the entire FoV, minus the source region.
The SXI event data were not merged before further reduction as the \hitomi\ team note in the analysis guide that cosmic ray echo effect varies between the ObsIDs and therefore separate RMF and ARF files should be created.
Accordingly, the data was reduced individually where only events detected during the non-``minus-Z day earth (MZDYE)" were selected to exclude light leakage affected events \citep{nakajima2018}.
Subsequently, spectra and responses were co-added using the ftool {\sf addascaspec}.
Likewise the HXI detectors, for the SXI observations the source and background regions as provided by the analysis guide were inspected and applied to the data. 
For this detector this implies the full FoV, with the calibration sources and their read-out streaks, some point sources, and the science source, excluded.

%%%%%%%%% X-RAY SPECTRUM PWN %%%%%%%%%%
\section{X-ray spectral analysis}
\label{sec:xray-analysis}

Since the source is a composite SNR, the X-ray spectrum of the PWN is superimposed on the emission arising from the SNR and central pulsar.
Only \chandra, with its superior angular resolution, is capable of spatially distinguishing the emission coming from each component (see Figure \ref{fig:g21_indiv_components}).
Recently, \cite{guest19} analyzed all \chandra\ data on this source to describe the spectrum of each substructure of the remnant.
To obtain the X-ray spectrum of the PWN observed with \nustar\ and \hitomi, we therefore include the obtained parameters of each substructure observed with \chandra\ (see Table \ref{table:chandra_guest}), leaving us with the `pure' PWN spectrum.
Here, we first report on the general X-ray analysis performed on all data, after which the results are presented. 

\begin{table}[tb]
%\begin{center}
\caption{Spectral parameters of the substructures observed in the remnant as derived by \citet{guest19}.}
\label{table:chandra_guest}
\begin{tabular}{lc}
\hline\hline
\multicolumn{2}{c}{Northern knot} \\
Photon Index ($\Gamma$) & 2.24 \\
Normalization & $2.51\times10^{-4}$ \\
\hline
\multicolumn{2}{c}{Eastern Limb}\\
Photon Index ($\Gamma$) & 2.22\\
Normalization & $3.76\times10^{-4}$\\
\hline
\multicolumn{2}{c}{PSR J$1833-1034$ (without Black body)}\\
Photon Index ($\Gamma$) & 1.54\\
Normalization & $8.34\times10^{-4}$\\
\hline
\multicolumn{2}{c}{PSR J$1833-1034$ (with Black body)} \\
Photon Index ($\Gamma$) & 1.35 \\
Normalization & $6.14\times10^{-4}$\\
kT (keV) & 0.43 \\
Normalization (BB) & $5.74\times10^{-6}$\\
\hline
\hline
\end{tabular}
%\end{center}
\end{table}

\subsection{X-ray Fitting Procedure}
\label{sec:x-ray_analysis}

After source extraction, each spectra was grouped to $>$20 counts per bin in the low energy range ($<$ 20\,keV) and to $>$100 counts per bin at higher energies ($>$ 20\,keV). Increasing the minimum grouping from 20 to 100 counts per bin at the higher energies had no effect on the fit parameters because of the robustness of the $C$ statistic in dealing with bins containing few counts \citep{cash}. 
The background and instrumental response corrected spectra were then analyzed using {\sf XSPEC} v12.10.1m \citep{xspec}. 

To obtain the spectrum of the PWN from \nustar\ and \hitomi\ spectra, we fit the source spectrum including the best-fit parameters of the substructures in \pwn\, reported by \cite{guest19}.
These components consist of the central pulsar PSR\,\psr, the limb-brightened eastern limb of the remnant, and the northern knot (see Table \ref{table:chandra_guest} for the spectral parameters of these components). As the source region for \chandra\ spectra include only the pulsar and PWN, the eastern limb and northern know components were not needed.   
For the source as a whole, the hydrogen column density is found to be $N_{\rm{H}} = 3.237\times 10^{22}$\,cm$^{-2}$ \citep{guest19}.

When fitting the pulsar component, \citet{guest19} find an improvement in their fit statistics when they include a black-body component {\sf bbody} to the power-law spectrum of the pulsar. 
However, since this improvement is marginal, we fit for the PWN spectrum both with and without including the pulsar black-body component. 

After the above mentioned components were fixed, the PWN spectrum was fit as a {\sf pegpwrlw} in which the photon index ($\Gamma$) and normalisation remain free. We chose the power-law model {\sf pegpwrlw} over the regular power-law model as it mitigates the issue of having a strong correlation between the photon index and normalization by using the unabsorbed flux between two specified energy ranges as its normalization \citep{pegpwrlw}. The absorption is treated using the Tuebingen-Bolder ISM absorption model, incorporated in {\sf XSPEC} as the {\sf tbabs} procedure, with solar abundances set to {\sf wilms} \citep{tbabs}. As mentioned at the beginning of this subsection, we set the fit statistic to {\sf cstat}. 

To obtain the uncertainties for the fit parameters (photon index and normalization) we opted to use {\sf XSPEC}'s Markov Chain Monte Carlo (MCMC) method. We followed the {\sf XSPEC} example of using the Goodman-Weare algorithm \citep{goodmanweare} with 8 walkers and a chain length of 10,000 steps.

\subsection{Piecewise Power-law Fits}
\label{sec:piecewise}
Theoretical models for the radiative evolution of a PWN (\citet{gelfand09}, \citet{torres14}) predict that the resultant spectrum is smoothly curving in the X-ray waveband.
As a result, while the broken power-law commonly used to describe this curvature does a reasonable job at indicating the turnover point in the spectrum it is not physically motivated. In addition, the location of this `break' is highly responsive to the boundaries of the observed energy range. This effect is demonstrated by the analysis of PWN \pwn\ where the different X-ray observatories, covering different energy ranges, report a different break energy (see Table \ref{table:prev_xray_obs}).
To better explore this curvature (i.e., change in photon index over the X-ray band), we propose to fit the PWN spectrum using {\it piecewise power laws} instead of the standard broken power law.
In this approach we split the total energy range we fit for into multiple contiguous and continuous energy bands. As we fit a power law in each energy band separately, we obtain a set of parameters and associated uncertainties in each energy band. We believe that in lieu of a PWN model that can accurately parametrize the smoothly curving nature of the spectrum, investigating the variation of the power-law parameters over distinct energy bands using the piecewise power-law approach is a valid and useful approach.  
 
We approach the piecewise power-law fitting by choosing energy bands that are roughly equal in log-space, contain sufficient counts, and are defined such that comparison between instruments is feasible.
We end up with the energy bands: 0.8--3.0\,keV (where \chandra\ and \hitomi\ SXI overlap), 3.0--8.0\,keV (where \chandra\, \hitomi\ SXI, \nustar\ detectors overlap), 8--20\,keV (where \nustar\ and \hitomi\ HXI detectors overlap), and 20--45\,keV (where \nustar\ and the \hitomi\ HXI detectors overlap). 

\begin{figure}[tb]
    \includegraphics[width=0.48\textwidth]{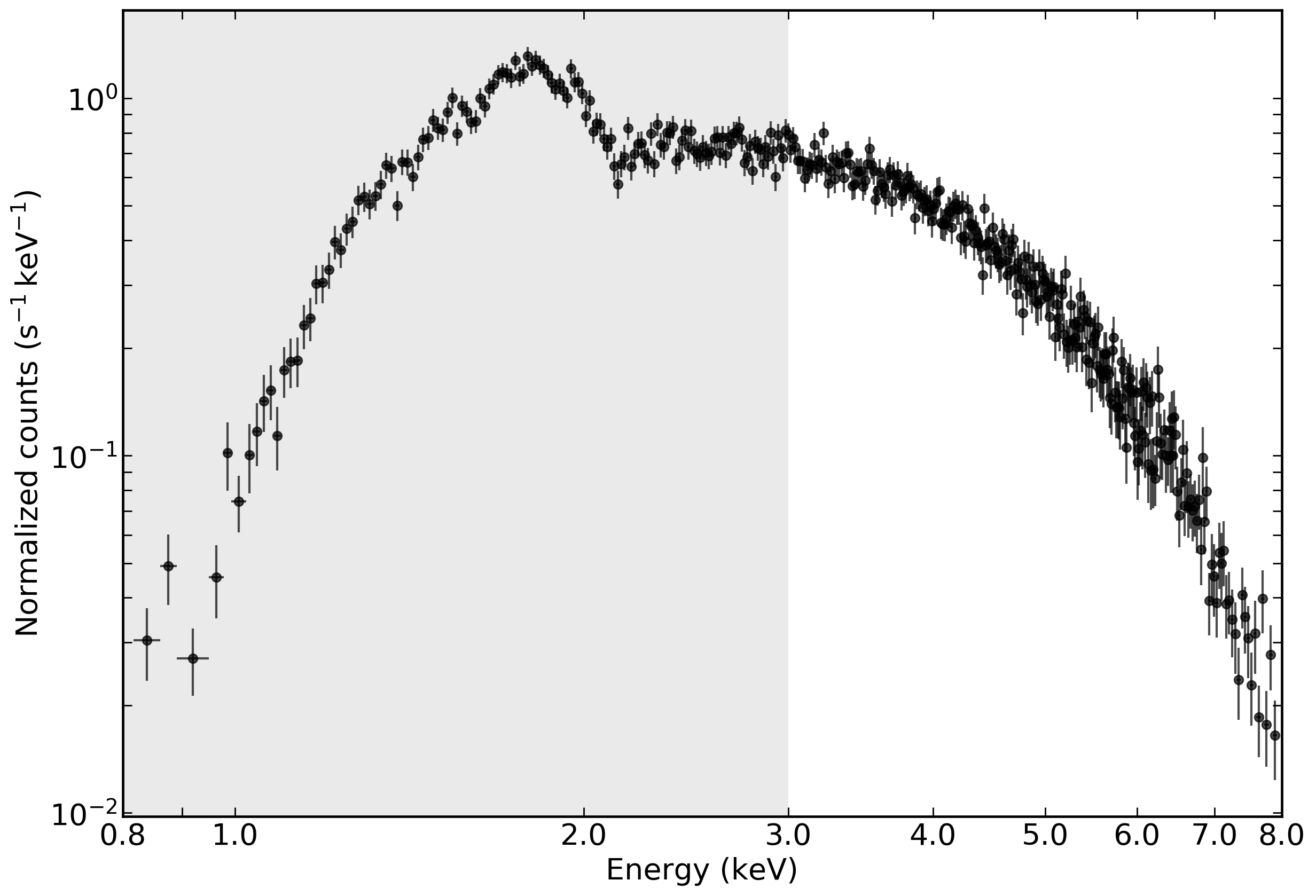}
    \caption{Spectrum of the \chandra\ ACIS detector in the 0.8--8 keV range. 
    Only the spectra from the longest observation (ObsID 1433) is shown.
    The grey and white shaded regions show the different fitting energy bands: 0.8--3\,keV, 3--8\,keV.}
    \label{fig:spectrum_chandra}
\end{figure}

The \chandra\ data were fit in the 0.8--3.0 and 3.0--8.0\,keV energy bands. The spectrum from the longest observation is shown in Figure \ref{fig:spectrum_chandra}. 

\begin{figure}[tb]
    \includegraphics[width=0.48\textwidth]{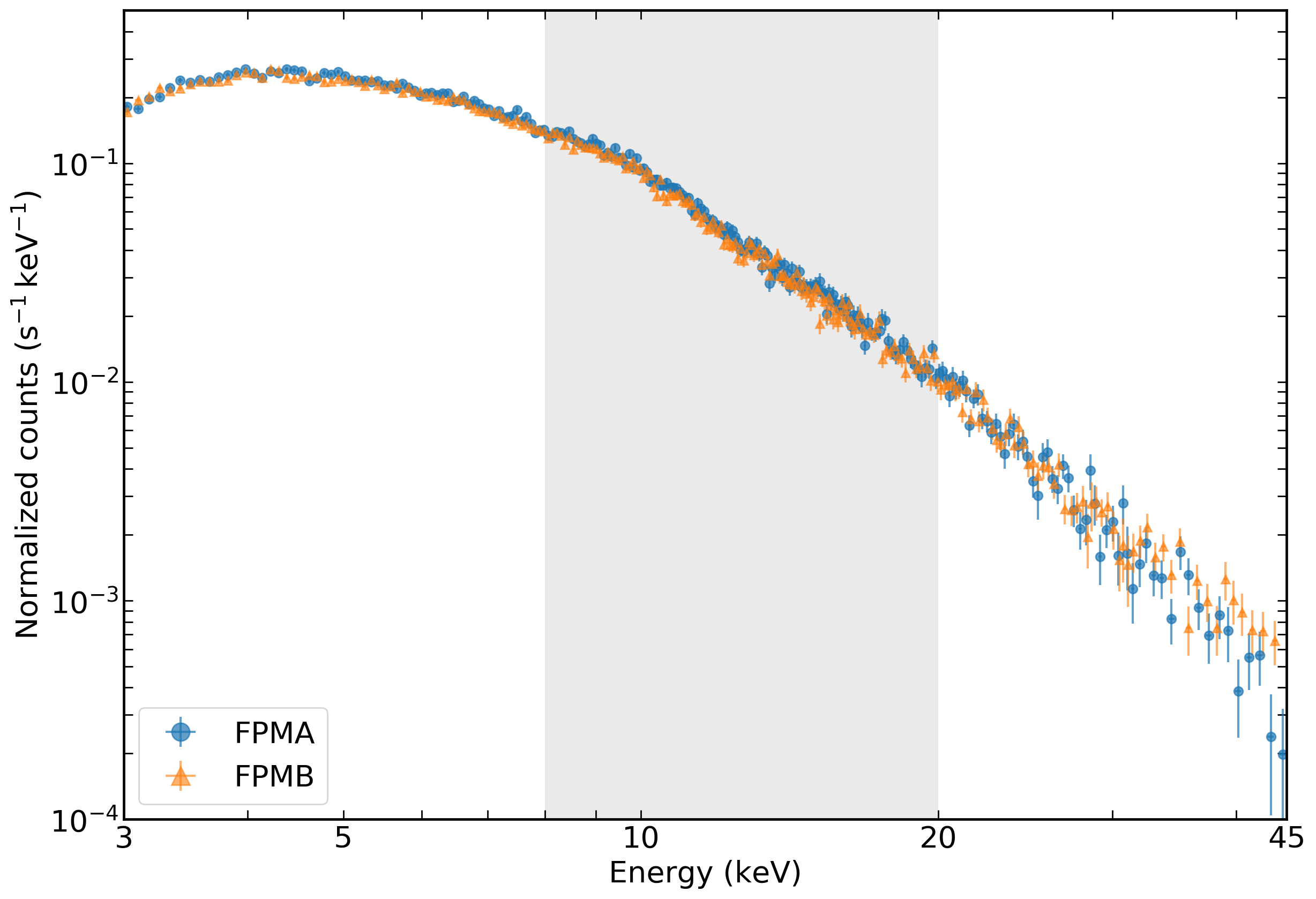}
    \caption{Wide-band spectra of the \nustar\ FPMA (blue) and FPMB (orange) detectors in the 3--45 keV range. 
    Only the spectra from the longest observation (ObsID 10002014005) are shown to prevent overcrowding the figure.
    The alternating white-grey bands show the different fitting energy bands: 3--8\,keV, 8--20\,keV (grey), 20--45\,keV.}
    \label{fig:spectrum_nustar}
\end{figure}

The \nustar\ data were fit in the 3.0--8.0\,keV, 8--20\,keV, and 20--45\,keV energy bands.  While \nustar\ operates in the 3--79\,keV range, the spectral fitting for this source was restricted to the 3--45\,keV range as the background dominates above 45\,keV. In addition, the spectra from FPMA and FPMB were fit separately as we noticed a consistent difference in the fit parameters when performing fits for each spectra independently. Specifically, we saw that the photon index $\Gamma$ was higher for FPMA spectra compared to FPMB spectra and that the unabsorbed flux was consistently higher for FPMB spectra compared to FPMA spectra. This discrepancy between the FPMA and FPMB spectra is described in Appendix \ref{sec:nustar_independent_fits}. 
This issue is unrelated to the discrepancy due to a thermal blanket tear for the FPMA detector causing an excess in low-energy photons \citep{madsen2020nustar} as the \nustar\ team believes the tear began in 2017, and all the observations analyzed in this study are from 2012 and 2013 (Table \ref{tab:nustar_obs}).
The spectra from the longest observation are shown in Figure \ref{fig:spectrum_nustar}.

\begin{figure}[tb]
  \centering
    \includegraphics[width=0.48\textwidth]{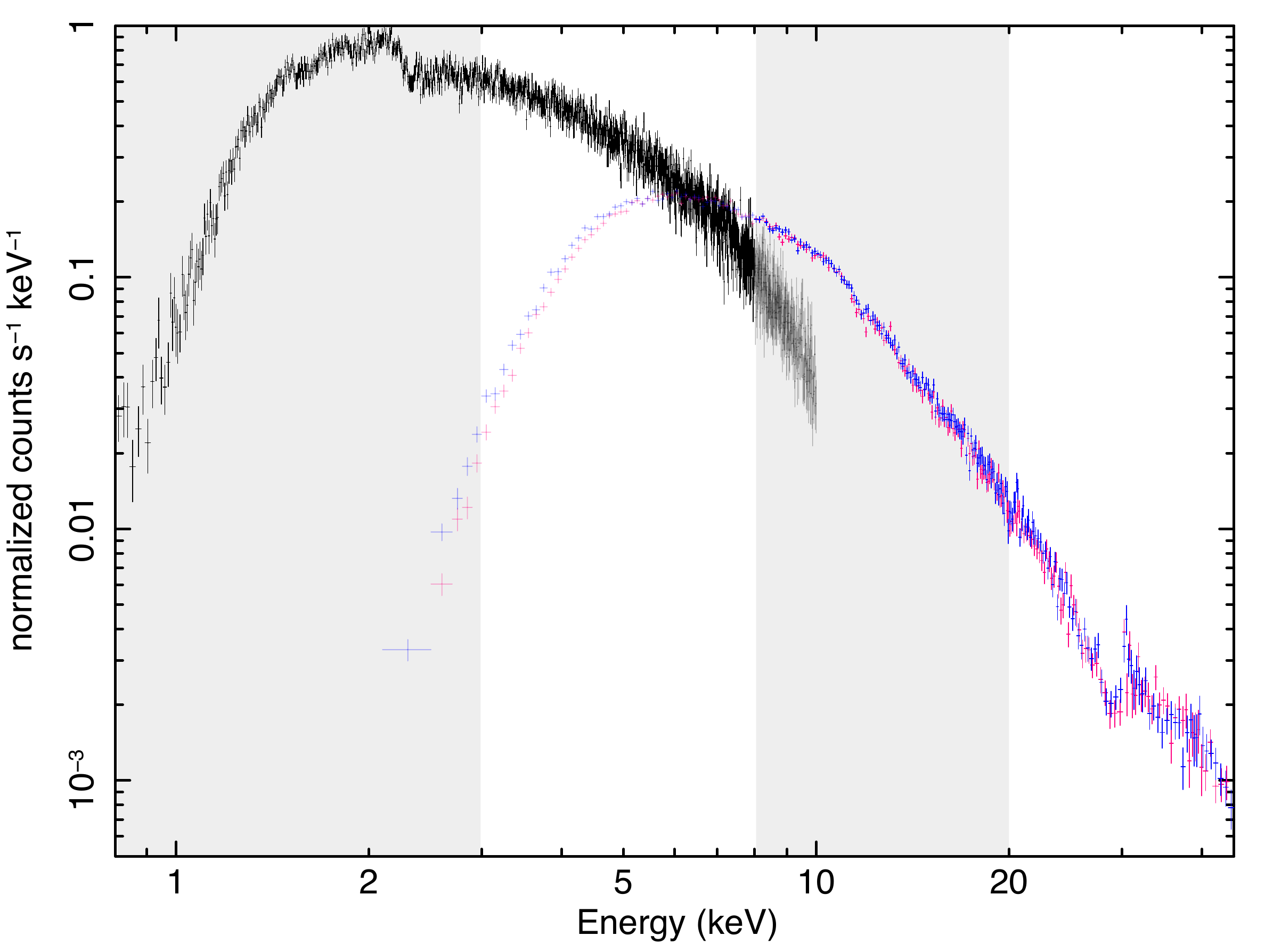}
   \caption{Wide-band spectra of the \hitomi\ SXI (black) and HXI detectors (HXI1 in blue and HXI2 in magenta) in the $0.8-45$\,keV energy range. The different fitting energy bands are indicated by the alternating grey bands. Data that is not fitted in a given energy band is made transparent. 
   \label{fig:spectrum_hitomi} }
\end{figure}

The data recorded by \hitomi\ spans the combined energy range of the \chandra\ and \nustar\ observations. Hence the \hitomi\ data were fit over all specified energy bands. 
Given that the full energy range of \hitomi\ is spread over two different type of detectors, we report on the results of each energy band for the respective detector sensitive to those energies. (see Tables \ref{table:gamma_all_instruments_bb}, \ref{table:gamma_all_instruments_noBB}, \ref{table:unabsorbed_flux_all_instruments_bb},  \ref{table:unabsorbed_flux_all_instruments_noBB}).
The observed spectra are shown in Figure \ref{fig:spectrum_hitomi}, where the detector spectrum relevant for each energy band is indicated.

\begin{figure*}[tbh]
    \centering
    \includegraphics[height=0.24\textheight]{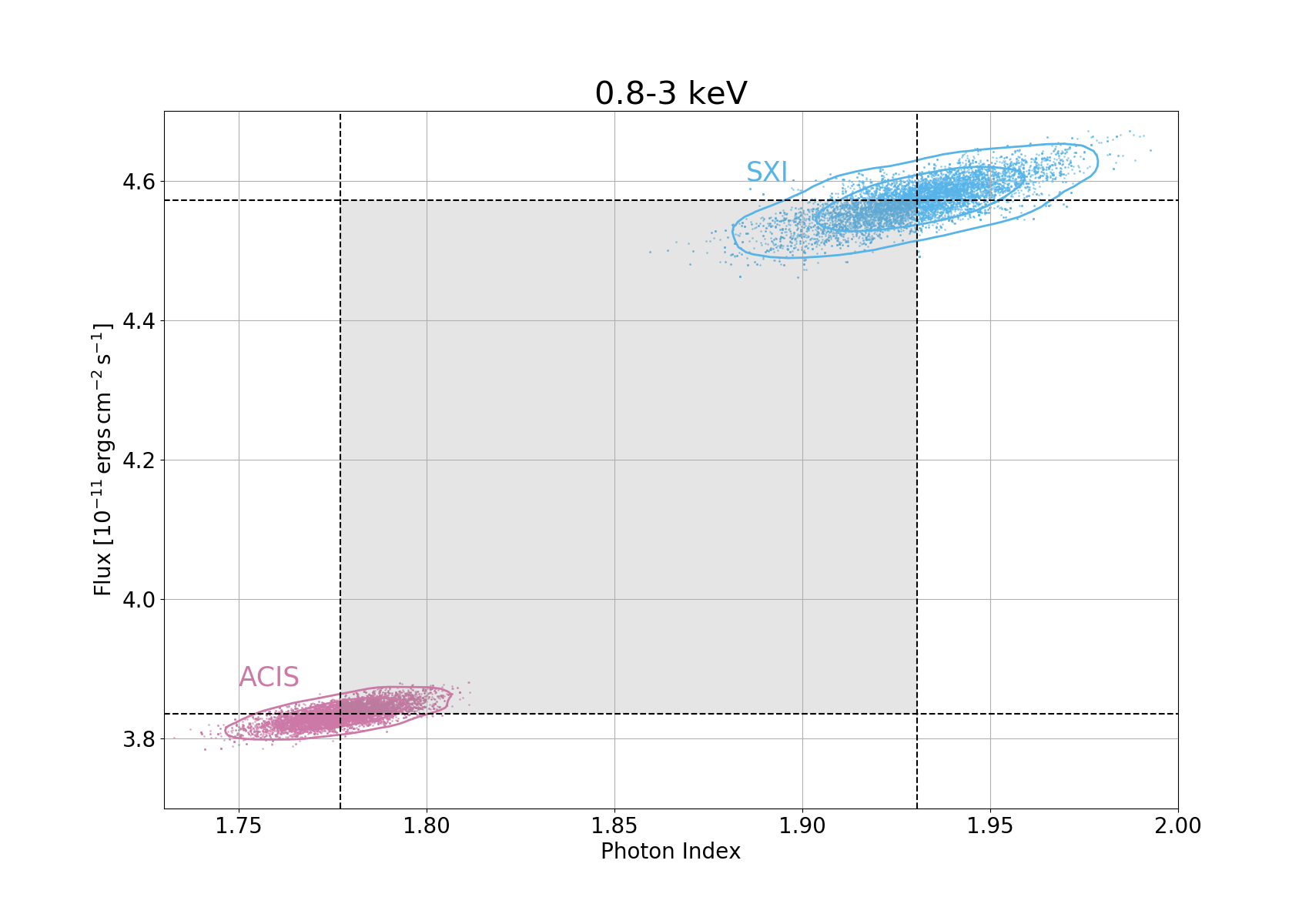}
    \includegraphics[height=0.24\textheight]{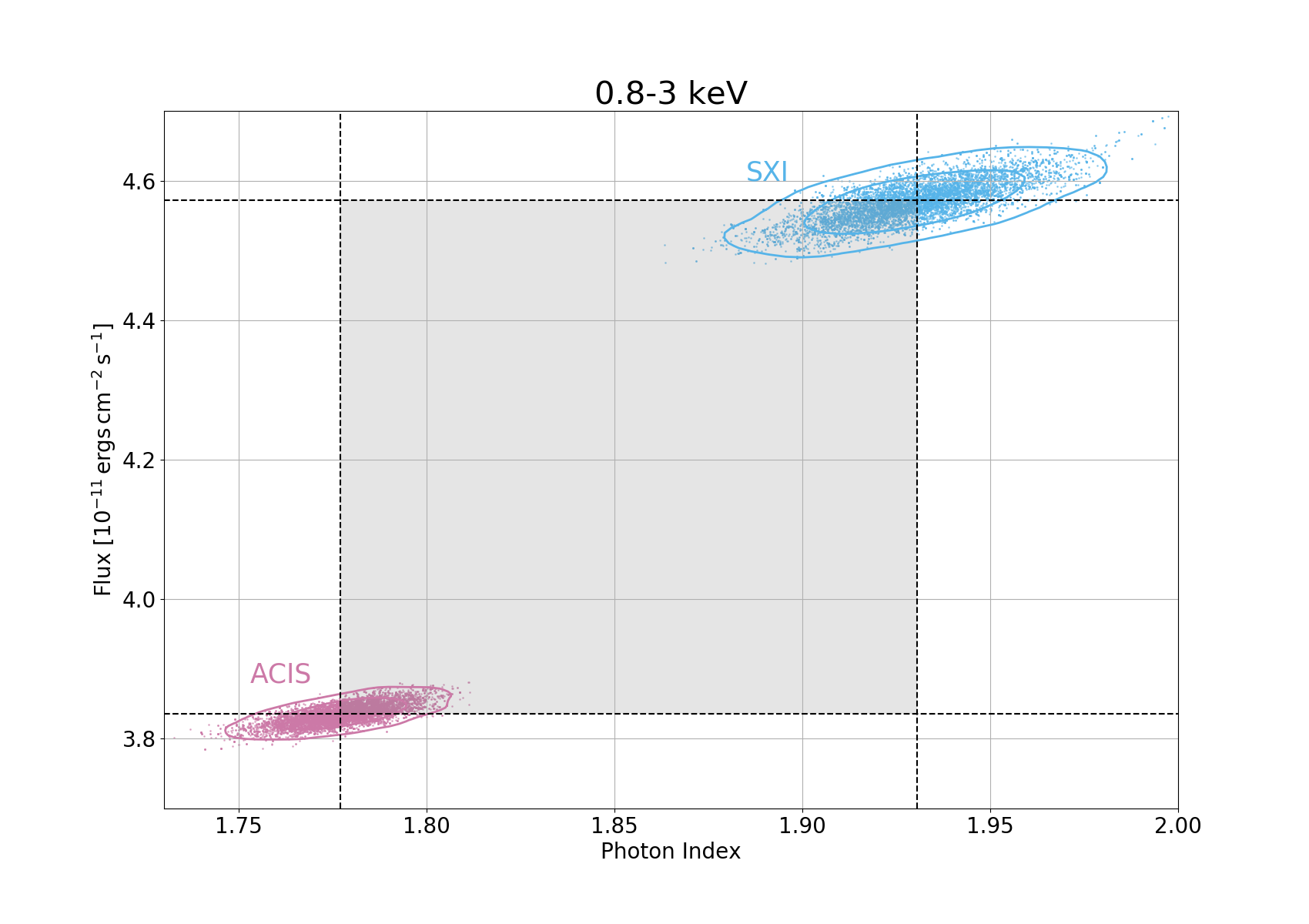} \\ 
    \includegraphics[height=0.24\textheight]{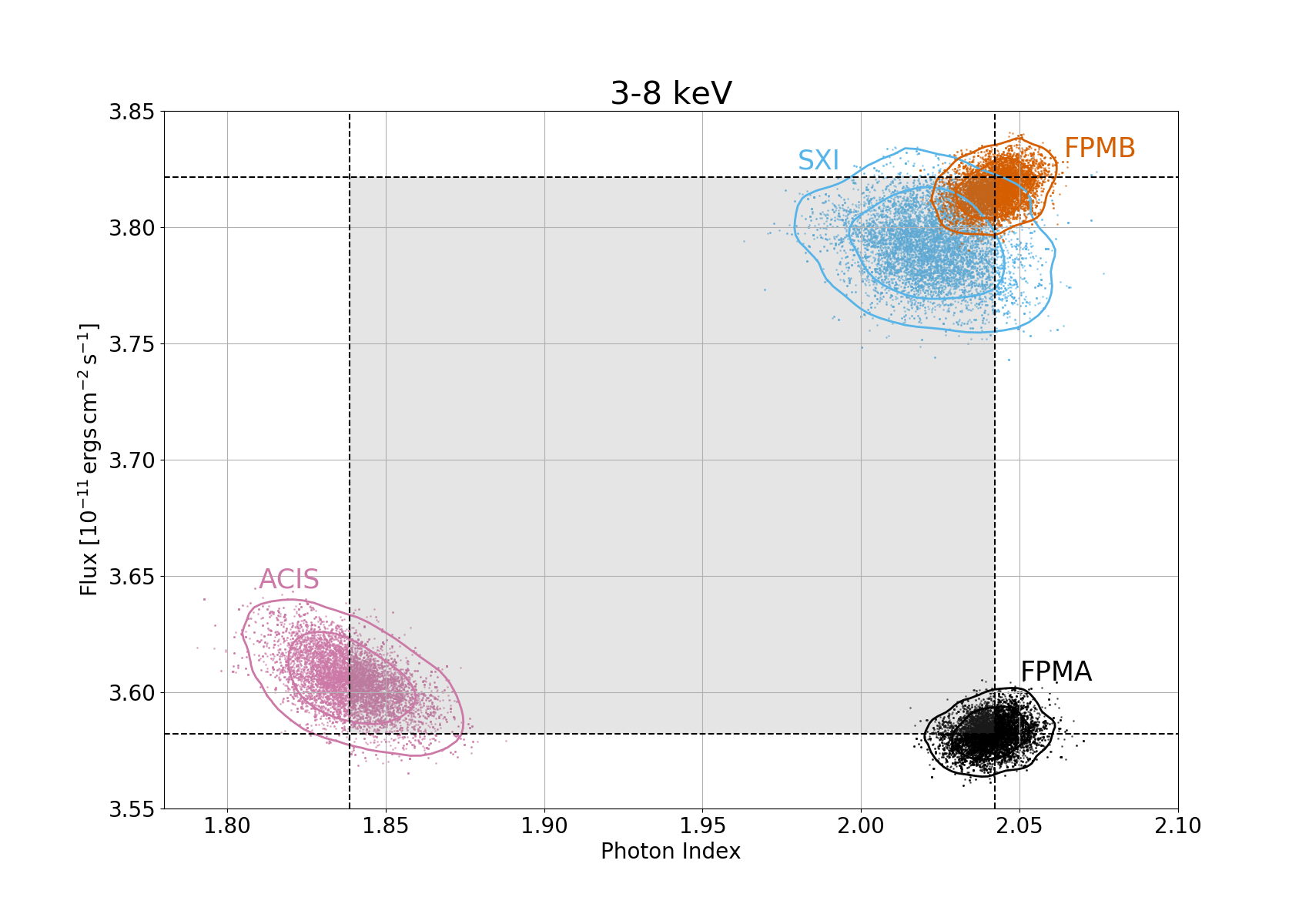}
    \includegraphics[height=0.24\textheight]{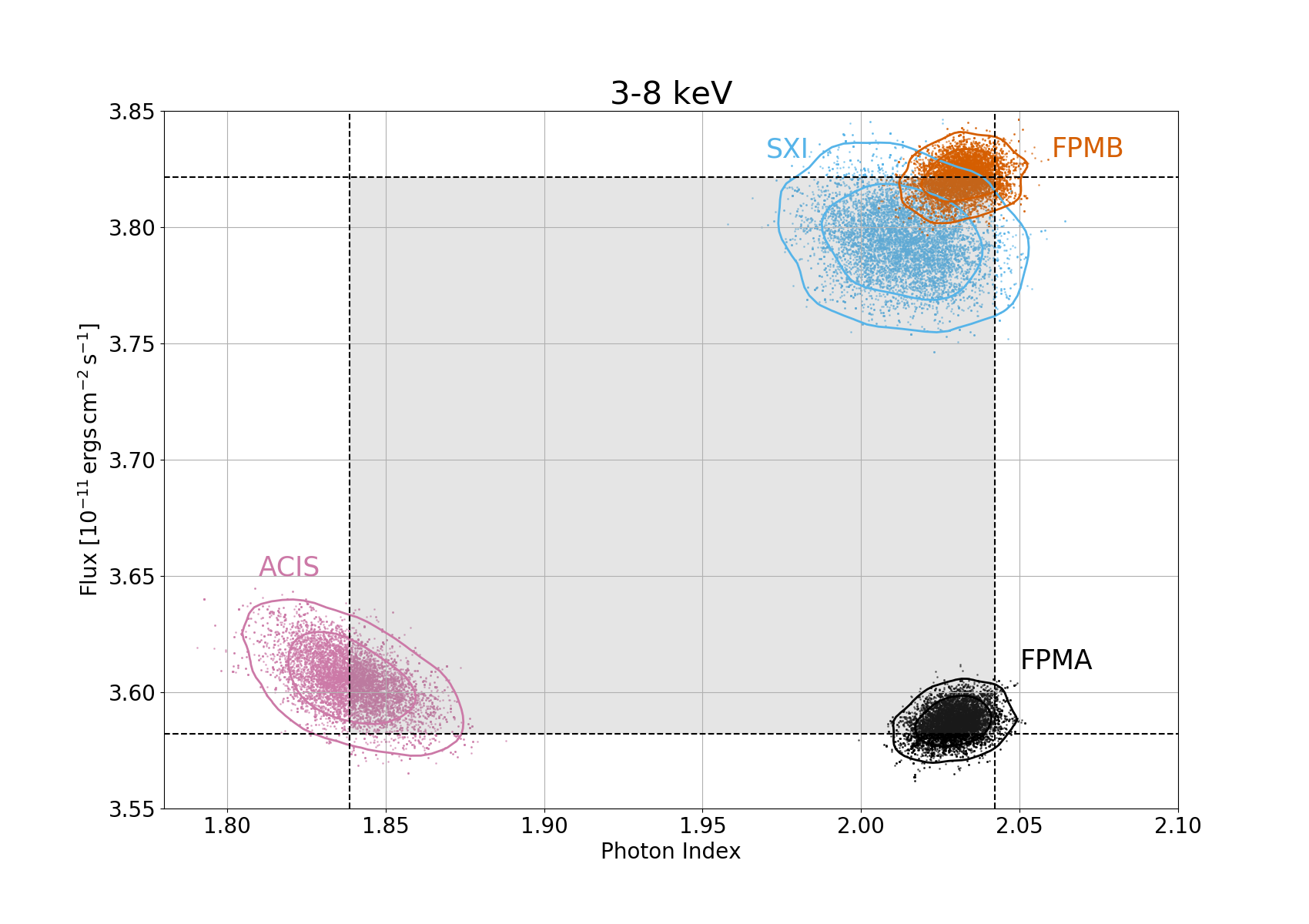} \\ 
    \includegraphics[height=0.24\textheight]{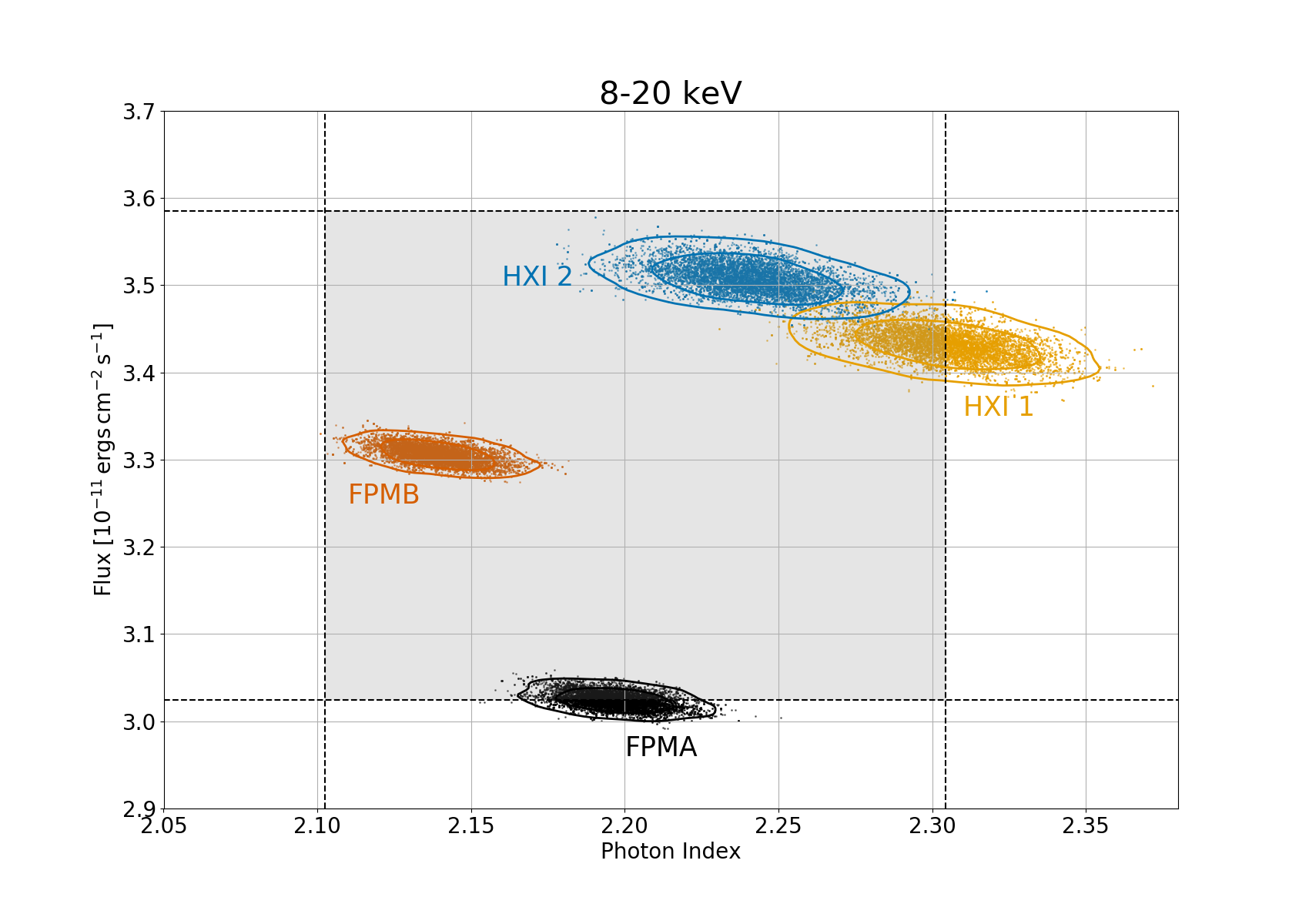}
    \includegraphics[height=0.24\textheight]{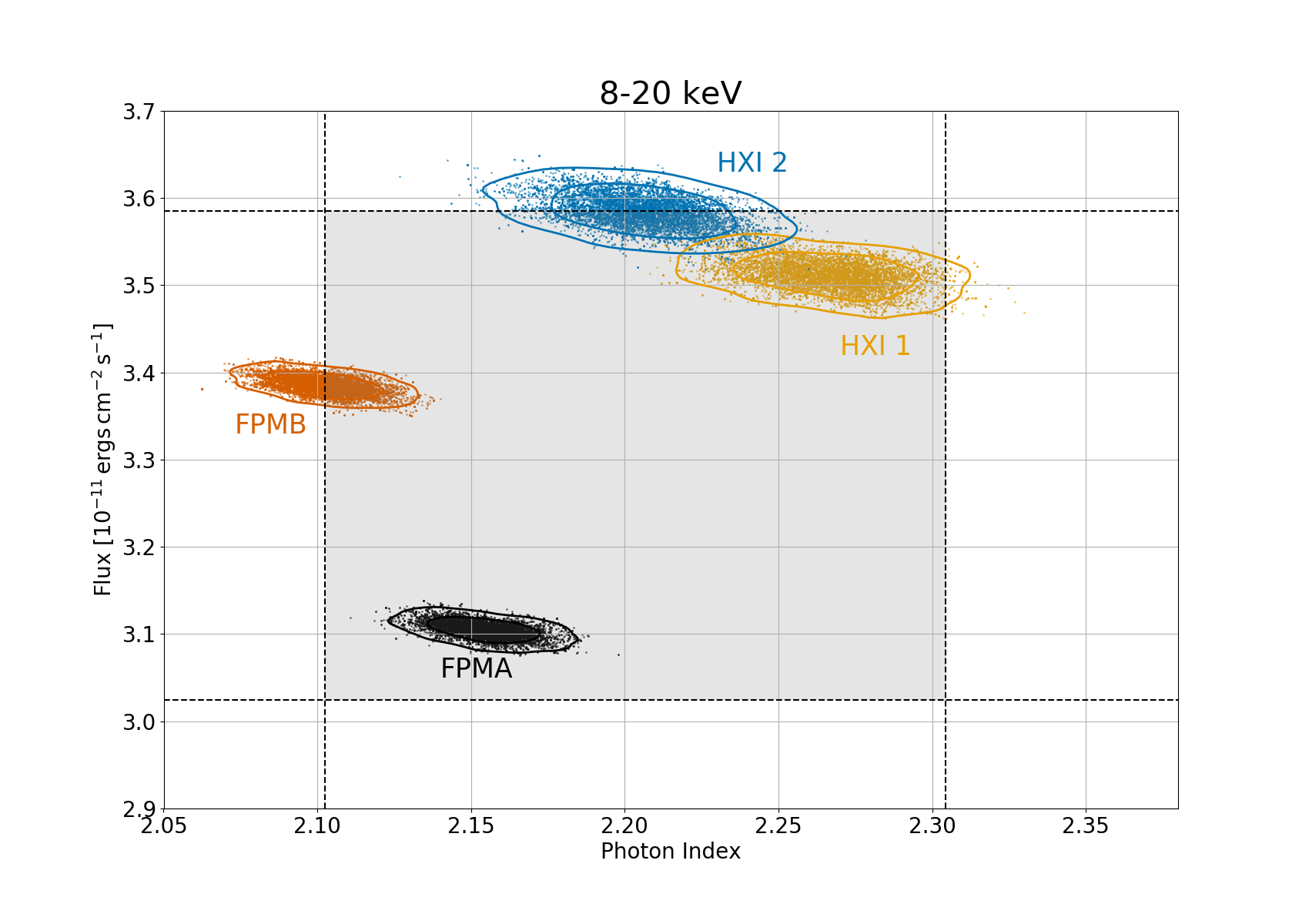} \\ 
    \includegraphics[height=0.24\textheight]{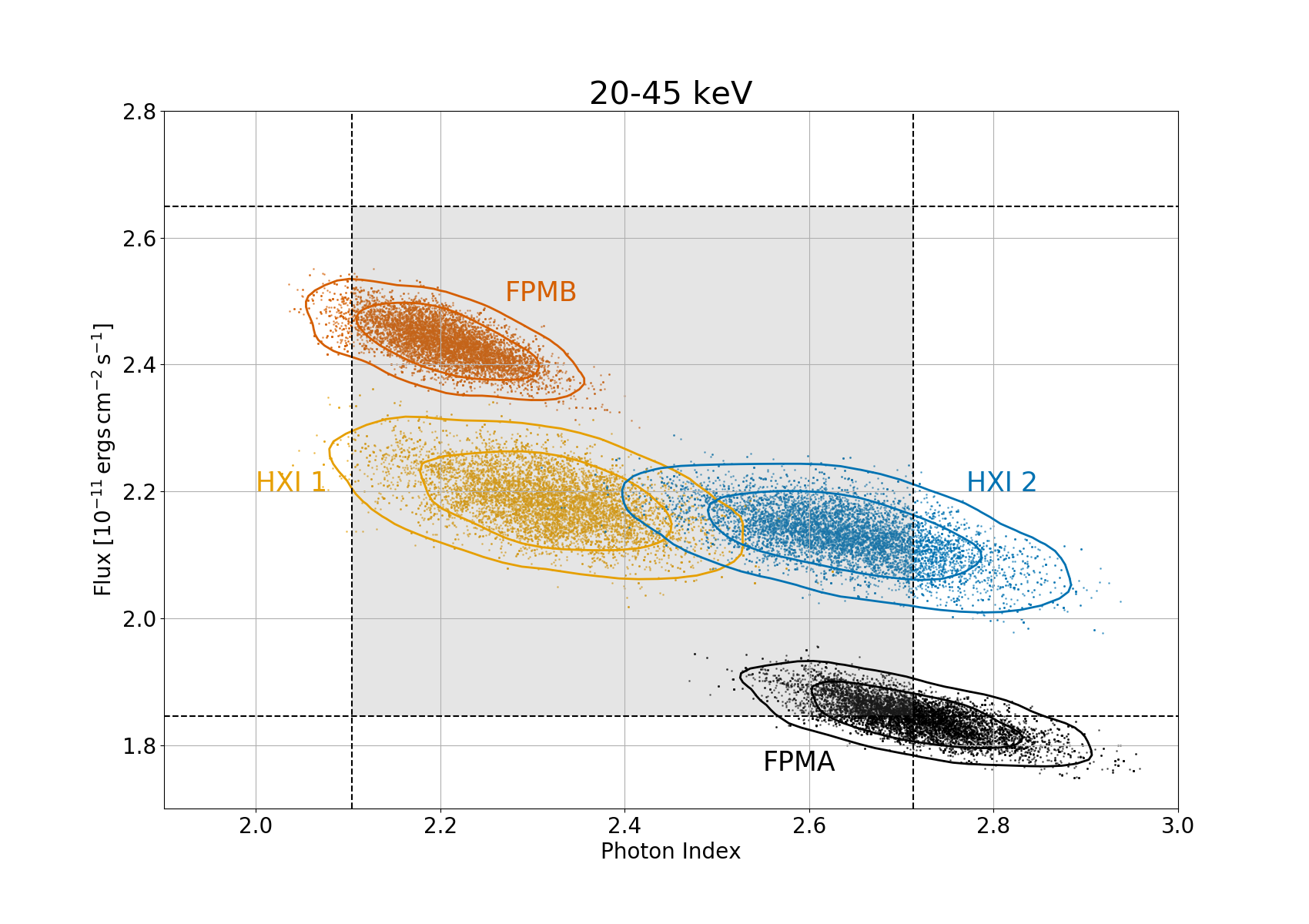}
    \includegraphics[height=0.24\textheight]{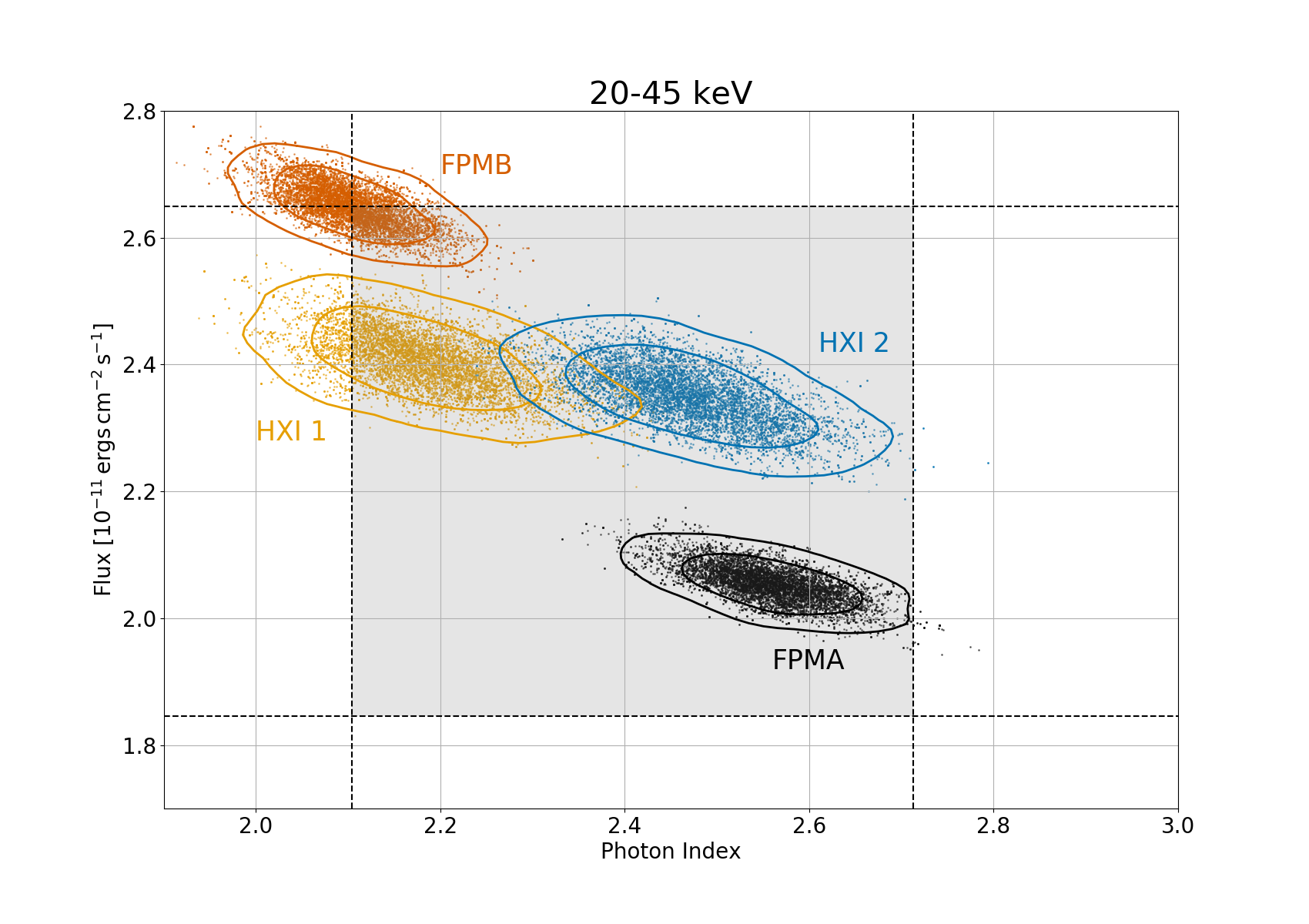} \\ 
    \caption{\replaced{
    Photon index $\Gamma$ and unabsorbed flux $F$ contour plots resulting from power-law fits to the X-ray spectrum.}{Scatter plots of the MCMC samples (Photon index $\Gamma$ and unabsorbed flux $F$).} The contours are drawn at the regions containing 68\% and 95\% of the samples. Each row shows a different energy range in increasing order from top to bottom. The left column is for the model with the pulsar black-body component and the right column is for the model without the black-body component. The \added{gray} shaded areas are consistent within a row and \replaced{its}{their} role is explained in. \ref{sec:pwnmodel}.}
    \label{fig:contour_plots}
\end{figure*}

\begin{figure*}[tbh]
    \centering
    \includegraphics[width=0.49\textwidth]{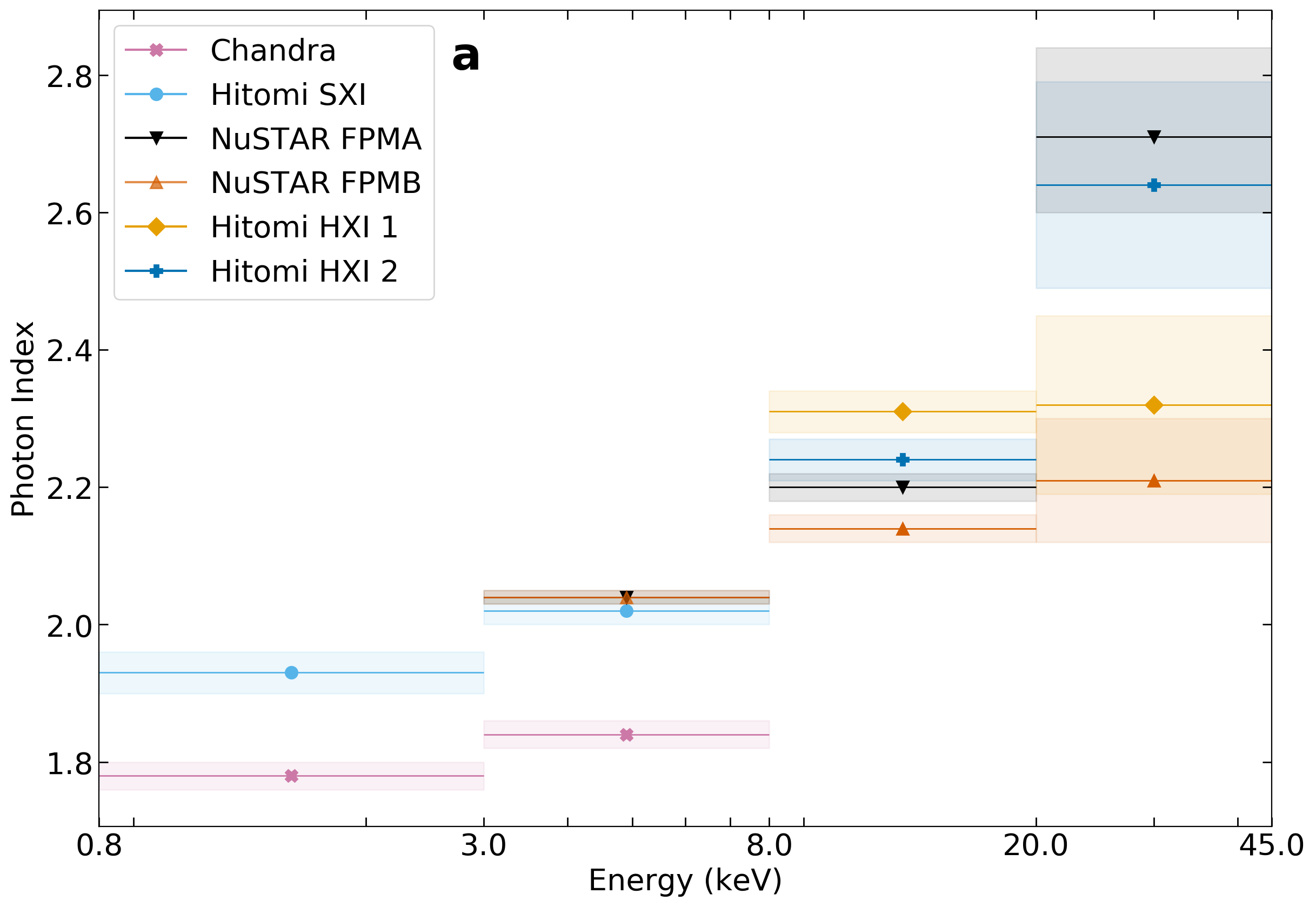}
    \includegraphics[width=0.49\textwidth]{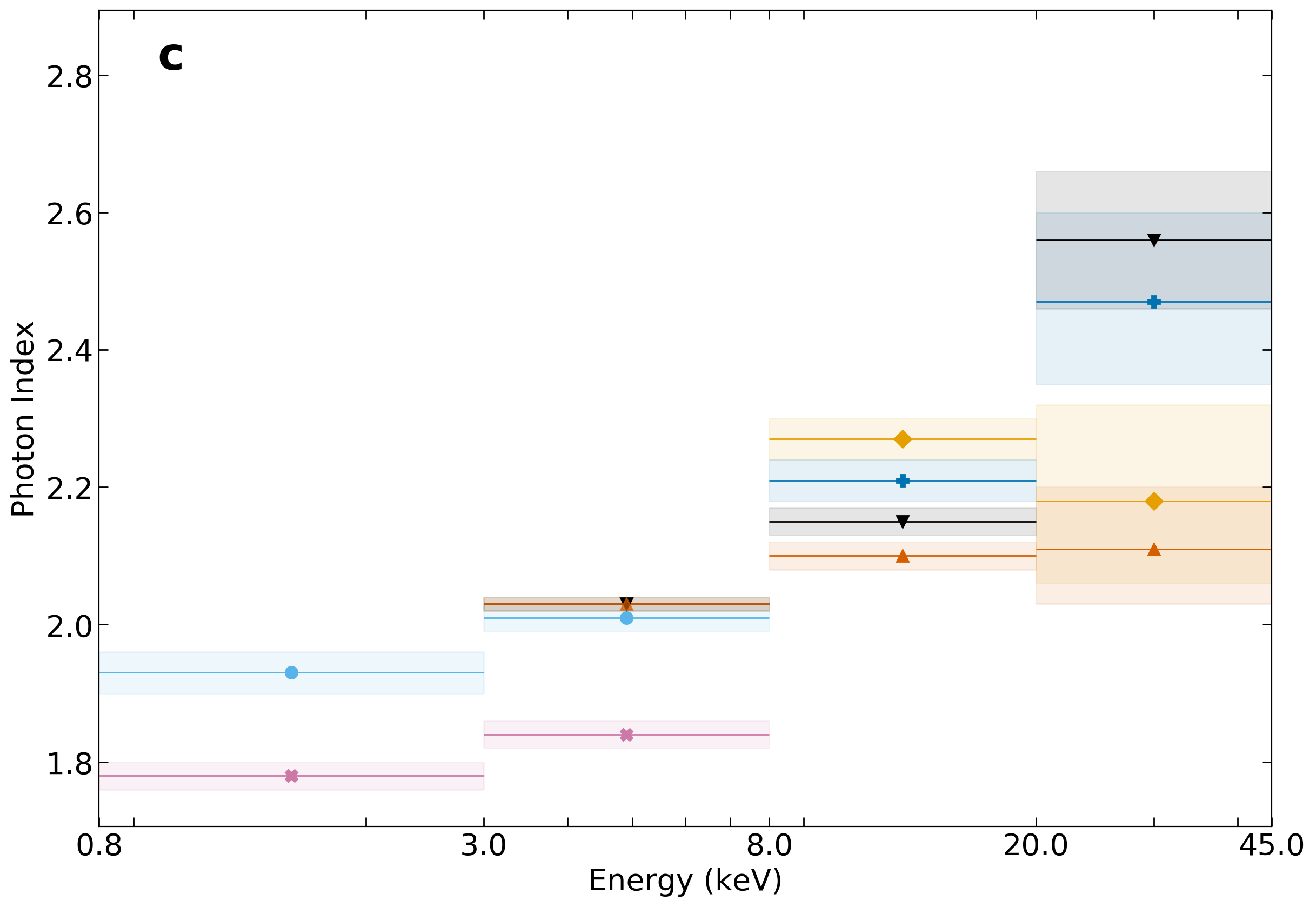} \\ 
    \includegraphics[width=0.49\textwidth]{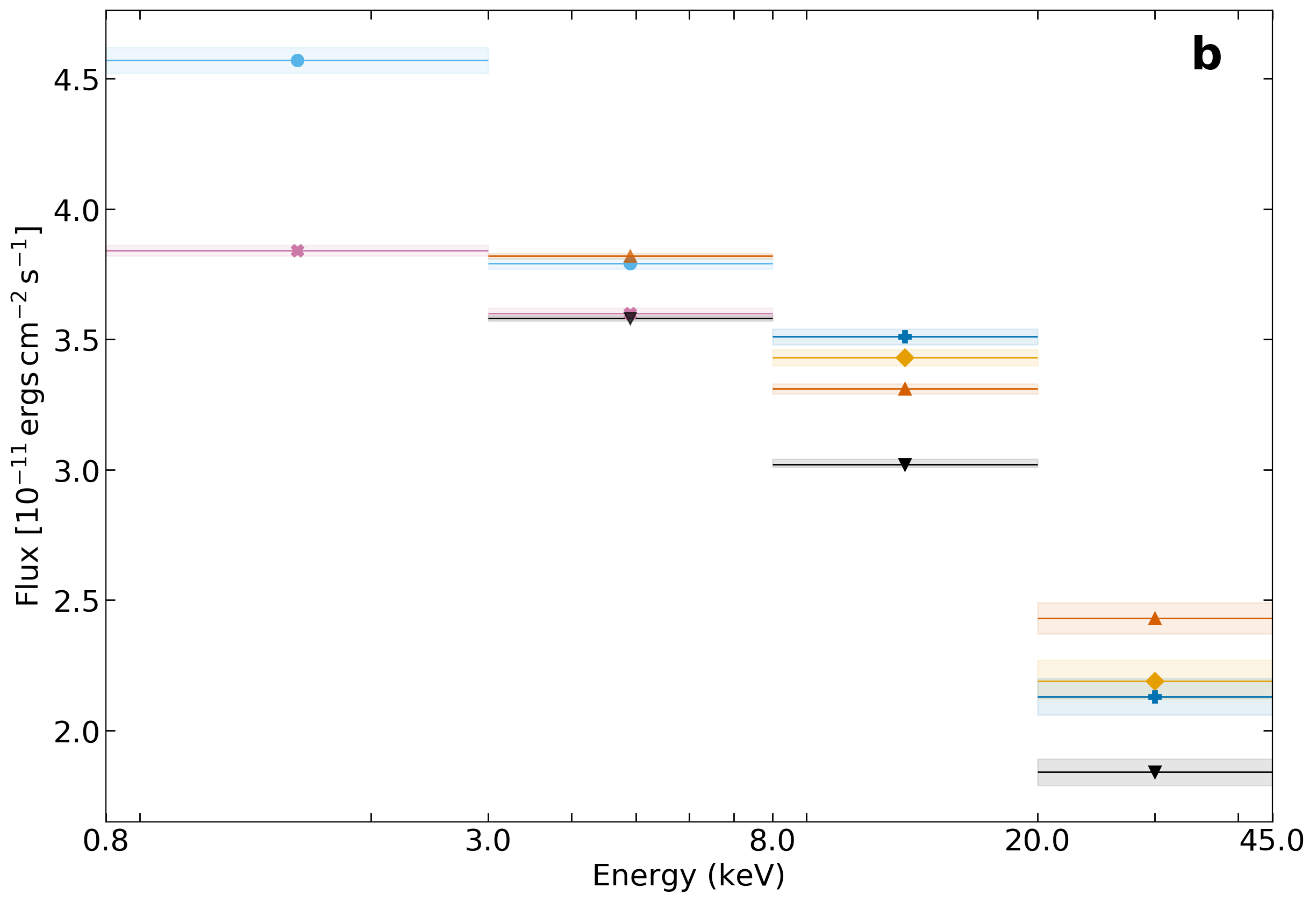}
    \includegraphics[width=0.49\textwidth]{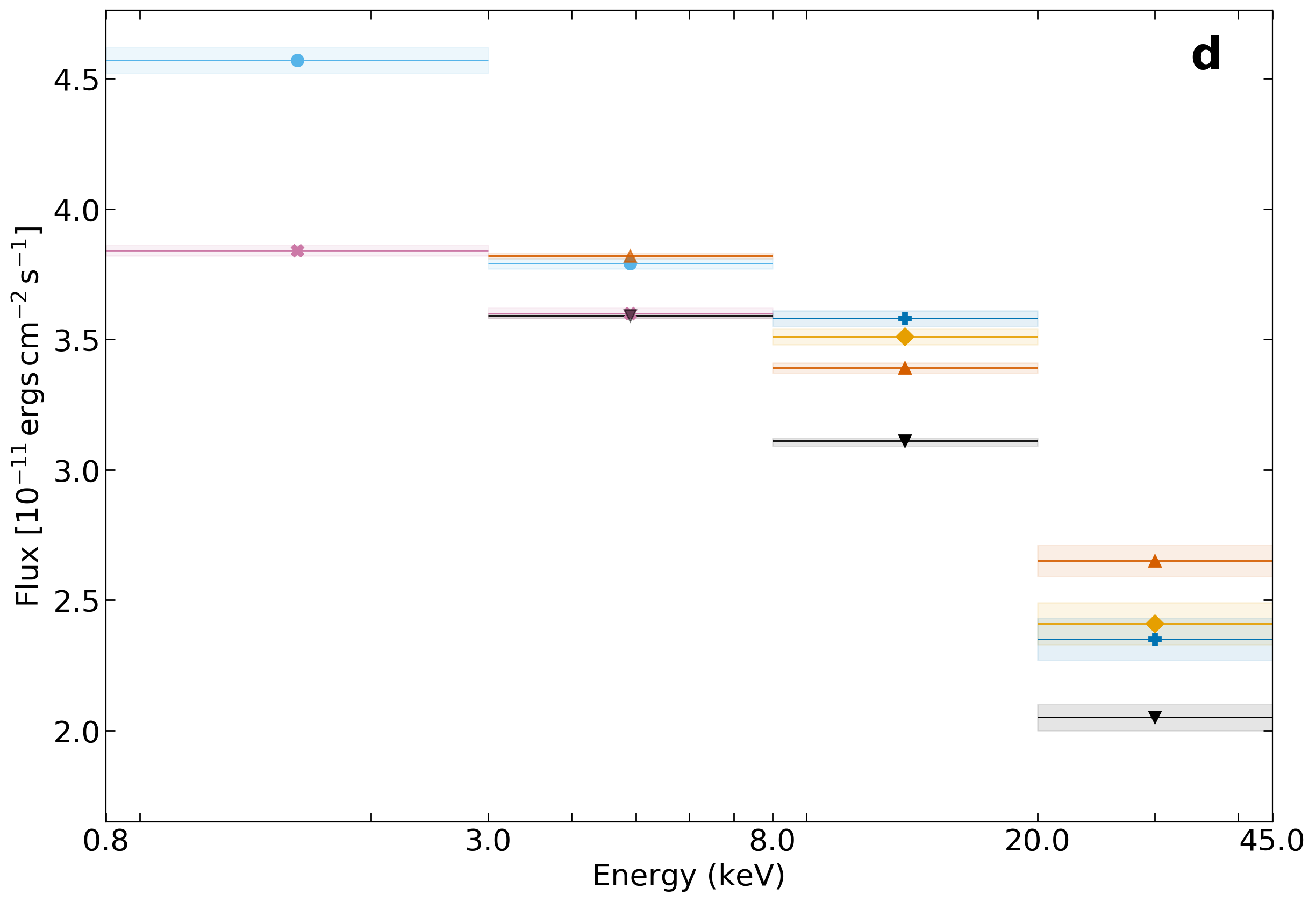} \\ 

    \caption{Plots showing the PWN fit parameter values for the detectors in each energy range. Plots a, b are the photon indices and \added{unabsorbed} flux values for the model {\it with} the black-body component, and plots c, d are the photon indices and \added{unabsorbed} flux values for the model {\it without} the black-body component. The shaded regions indicate the 90\% confidence intervals consistent with those given in the tables.}
    \label{fig:par_vs_energy}
\end{figure*}

\begin{table*}[t]
% \caption{The photon index $\Gamma$ per energy range as determined for the different detectors. Values are obtained following our piece-wise modelling and applying the model {\it with} the pulsar black-body component. Errors reflect the 90\% confidence intervals.}
\caption{PWN Photon index $\Gamma$ per energy range for each detector when fitting the model \textit{with} a pulsar black-body component\tablenotemark{\scriptsize{a}}}
\label{table:gamma_all_instruments_bb}
\vspace*{-0.5cm}
\begin{center}
\begin{tabular}{ccccccc}
\hline
\hline
Energy Range [keV] & Chandra & Hitomi SXI & Hitomi HXI 1 & Hitomi HXI 2 & NuSTAR FPMA & NuSTAR FPMB \\ \hline
0.8 -- 3.0 & $1.78\substack{+0.02 \\ -0.02}$ & $1.93\substack{+0.03 \\ -0.03}$ & -- & -- & -- & -- \\
3--8 & $1.84\substack{+0.02 \\ -0.02}$ & $2.02\substack{+0.02 \\ -0.02}$ & -- & -- & $2.04\substack{+0.01 \\ -0.01}$ & $2.04\substack{+0.01 \\ -0.01}$ \\
8--20 & -- & -- & $2.31\substack{+0.03 \\ -0.03}$ & $2.24\substack{+0.03 \\ -0.03}$ & $2.20\substack{+0.02 \\ -0.02}$ & $2.14\substack{+0.02 \\ -0.02}$ \\
20--45 & -- & -- & $2.32\substack{+0.13 \\ -0.14}$ & $2.64\substack{+0.15 \\ -0.15}$ & $2.71\substack{+0.13 \\ -0.11}$ & $2.21\substack{+0.09 \\ -0.09}$\\ 
\hline
\hline
\end{tabular}
\end{center}
\tablenotetext{\scriptsize{a}}{\scriptsize{Errors indicate 90\% confidence intervals}}
\end{table*}

\begin{table*}[t]
% \caption{The photon index $\Gamma$ per energy range as determined for the different detectors. Values are obtained following our piece-wise modelling and applying the model {\it without} the pulsar black-body component. Errors reflect 90\% confidence.}
\caption{PWN Photon index $\Gamma$ per energy range for each detector when fitting the model \textit{without} a pulsar black-body component\tablenotemark{\scriptsize{a}}}
\label{table:gamma_all_instruments_noBB}
\vspace*{-0.5cm}
\begin{center}
\begin{tabular}{ccccccc}
\hline
\hline
Energy Range {[}keV{]} & Chandra & Hitomi SXI & Hitomi HXI 1 & Hitomi HXI 2 & NuSTAR FPMA & NuSTAR FPMB \\
\hline
0.8--3.0 & $1.78\substack{+0.02 \\ -0.02}$ & $1.93\substack{+0.03 \\ -0.03}$ & -- & -- & -- & -- \\
3--8 & $1.84\substack{+0.02 \\ -0.02}$ & $2.01\substack{+0.02 \\ -0.02}$ & -- & -- & $2.03\substack{+0.01 \\ -0.01}$ & $2.03\substack{+0.01 \\ -0.01}$ \\
8--20 &  & -- & $2.27\substack{+0.03 \\ -0.03}$ & $2.21\substack{+0.03 \\ -0.03}$ & $2.15\substack{+0.02 \\ -0.02}$ & $2.10\substack{+0.02 \\ -0.02}$ \\
20--45 &  & -- & $2.18\substack{+0.14 \\ -0.12}$ & $2.47\substack{+0.13 \\ -0.12}$ & $2.56\substack{+0.10 \\ -0.10}$ & $2.11\substack{+0.09 \\ -0.08}$\\ 
\hline
\hline
\end{tabular}
\end{center}
\tablenotetext{\scriptsize{a}}{\scriptsize{Errors indicate 90\% confidence intervals}}
\end{table*}

\begin{table*}[t]
% \caption{The unabsorbed flux (given in units of $10^{-11} \mathrm{erg~s^{-1}~cm^{-2}}$) for the PWN component 
% %per energy range as determined for the different detectors. Values are obtained 
% following our piece-wise modelling and applying the model {\it with} the pulsar black-body component. Errors reflect 90\% confidence.}
\caption{PWN unabsorbed flux\tablenotemark{$\scriptstyle{\dagger}$} per energy range for each detector when fitting the model \textit{with} a pulsar black-body component\tablenotemark{\scriptsize{a}}}
\label{table:unabsorbed_flux_all_instruments_bb}
\vspace*{-0.5cm}
\begin{center}
\begin{tabular}{ccccccc}
\hline
\hline
Energy Range {[}keV{]} & Chandra & Hitomi SXI & Hitomi HXI 1 & Hitomi HXI 2 & NuSTAR FPMA & NuSTAR FPMB \\
\hline
0.8--3.0 & $3.84\substack{+0.02 \\ -0.02}$ & $4.57\substack{+0.05 \\ -0.05}$ & -- & -- & -- & -- \\
3--8 & $3.60\substack{+0.02 \\ -0.02}$ & $3.79\substack{+0.02 \\ -0.02}$ & -- & -- & $3.58\substack{+0.01 \\ -0.01}$ & $3.82\substack{+0.01 \\ -0.01}$ \\
8--20 & -- & -- & $3.43\substack{+0.03 \\ -0.03}$ & $3.51\substack{+0.03 \\ -0.03}$ & $3.02\substack{+0.02 \\ -0.01}$ & $3.31\substack{+0.02 \\ -0.02}$ \\
20--45 & -- & -- & $2.19\substack{+0.08 \\ -0.07}$ & $2.13\substack{+0.07 \\ -0.07}$ & $1.84\substack{+0.05 \\ -0.05}$ & $2.43\substack{+0.06 \\ -0.06}$ \\ 
\hline
\hline
\end{tabular}
\end{center}
\tablenotetext{\dagger}{\scriptsize{Units of $10^{-11}$\,\flux}}
\tablenotetext{\scriptsize{a}}{\scriptsize{Errors indicate 90\% confidence intervals}}
\end{table*}

\begin{table*}[th]
% \caption{The unabsorbed flux (given in units of $10^{-11}~\mathrm{erg~s^{-1}~cm^{-2}}$) for the PWN component
% %per energy range as determined for the different detectors. Values are obtained 
% following our piece-wise modelling and applying the model {\it without} the pulsar black-body component. Errors reflect 90\% confidence.}
\caption{PWN unabsorbed flux\tablenotemark{$\scriptstyle{\dagger}$} per energy range for each detector when fitting the model \textit{without} a pulsar black-body component\tablenotemark{\scriptsize{a}}}
\label{table:unabsorbed_flux_all_instruments_noBB}
\vspace*{-0.5cm}
\begin{center}
\begin{tabular}{ccccccc}
\hline
\hline
Energy Range {[}keV{]} & Chandra & Hitomi SXI & Hitomi HXI 1 & Hitomi HXI 2 & NuSTAR FPMA & NuSTAR FPMB \\
\hline
0.8--3.0 & $3.84\substack{+0.02 \\ -0.02}$ & $4.57\substack{+0.05 \\ -0.05}$ & -- & -- & -- & -- \\
3--8 & $3.60\substack{+0.02 \\ -0.02}$ & $3.79\substack{+0.03 \\ -0.02}$ & -- & -- & $3.59\substack{+0.01 \\ -0.01}$ & $3.82\substack{+0.01 \\ -0.01}$ \\
8--20 & -- & -- & $3.51\substack{+0.03 \\ -0.03}$ & $3.58\substack{+0.03 \\ -0.03}$ & $3.11\substack{+0.01 \\ -0.02}$ & $3.39\substack{+0.02 \\ -0.02}$ \\
20--45 & -- & -- & $2.41\substack{+0.08 \\ -0.08}$ & $2.35\substack{+0.08 \\ -0.08}$ & $2.05\substack{+0.05 \\ -0.05}$ & $2.65\substack{+0.06 \\ -0.06}$ \\ 
\hline
\hline
\end{tabular}
\end{center}
\tablenotetext{\dagger}{\scriptsize{Units of $10^{-11}$\,\flux}}
\tablenotetext{\scriptsize{a}}{\scriptsize{Errors indicate 90\% confidence intervals}}
\end{table*}

\subsection{Results}
\label{sec:results}
As mentioned in \ref{sec:x-ray_analysis}, we fit two different models: one model incorporating a pulsar black-body component and another model without a pulsar black-body component, to the X-ray spectra. 
We show the results of the fits for the photon index and normalization (i.e., unabsorbed flux) parameters for the PWN spectra in Figure \ref{fig:contour_plots}; Tables \ref{table:gamma_all_instruments_bb}, \ref{table:gamma_all_instruments_noBB}, \ref{table:unabsorbed_flux_all_instruments_bb}, \ref{table:unabsorbed_flux_all_instruments_noBB}; and Figure \ref{fig:par_vs_energy}. Figure \ref{fig:contour_plots} is a \replaced{
contour plot showing the joint probability distributions of the two fit parameters.}{collection of scatter plots showing the MCMC samples.} The contours indicate regions containing 68\% and 95\% of the samples, respectively. The results are also tabulated in Tables \ref{table:gamma_all_instruments_bb}, \ref{table:gamma_all_instruments_noBB}, \ref{table:unabsorbed_flux_all_instruments_bb}, \ref{table:unabsorbed_flux_all_instruments_noBB}. The uncertainties in the tables are the one-dimensional 90\% confidence intervals. Figure \ref{fig:par_vs_energy} shows the two fit parameters separately with energy on the x-axis to highlight differences between energy bands.

As shown in Table \ref{table:chandra_guest}, the photon index for the pulsar component is $\Gamma=1.35$ when {\it including} the black-body, while $\Gamma=1.54$ when {\it excluding} the black-body component. This increase of $\sim0.2$ in $\Gamma$ when excluding the black-body model assumes a softer spectrum for the pulsar component. As a softer spectrum for the pulsar component implies that the pulsar's emission is more concentrated at lower energies and does not extend to higher energies, a larger fraction of the overall emission at higher energies is attributed to the PWN. This effect results in a harder spectrum for the PWN (i.e., smaller $\Gamma$). This difference in $\Gamma$ for the PWN component between the two models is larger for higher energies.  

While the photon index $\Gamma$ from different detectors in the same energy band do not necessarily agree (Figure \ref{fig:contour_plots}), overall we see a general spectral softening (i.e., increase in $\Gamma$) over the four energy bands (Figure \ref{fig:par_vs_energy}). An exception to this trend is \hitomi's HXI 1 detector when fitting with the model that {\it does not} contains the pulsar black-body component (Table \ref{table:gamma_all_instruments_noBB}) in the 20--45 keV band. However, while the best-fit photon index changes from $\Gamma = 2.27$ in the 8--20 keV band to $\Gamma = 2.18$ in the 20--45 keV band, the uncertainty in the 20--45 keV band is large enough to make a softening plausible.

Regarding the normalization (i.e., unabsorbed flux) parameter, while the values of the normalization parameter decrease for each detector as we go from lower energy bands to higher energy bands, similar to the photon index\added{,} within the same energy band the values from different detectors disagree.

Next, we discuss how we incorporate the uncertainties in the fit parameters, due to the disagreement between detectors and the choice to include/exclude the pulsar black-body component, into our PWN modelling.

\section{Discussion}
\label{sec:discussion}

Here we discuss the results of modelling the PWN while taking into consideration the above IR and X-ray analysis. 
In \S\ref{dust} we discuss the potential origin of the observed IR emission. We then use this information and our updated X-ray analysis to determine the observed properties which should be reproduced by a physical model for the evolution of a PWN inside a SNR.  In \S\ref{sec:pwnmodel}, we describe such an evolutionary model and the method by which we identified the combination of input parameters that best reproduces the observed properties of this system.  We further discuss the implications of the derived values for the model parameters in \S\ref{sec:pwnmodel}, and use them to constrain structures in the surrounding ISM needed to reproduce the morphology of the surrounding SNR shell in \S\ref{sec:snrhydro}.

\subsection{Infrared Emission}
\label{dust}

As mentioned in \S\ref{sec:intro}, previous attempts to model the emission of this PWN were unable to simultaneously reproduce the obsreved IR and X-ray spectrum assuming both were synchrotron emission from the PWN (e.g., \citealt{torres14,hitomi18}).  While the power-law spectrum for the X-ray emission derived in our analysis (\S\ref{sec:results}) strongly suggests this is correct, below we evaluate if the IR emission is also synchrotron radiation from high-energy leptons in the PWN.

The total IR flux densities of the PWN region in \pwn\ are listed in Table~\ref{tab:ir_fluxes} and plotted in Figure~\ref{fig:model_sed}.  Figure~\ref{fig:irfig} shows the \ooneline\ and \ctwoline\ line maps in left and middle panels, as well as the MIPS~24~\micron\ of the PWN region in the right panel for comparison. These lines contribute to the emission seen in the \replaced{\textit{Herchel}}{\textit{Herschel}} PACS 70 and 160~\micron\ images shown in Figure~\ref{fig:irfig2}. Since the emission at 24~\micron\ has a similar filamentary morphology as the \ooneline\ map, it is not unreasonable to assume that a significant part of the emission seen at 24~\micron\ arises from \ion{O}{4} or \ion{Fe}{2} ejecta lines that fall in the wavelength range of the MIPS bandpass. 

As a result, it is likely that a significant fraction of the IR emission detected from this source is produced by dust and gas that resides in the ejecta filaments.  In order to use the IR properties of this sources to study the innermost dust and gas inside the SNR, it is first necessary to quantify the contribution from the PWN.  Often this is done by simply extrapolating a power-law fit to the spectrum at higher (typically X-ray) or lower (GHz radio) frequencies (e.g., \citealt{koo16}).  However, the modeling described below in \S\ref{sec:pwnmodel} potentially provides a more accurate way of estimating the synchrotron IR emission from the PWN.

\subsection{Modeling of PWN}
\label{sec:pwnmodel}

As discussed in \S\ref{sec:intro}, the properties of a PWN inside a SNR provide invaluable information on the progenitor star and supernova explosion, the birth properties of the neutron star, and the content of its pulsar wind.  Currently, one of the best methods of obtaining these properties is to use a (time-dependent) model for the evolution of a PWN in a SNR to reproduce the dynamical and broadband spectral energy distribution (SED) of a particular system (see recent reviews by \citealt{gelfand17} and \citealt{slane17} as well as references therein).  Here, we use the evolutionary model described by \citet{gelfand09} to reproduce the properties listed in Table \ref{tab:model_obs}, as we have previously done for the PWNe in G54.1+0.3 \citep{gelfand15}, HESS J1640$-$465 \citep{gotthelf14}, and Kes 75 \citep{gelfand13}.

\begin{table*}
    \caption{Observed properties of \pwn\ used in the modeling of this source}
    \label{tab:model_obs}
    \resizebox{0.95\linewidth}{!}{%
    \centering
    \begin{tabular}{ccccc}
    \hline
    \hline 
    {\sc Property} & {\sc Observed} & \multicolumn{2}{c}{\sc ``Best Fit" Values} & {\sc Citation} \\
    $\cdots$ & $\cdots$ & Variable $p$ & $p\equiv1.85690$ & $\cdots$\\
    \hline
    \multicolumn{5}{c}{\it PSR J1833$-$1034} \\
    Current spin-down luminosity $\dot{E}$ & $3.37\times10^{37}~\frac{\rm erg}{\rm s}$ & $\cdots$ & $\cdots$ & \citet{camilo06}\\
    Current characteristic age $t_{\rm ch}$ & 4850~{\rm years} & $\cdots$ & $\cdots$ & \citet{camilo06} \\
    \multicolumn{5}{c}{\it Pulsar Wind Nebula} \\
    Angular radius $\theta_{\rm pwn}$ & $40\arcsec \pm 4\arcsec$ & $42\farcs6$ & $40\farcs5$ & \citet{matheson10} \\
    Angular expansion rate $\dot{\theta}_{\rm pwn}$ & $(0.11\pm0.02)\frac{\%}{\rm year}$ & $0.07 \frac{\%}{\rm year}$ & $0.07 \frac{\%}{\rm year}$ & \cite{bietenholz08} \\
    327~MHz Flux Density & $7.3\pm0.7$~Jy & 5.8~Jy & 4.9~Jy & \citet{bietenholz11}\\
    1.43~GHz Flux Density & $7.0\pm0.4$~Jy & 7.2~Jy & 6.4~Jy & \citet{bietenholz11} \\
    4.8~GHz Flux Density & $6.5\pm0.4$~Jy & 7.5~Jy & 6.9~Jy & \citet{sun11} \\
    4.49 -- 7.85 GHz Spectral Index\tablenotemark{a} & $-0.12\pm0.03$ & $-0.06 \pm 0.01$ & $-0.03 \pm 0.01$ & \citet{bhatnagar11} \\
    70~GHz Flux Density & $4.3\pm0.6$~Jy & 3.7~Jy & 3.7~Jy & \citet{planck16} \\
    84.2~GHz Flux Density & $3.9\pm0.7$~Jy & 3.5~Jy & 3.5~Jy & \citet{salter89a} \\
    90.7~GHz Flux Density & $3.8\pm0.4$~Jy & 3.2~Jy & 3.3~Jy & \citet{salter89b} \\
    94~GHz Flux Density & $3.5\pm0.4$~Jy & 3.2~Jy & 3.3~Jy & \citet{bock01} \\
    100~GHz Flux Density & $2.7\pm0.5$~Jy & 3.0~Jy & 3.1~Jy & \citet{planck16} \\
    141.9~GHz Flux Density & $2.5\pm1.2$~Jy & 2.4~Jy & 2.5~Jy & \citet{salter89b} \\
    143~GHz Flux Density & $3.0\pm0.4$~Jy & 2.4~Jy & 2.5~Jy & \citet{planck16} \\
    $0.8-3.0$~keV Unabsorbed Flux & $(3.84_{-0.02}-4.57^{+0.05})\times10^{-11}~\frac{\rm ergs}{\rm cm^2~s}$ & $4.46\times10^{-11}~\frac{\rm ergs}{\rm cm^2~s}$ & $4.46\times10^{-11}~\frac{\rm ergs}{\rm cm^2~s}$ & Tables \ref{table:unabsorbed_flux_all_instruments_bb} \& \ref{table:unabsorbed_flux_all_instruments_noBB}\\
    $0.8-3.0$~keV Photon Index & $1.78_{-0.02}-1.93^{+0.03}$ & $1.84\pm0.01$ & $1.84\pm0.01$ & Tables \ref{table:gamma_all_instruments_bb} \& \ref{table:gamma_all_instruments_noBB} \\
    $3.0-8.0$~keV Unabsorbed Flux & $(3.58_{-0.01}-3.82^{+0.01})\times10^{-11}~\frac{\rm ergs}{\rm cm^2~s}$ & $3.82\times10^{-11}~\frac{\rm ergs}{\rm cm^2~s}$ & $3.82\times10^{-11}~\frac{\rm ergs}{\rm cm^2~s}$ & Tables \ref{table:unabsorbed_flux_all_instruments_bb} \& \ref{table:unabsorbed_flux_all_instruments_noBB}\\
    $3.0-8.0$~keV Photon Index & $1.84_{-0.02}-2.04^{+0.01}$ & $2.01\pm0.02$ & $2.00\pm0.02$ & Tables \ref{table:gamma_all_instruments_bb} \& \ref{table:gamma_all_instruments_noBB} \\
    $8.0-20.0$~keV Unabsorbed Flux & $(3.02_{-0.02}-3.58^{+0.04})\times10^{-11}~\frac{\rm ergs}{\rm cm^2~s}$ & $3.02\times10^{-11}~\frac{\rm ergs}{\rm cm^2~s}$ & $3.04\times10^{-11}~\frac{\rm ergs}{\rm cm^2~s}$ & Tables  \ref{table:unabsorbed_flux_all_instruments_bb} \& \ref{table:unabsorbed_flux_all_instruments_noBB}\\
    $8.0-20.0$~keV Photon Index & $2.10_{-0.02}-2.31^{+0.03}$ & $2.19\pm0.02$ & $2.17\pm0.02$ & Tables \ref{table:gamma_all_instruments_bb} \& \ref{table:gamma_all_instruments_noBB} \\
    $20.0-45.0$~keV Unabsorbed Flux & $(1.84_{-0.05}-2.65^{+0.06})\times10^{-11}~\frac{\rm ergs}{\rm cm^2~s}$ & $2.05\times10^{-11}~\frac{\rm ergs}{\rm cm^2~s}$ & $2.11\times10^{-11}~\frac{\rm ergs}{\rm cm^2~s}$ & Tables \ref{table:unabsorbed_flux_all_instruments_bb} \& \ref{table:unabsorbed_flux_all_instruments_noBB}\\
    $20.0-45.0$~keV Photon Index & $2.10_{-0.08}-2.71^{+0.13}$ & $2.65\pm0.03$ & $2.61\pm0.03$ & Tables \ref{table:gamma_all_instruments_bb} \& \ref{table:gamma_all_instruments_noBB} \\
    $10-20$ GeV Photon Flux & $8.6_{-2.2}^{+2.5}\times10^{-11}~\frac{\rm photons}{\rm cm^2~s}$ & $6.36\times10^{-11}~\frac{\rm photons}{\rm cm^2~s}$ & $6.15\times10^{-11}~\frac{\rm photons}{\rm cm^2~s}$ & \cite{3fhl} \\
    $20-50$ GeV Photon Flux & $1.85_{-0.93}^{+1.23}\times10^{-11}~\frac{\rm photons}{\rm cm^2~s}$ & $3.05\times10^{-11}~\frac{\rm photons}{\rm cm^2~s}$ & $2.94\times10^{-11}~\frac{\rm photons}{\rm cm^2~s}$ & \cite{3fhl} \\
    $150-500$ GeV Photon Flux & $<6.31\times10^{-12}~\frac{\rm photons}{\rm cm^2~s}$ (3$\sigma$) & $2.75\times10^{-12}~\frac{\rm photons}{\rm cm^2~s}$ & $2.61\times10^{-12}~\frac{\rm photons}{\rm cm^2~s}$ & \cite{3fhl} \\
    $0.5-1.2$ TeV Photon Flux & $<5.34\times10^{-12}~\frac{\rm photons}{\rm cm^2~s}$ (3$\sigma$) & $6.59\times10^{-13}~\frac{\rm photons}{\rm cm^2~s}$ & $6.15\times10^{-13}~\frac{\rm photons}{\rm cm^2~s}$ & \cite{3fhl} \\
    $1-10$ TeV Flux & $(1.29\pm0.25)\times10^{-12}~\frac{\rm ergs}{\rm cm^2~s}$ & $1.21\times10^{-12}~\frac{\rm ergs}{\rm cm^2~s}$ & $1.10\times10^{-12}~\frac{\rm ergs}{\rm cm^2~s}$ & \citet{hgps} \\
    $1-10$ TeV Photon Index & $2.4\pm0.2$ & $2.25\pm0.01$ & $2.27\pm0.01$ & \citet{hgps} \\
    \multicolumn{5}{c}{\it Supernova Remnant} \\
    Angular radius $\theta_{\rm snr}$ & $2\farcm44 \pm 0\farcm24$ & $2\farcm55$ & $2\farcm56$ & \citet{guest19} \\
    $v_{\rm ej}(R_{\rm pwn})$ & $350-1000~\frac{\rm km}{\rm s}$ & $525~\frac{\rm km}{\rm s}$ & $525~\frac{\rm km}{\rm s}$ & \S\ref{sec:ir} \\
    Distance & 4.4~kpc & $\cdots$ & $\cdots$ & \citet{ranasinghe18}\\
    \hline
    \hline
    \end{tabular}
    }
    \tablenotetext{a}{Spectral index $\alpha$ is defined as flux density $S_\nu \propto \nu^\alpha$.}
\tablecomments{For upper limits, their statistical significance is indicated next to the Observed value.  Properties with no ``predicted" values were fixed in this modeling, as described in \S\ref{sec:pwnmodel}.}
\end{table*}

The input parameters to this model are listed in Table \ref{tab:model_pars}.  As in the past analyses listed above, we make the following assumptions:
\begin{enumerate}
    \item Assume that the density profile of the unshocked SN ejecta consists of a uniform density $(\rho \propto r^0)$ core surrounded by a $\rho \propto r^{-9}$ envelope.  While this assumption is common in this field (see \citealt{gelfand17} for a recent review), as discussed by \citet{chevalier05} different supernova progenitors will likely have different ejecta density profiles.
    \item Assume that the supernova ejecta with mass $M_{\rm ej}$ and initial kinetic energy $E_{\rm sn}$ is expanding into a medium with uniform density $n_{\rm ism}$.  As discussed in \S\ref{sec:snrhydro}, the X-ray morphology of the SNR shell strongly suggests a density enhancement North of the explosion site.  However, as this enhancement has only impacted a small fraction of the shell -- not affecting the average SNR radius $\theta_{\rm snr}$ used in our modeling -- nor caused the SN reverse shock to collide with any part of the PWN, this has a minimal effect on the results of our modeling.
    \item Calculate the age $t_{\rm age}$ and initial spin-down luminosity $\dot{E}_0$ of associated PSR~J1833$-$1034 for a particular (assumed constant) pulsar braking index $p$ and spin-down timescale $\tau_{\rm sd}$ using the characteristic age $t_{\rm ch}$ and current spin-down luminosity $\dot{E}$ (given in Table \ref{tab:model_obs}) inferred from the measured period $P$ and period-derivative $\dot{P}$ of the PSR (e.g., \citealt{gelfand15}):
    \begin{eqnarray}
    \label{eqn:tage}
    t_{\rm age} & = & \frac{2 t_{\rm ch}}{p-1}-\tau_{\rm sd} \\
    \label{eqn:e0dot}
    \dot{E}_0 & = & \dot{E} \left(1+\frac{t_{\rm age}}{\tau_{\rm sd}} \right)^{+\frac{p+1}{p-1}}
    \end{eqnarray}
    \added{Analysis of 5.5 years of timing observations of 
    PSR J1833$-$1034 recently measured the braking index of this pulsar to be $p=1.8569\pm0.0006$ \citep{roy12}.  However, the result is sensitive to the treatment of the ``glitches" which occurred during this campaign.  Analysis of the timing properties of this pulsar during the first $\sim1.5$ years of this campaign, during which no significant glitches were detected, yielded  $p=2.168\pm0.008$ \citep{roy12}.  We therefore model the properties of this PWN for two cases: $p\equiv 1.8569$ and $p$ unconstrained.}
    
    \item Assume the entire spin-down luminosity $\dot{E}$ is injected into the PWN as either magnetic fields $\dot{E}_{\rm B}$ or the kinetic energy $\dot{E}_{\rm p}$ of relativistic leptons ($e^\pm$), such that:
    \small
    \begin{eqnarray}
    \dot{E}_{\rm B}(t) & = & \eta_{\rm B} \dot{E}(t) = \eta_{\rm B} \dot{E}_0 \left(1+\frac{t}{\tau_{\rm sd}}\right)^{-\frac{p+1}{p-1}} \\
    \dot{E}_{\rm p}(t) & = & (1-\eta_{\rm B}) \dot{E}(t) = (1-\eta_{\rm B}) \dot{E}_0 \left(1+\frac{t}{\tau_{\rm sd}}\right)^{-\frac{p+1}{p-1}}
    \end{eqnarray}
    \normalsize
    where $\eta_{\rm B}$ is constant with time.  While the pulsed $\gamma$-ray luminosity of some pulsars can be a significant fraction of $\dot{E}$, the observed pulsed $\gamma$-ray luminosity of PSR J1833$-$1034 is $\approx0.005 \dot{E}$ \citep{2fpc}.
    \item Assume that the spectrum of particles injected into the PWN is well-described by a broken power-law of the form:
    \small
    \begin{eqnarray}
    \label{eqn:bpl}
    \frac{d\dot{N}}{dE} & = & \left\{ \begin{array}{cc}
    \dot{N}_{\rm break} \left(\frac{E}{E_{\rm break}} \right)^{-p_1} & E_{\rm min} < E < E_{\rm break} \\
    \dot{N}_{\rm break} \left(\frac{E}{E_{\rm break}} \right)^{-p_2} & E_{\rm break} < E < E_{\rm max} \\
    \end{array} \right.
    \end{eqnarray}
    \normalsize
    where the five free parameters ($E_{\rm min}$, $E_{\rm break}$, $E_{\rm max}$, $p_1$, and $p_2$) in Equation \ref{eqn:bpl} are assumed to be constant with time and the normalization $\dot{N}_{\rm break}$ is calculated by requiring that:
    \begin{eqnarray}
    \dot{E}_{\rm p} & = & \int\limits_{E_{\rm min}}^{E_{\rm max}} E \frac{d\dot{N}}{dE} dE
    \end{eqnarray}
    at all times $t$.
    \item Assume that only radiative losses suffered by particles trapped within the PWN are the result of synchrotron and inverse Compton (IC) emission.  When calculating synchrotron losses, we assume the PWN's magnetic field has a uniform strength $B_{\rm pwn}(t)$ (whose evolution is calculated using the procedure described by \citealt{gelfand09}) and that the particle pitch angles (i.e., the angle between their velocity $\vec{v}$ and local magnetic field $\vec{B}$) is randomly distributed.  For IC emission, we consider particles scattering photons emitted by the Cosmic Microwave Background (temperature $T_{\rm cmb} = 2.7255~{\rm K}$; \citealt{fixsen09}) as well as an additional background field which has a blackbody spectrum with temperature $T_{\rm ic}$ and normalization $K_{\rm ic}$, such that this photon field has an energy density:
    \begin{eqnarray}
    u_{\rm ic} & = & K_{\rm ic} a_{\rm bb} T_{\rm ic}^4,
    \end{eqnarray}
    where $a_{\rm bb} \approx 7.5657\times10^{-15}~\frac{\rm ergs}{\rm cm^3 K^4}$.  We do not consider Synchrotron Self-Compton (SSC) emission, since previous theoretical work have found that SSC emission significantly contributes to the total IC emission only at extremely early times (e.g., \citealt{gelfand09, martin12}).
\end{enumerate}
To convert the physical quantities predicted by our model to the observed properties of this system, we assume a distance $d\equiv4.4~{\rm kpc}$ -- the central value derived from a recent study of its H{\sc i} emission ($d=4.4\pm0.2$~kpc; \citealt{ranasinghe18}). 

\small
\begin{table}
\caption{PWN model parameters which best reproduces the properties of \pwn }
\label{tab:model_pars}
\resizebox{0.95\columnwidth}{!}{%
\centering
\begin{tabular}{ccc}
\hline
\hline
Model Parameter & Variable $p$ & $p\equiv1.85690$ \\
\hline
Supernova Explosion Energy $E_{\rm sn}$ & $1.2\times10^{50}~{\rm ergs}$ & $1.2\times10^{50}~{\rm ergs}$ \\
Supernova Ejecta Mass $M_{\rm ej}$ & 11.32~M$_\odot$ & 11.33~M$_\odot$ \\
ISM Density $n_{\rm ism}$ & 0.2~cm$^{-3}$ & 0.2~cm$^{-3}$ \\
Pulsar Braking Index $p$ & 3.126 & $\equiv1.85690$\\
Pulsar Spindown Timescale $\tau_{\rm sd}$ & 2900~years & 9600~years \\
Wind Magnetization $\eta_{\rm B}$ & $3.2\times10^{-3}$ & $3.5\times10^{-3}$\\
Minimum Energy of Injected Leptons $E_{\rm min}$ & 12.5~GeV & 12.5~GeV \\
Break Energy of Injected Leptons $E_{\rm break}$ & 1.0~TeV & 1.0~TeV \\
Maximum Energy of Injected Leptons $E_{\rm max}$ & 0.26~PeV & 0.18~PeV \\
Low-Energy Particle Index $p_1$ & 2.86 & 2.86 \\
High-Energy Particle Index $p_2$ & 2.51 & 2.51 \\
Temperature of External Photon Field $T_{\rm ic}$ & 1700~K & 1700~K \\
Normalization of External Photon Field $K_{\rm ic}$ & $3.5\times10^{-10}$ & $3.8\times10^{-10}$\\
\hline
$\chi^2$ / degrees of freedom & 30 / 16 & 37 / 17 \\
\hline
\hline
\end{tabular}%
}
\tablecomments{The \deleted{13} free parameters in the physical \replaced{model}{models} used to reproduce the observed properties of \pwn\ listed in Table \ref{tab:model_obs}.  The reported values are the combination which had the highest likelihood ${\mathcal L}$\replaced{, which corresponds to $\chi^2 \approx 30 $ for 16 degrees of freedom.}{which corresponds to the given $\chi^2$. }} 
\end{table}
\normalsize

We used a Metropolis MCMC algorithm (\citealt{metropolis85}; see \S3.2 of \citet{gelfand15} for a detailed description) to identify the combination of the 13 model input parameters $\Theta$ listed in Table \ref{tab:model_pars} which best reproduce the 29 observed properties ${\mathcal D}$ of \pwn\ listed in Table \ref{tab:model_obs}.  This is accomplished by the maximum likelihood estimation method, in which we find the combination $\Theta$ whose predicted values of the observed properties ${\mathcal M}$ maximizes the likelihood ${\mathcal L}({\mathcal D}|\Theta)$:
\begin{eqnarray}
\label{eqn:likelihood}
{\mathcal L} & \equiv & \prod\limits_{i=1}^{29} {\mathcal L}({\mathcal D}_i|\Theta) \\
\label{eqn:lnlikelihood}
\ln {\mathcal L} & = & \sum\limits_{i=1}^{29} \ln {\mathcal L}({\mathcal D}_i|\Theta)
\end{eqnarray}
As listed in Table \ref{tab:model_obs}, there are three types of observed quantities ${\mathcal D}_i$, those:
\begin{enumerate}
    \item whose measured error $\sigma_i$ is Gaussian in nature (indicated by $\pm$ in Table \ref{tab:model_obs}), 
    \item constrained to be below some value ${\mathcal D}_i < {\mathcal D}_i^{\rm up}$ (indicated by $<$ in Table \ref{tab:model_obs}), and
    \item whose true value is believed to lie within a range ${\mathcal D}_i^{\rm lo} < {\mathcal D}_i < {\mathcal D}_i^{\rm hi}$.
\end{enumerate}   
The likelihood ${\mathcal L}({\mathcal D}_i|\Theta)$ is defined differently for these three cases, as described below.

In the first case where the errors are Gaussian, we define ${\mathcal L}({\mathcal D}_i|\Theta)$ to be:
\begin{eqnarray}
\label{eqn:like1}
{\mathcal L}({\mathcal D}_i|\Theta) & = & \frac{1}{\sigma_i\sqrt{2\pi}} e^{-\frac{1}{2}\left(\frac{{\mathcal D}_i - {\mathcal M}_i}{\sigma_i}\right)^2} \\
\label{eqn:lnlike1}
\ln {\mathcal L}({\mathcal D}_i|\Theta) & = & C - \ln \sigma_i - \frac{1}{2}\left(\frac{{\mathcal D}_i - {\mathcal M}_i}{\sigma_i}\right)^2 \\
\label{eqn:chi1}
\chi^2_i & = & \left(\frac{{\mathcal D}_i - {\mathcal M}_i}{\sigma_i}\right)^2
\end{eqnarray}
where $C \equiv -\frac{1}{2}\ln(2\pi)$ and ${\mathcal M}_i$ is the value for ${\mathcal D}_i$ predicted by the model for a particular combination of input parameters $\Theta$. 

For the second case which measurements have only yielded upper-limits (i.e., observable ${\mathcal D}_i < {\mathcal D}_{i}^{\rm up}$ where ${\mathcal D}_{i}^{\rm up}$ is $N\sigma$ above the background), we define:
\small
\begin{eqnarray}
\label{eqn:like2}
{\mathcal L}({\mathcal D}_i|\Theta) & = & \left\{ \begin{array}{cc}
1 & {\scriptstyle {\mathcal M}_i < {\mathcal D}_i^{\rm up}} \\
\frac{1}{\sigma_i\sqrt{2\pi}} e^{-\frac{1}{2}\left(\frac{{\mathcal M}_i - {\mathcal D}_i^{\rm up}}{\sigma_i}\right)^2} & {\scriptstyle {\mathcal M}_i > {\mathcal D}_i^{\rm up}} \\
\end{array}
\right. \\
\label{eqn:lnlike2}
\ln {\mathcal L}({\mathcal D}_i|\Theta) & = & \left\{  \begin{array}{cc}
0 & {\scriptstyle {\mathcal M}_i < {\mathcal D}_i^{\rm up}} \\
C - \ln \sigma_i - \frac{1}{2}\left(\frac{{\mathcal D}_i - {\mathcal M}_i}{\sigma_i}\right)^2 & {\scriptstyle {\mathcal M}_i > {\mathcal D}_i^{\rm up}} \\
\end{array}
\right. \\
\label{eqn:chi2}
\chi^2_i & = & \left\{  \begin{array}{cc}
0 & {\scriptstyle {\mathcal M}_i < {\mathcal D_i}^{\rm up}} \\
\left(\frac{{\mathcal D}_i - {\mathcal M}_i}{\sigma_i}\right)^2 & {\scriptstyle {\mathcal M}_i > {\mathcal D_i}^{\rm up}}  \\
\end{array} \right.
\end{eqnarray}
\normalsize
where $C\equiv -\frac{1}{2}\ln(2\pi)$ and $\sigma_i \equiv \frac{{\mathcal D}_{i}^{\rm up}}{N}$.

The third case is applied to the unabsorbed fluxes and photon indices of the PWN in the X-ray band.  Unfortunately, measurements of these parameters are strongly dependent on the (assumed) model for the pulsar's X-ray emission as well as the instrument used to make the measurement.  As listed in Tables \ref{table:gamma_all_instruments_bb}--\ref{table:unabsorbed_flux_all_instruments_noBB}, the measured values for these quantities span a range ${\mathcal D}_i^{\rm lo} - {\mathcal D}_i^{\rm hi}$ significantly larger than the statistical errors of an individual measurement (Figure \ref{fig:contour_plots}).  Since resolving these fundamentally `systematic' uncertainties is beyond the scope of this work, when determining the likelihood that the predicted value ${\mathcal M}_i$ is consistent with measured value ${\mathcal D}_i$, we adopt:
\small
\begin{eqnarray}
\label{eqn:like3}
{\mathcal L}_i({\mathcal D}_i|\Theta) & = & \left\{ \begin{array}{cc}
\frac{1}{\sigma_i^{\rm lo} \sqrt{2\pi}} e^{-\frac{1}{2}\left(\frac{{\mathcal D}_i^{\rm lo} - {\mathcal M}_i}{\sigma_i^{\rm lo}}\right)^2} & {\scriptstyle {\mathcal M}_i < {\mathcal D}_{i}^{\rm lo}} \\
1 & {\scriptscriptstyle {\mathcal D}_{i}^{\rm lo} < {\mathcal M}_i < {\mathcal D}_{i}^{\rm hi}} \\
\frac{1}{\sigma_i^{\rm hi} \sqrt{2\pi}} e^{-\frac{1}{2}\left(\frac{{\mathcal D}_i^{\rm hi} - {\mathcal M}_i}{\sigma_i^{\rm hi}}\right)^2} & {\scriptstyle {\mathcal M}_i > {\mathcal D}_{i}^{\rm hi}} \\
\end{array}
\right. \\
\label{eqn:lnlike3}
\ln {\mathcal L}_i({\mathcal D}_i|\Theta) & = & \left\{ \begin{array}{cc}
C - \ln \sigma_i^{\rm lo} - \frac{1}{2}\left(\frac{{\mathcal D}_i^{\rm lo} - {\mathcal M}_i}{\sigma_i^{\rm lo}}\right)^2 & {\scriptstyle {\mathcal M}_i < {\mathcal D}_i^{\rm lo}} \\
0 & {\scriptscriptstyle {\mathcal D}_{i}^{\rm lo} < {\mathcal M}_i < {\mathcal D}_{i}^{\rm hi}} \\
C - \ln \sigma_i^{\rm hi} - \frac{1}{2}\left(\frac{{\mathcal D}_i^{\rm hi} - {\mathcal M}_i}{\sigma_i^{\rm hi}}\right)^2 & {\scriptscriptstyle {\mathcal M}_i > {\mathcal D}_i^{\rm hi}} \\
\end{array}
\right. \\
\label{eqn:chi3}
\chi^2_i & = & \left\{ \begin{array}{cc}
\left(\frac{{\mathcal D}_i^{\rm lo} - {\mathcal M}_i}{\sigma_i^{\rm lo}}\right)^2 & {\scriptstyle {\mathcal M}_i < {\mathcal D}_i^{\rm hi}} \\
0 & {\scriptscriptstyle {\mathcal D}_{i}^{\rm lo} < {\mathcal M}_i < {\mathcal D}_{i}^{\rm hi}} \\
\left(\frac{{\mathcal D}_i^{\rm hi} - {\mathcal M}_i}{\sigma_i^{\rm hi}}\right)^2 & {\scriptstyle {\mathcal M}_i > {\mathcal D}_i^{\rm hi}} \\
\end{array}
\right.
\end{eqnarray}
\normalsize
where $\sigma_i^{\rm lo}$ is the lower error on the lowest measurement of ${\mathcal D}_i$, $\sigma_i^{\rm hi}$ is the upper error on the highest measurement of ${\mathcal D}_i$, and $C \equiv -\frac{1}{2}\ln(2\pi)$. Since it is difficult to interpret the quality of a fit based on the value of ${\mathcal L}$ or $\ln {\mathcal L}$, we also calculate a representative $\chi^2 = \sum \chi^2_i$ defined in Equations \ref{eqn:chi1}, \ref{eqn:chi2}, \& \ref{eqn:chi3}.

\begin{figure}
    \centering
    \includegraphics[width=0.475\textwidth]{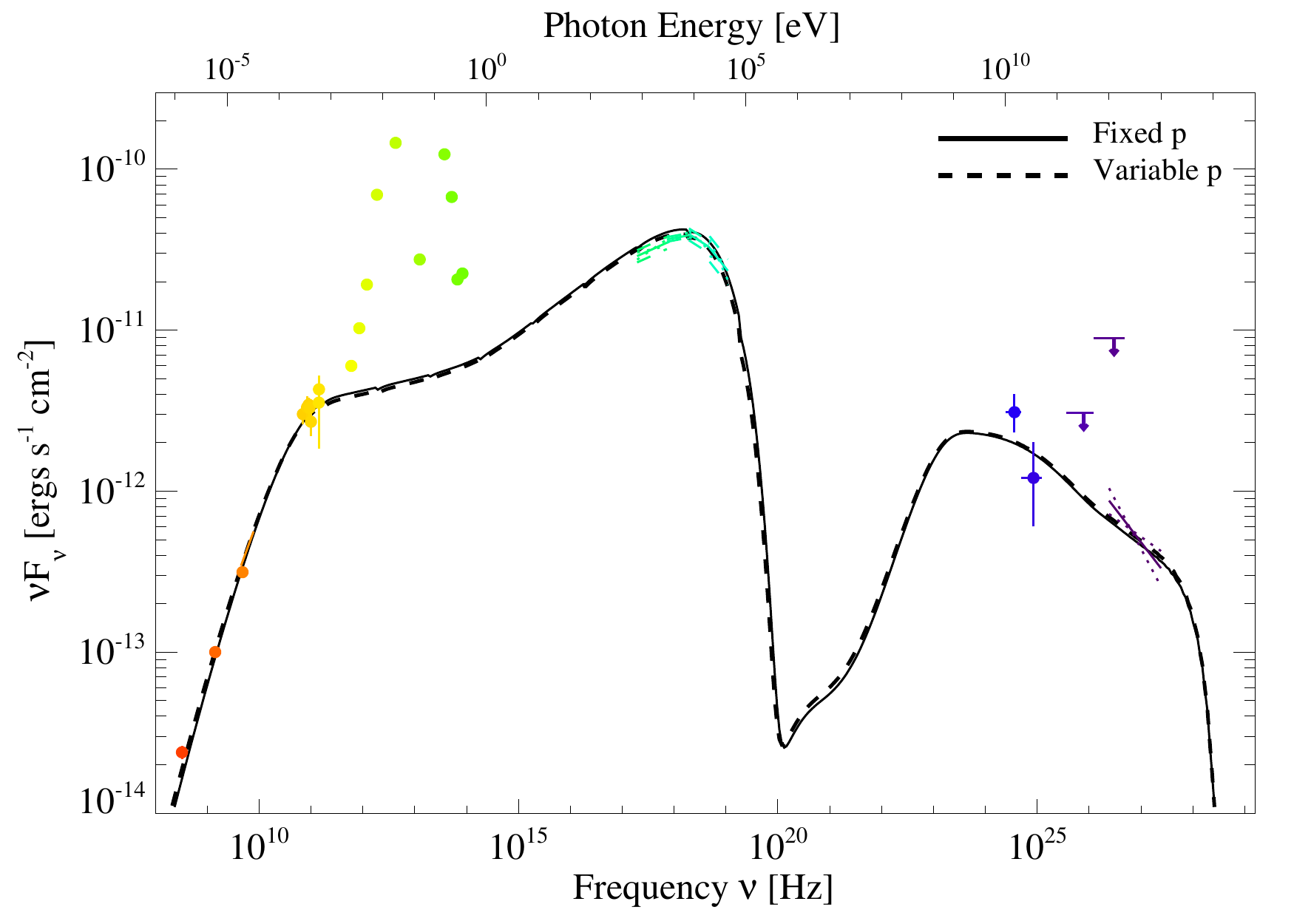} 
    \caption{SED of PWN \pwn\ predicted for the ``most likely" combination of model input parameter $\Theta$ for \added{both a fixed ({\it solid}) and variable ({\it dashed}) braking index, as} given in Table \ref{tab:model_pars}. The observed data points are given in Table \ref{tab:model_obs}. }
    \label{fig:model_sed}
\end{figure}

The combination of input parameters $\Theta$ which resulted in the largest ${\mathcal L}$ of our MCMC run is given in Table \ref{tab:model_pars}, with the value predicted by this combination for each observable given in Table \ref{tab:model_pars} and the predicted SED shown in Figure \ref{fig:model_sed}.  \added{For both a fixed and variable braking index $p$}, our model \replaced{is able to reproduce}{reproduces} most of the properties of \pwn\ to within $<1\sigma$ of the observed values, with \replaced{the most significant deviations being $\dot{\theta}_{\rm pwn}$ (where the predicted values is $\approx2\sigma$ away from the observed value), the 327 MHz flux density ($2.2\sigma$ discrepancy), 4.8 GHz flux density ($2.6\sigma$ discrepancy), and the 4.49$-$7.86~GHz spectral index ($1.9\sigma$ discrepancy)}{the value predicted by the model deviating by $\sim1-3\sigma$ from the observed values of $\dot{\theta}_{\rm pwn}$, 327 MHz flux density, 4.8 GHz flux density, and 4.49$-$7.86~GHz spectral index}.  Notably, this model successfully reproduces the unabsorbed flux and photon index measured in each of the four X-ray bands -- unlike many previous attempts of modeling the SED of this source (e.g., \citealt{tanaka11, torres14, hitomi18}).  We note that we did not attempt to reproduce the IR flux density observed from this PWN since, as described in \S\ref{sec:ir}, emission from surrounding gas dust is likely a significant contributor in this band.  As shown in Figure \ref{fig:model_sed}, the predicted flux density of the PWN's synchrotron emission in this band is significantly lower than the observed values (\S\ref{sec:ir}; Table \ref{tab:ir_fluxes}).  Furthermore, the expected PWN contribution is not well described by a simple power-law extrapolation of the measured radio or X-ray flux densities.
  
\begin{table}
    \caption{Physical properties of \pwn\ predicted by the \replaced{model}{models} whose parameters are given in Table \ref{tab:model_pars}.}
    \resizebox{0.975\columnwidth}{!}{%
    \centering
    \begin{tabular}{cccc}
    \hline
    \hline
    {\sc Property} & Variable $p$ & $p\equiv1.85690$ \\
    \hline
    Pulsar Age $t_{\rm age}$ & 1700~years & 1700~years \\
    Pulsar Initial Spindown Luminosity $\dot{E}_0$ & $8.3\times10^{37}~\frac{\rm ergs}{\rm s}$ & $5.8\times10^{37}~\frac{\rm ergs}{\rm s}$ \\
    Mass of Ejecta Swept-up by PWN $M_{\rm sw,pwn}$ & 0.85~M$_\odot$ & 0.73~M$_\odot$ \\
    PWN Expansion Velocity $v_{\rm pwn}$ & $\approx 610~\frac{\rm km}{\rm s}$ & $\approx 590~\frac{\rm km}{\rm s}$ \\
    Ejecta Speed just outside PWN $v_{\rm ej}(R_{\rm pwn})$ & $\approx 525~\frac{\rm km}{\rm s}$  & $\approx 500~\frac{\rm km}{\rm s}$ \\
    Pulsar Initial Spin Period $P_0$ & $\approx50~{\rm ms}$ & $\approx51~{\rm ms}$ \\
    PWN Magnetic Field Strength $B_{\rm pwn}$ & $\approx31~\mu{\rm G}$ & $\approx33~\mu{\rm G}$ \\
    \hline
    \hline
    \end{tabular}%
    }
    \label{tab:model_physprop}
\end{table}

While we have not extensively sampled the possible parameter space and obtained formal uncertainties on the parameters, as done for G54.1+0.3 \citep{gelfand15}, the ``most likely" parameters identified in our analysis can be used to derive information regarding the formation and underlying physics of this system.  As indicated in Table \ref{tab:model_pars}, our modeling suggests the progenitor supernova ejected $M_{\rm ej} \approx 11~{M}_\odot$ of material with a rather low initial kinetic energy $E_{\rm sn} \approx 1.2\times10^{50}~{\rm ergs}$ -- a situation where current simulations for core-collapse supernova favor the creation of stellar-mass black hole (e.g., \citealt{sukhbold16}), not a neutron star as observed here. It is important to note that we did center numerous MCMC chains (each consisting of $\approx50,000$ samples) around a canonical supernova explosion of $M_{\rm ej} \sim 8~{M}_\odot$ and $E_{\rm sn} \sim 10^{51}~{\rm ergs}$ and were unable to reproduce the observed properties of this system in this region of parameter space.  As a result, our modeling strongly suggest this system is the result of a low energy, high mass supernova explosion.

This conclusion can be tested by measuring the properties of the supernova ejecta.  This is best done by detected thermal X-rays from ejecta heated by the reverse shock.  Unfortunately, our results suggest that very little ejecta has interacted with the reverse shock (\S\ref{sec:snrhydro}).  However, as the PWN expands it sweeps up and shocks the inner-most ejecta.  For the most likely set of parameters given in Table \ref{tab:model_pars}, we find that the PWN has swept-up $M_{\rm sw, pwn} \approx 0.7-0.9~{M}_\odot$ of ejecta, and is currently expanding ($v_{\rm pwn} - v_{\rm ej}(R_{\rm pwn})) \sim 85-90~\frac{\rm km}{\rm s}$ faster than its surroundings \added{(Table \ref{tab:model_physprop})}. These predictions can be tested with future analysis of the IR emission of this source.

\begin{table*}
\caption{Properties of PSR J1833$-$1034 \deleted{, which is} associated with \pwn\ and \replaced{PSR J1640$-$4631, the first pulsar with a measured braking index $p>3$}{other pulsars with either changing braking indices or $p>3$}.}
    \resizebox{0.93\linewidth}{!}{%
    \centering
    \begin{tabular}{ccccc}
    \hline
    \hline
    Property & PSR J1833$-$1034 & PSR J1846$-$0258 & PSR B0540-69 & PSR J1640$-$4631 \\
    \hline
    Period $P$ & $\approx 61.8$~ms\tablenotemark{a} & $\approx327$~ms\tablenotemark{c} & $\approx50.5$~ms\tablenotemark{d} & $\approx 206$~ms\tablenotemark{f}  \\
    Period-Derivative $\dot{P}$ & $2.02\times10^{-13}~\frac{\rm s}{\rm s}$\tablenotemark{a} & $7.11\times10^{-12}~\frac{\rm s}{\rm s}$\tablenotemark{c} & $4.78\times10^{-13}~\frac{\rm s}{\rm s}$\tablenotemark{d} & $9.76\times10^{-13}~\frac{\rm s}{\rm s}$\tablenotemark{f} \\
    Spin-down Luminosity $\dot{E}$ & $3.4\times10^{37}~\frac{\rm ergs}{\rm s}$\tablenotemark{a} & $8.1\times10^{36}~\frac{\rm ergs}{\rm s}$\tablenotemark{c} & $1.5\times10^{38}~\frac{\rm ergs}{\rm s}$\tablenotemark{d} & $4.4\times10^{36}~\frac{\rm ergs}{\rm s}$\tablenotemark{f} \\
    Characteristic Age $t_{\rm ch}$ & $\approx4850$~years\tablenotemark{a} & $\approx730$~years\tablenotemark{c} & $\approx1670$~years\tablenotemark{d} & $\approx3350$~years\tablenotemark{f} \\
    Surface Dipole Magnetic Field $B_{\rm ns}$ & $3.6\times10^{12}~{\rm G}$\tablenotemark{a} & $4.9\times10^{13}~{\rm G}$\tablenotemark{c} &  $5.0\times10^{12}~{\rm G}$\tablenotemark{d} & $1.4\times10^{13}~{\rm G}$\tablenotemark{f} \\
    Braking Index $p$ & $\approx 3.1 / \equiv 1.8659$\tablenotemark{b} & $2.65 / 2.16$\tablenotemark{c} & $2.13 / 0.03 - 0.9$ \tablenotemark{d,e} & $3.15\pm0.03$\tablenotemark{g}  \\ 
    \hline
    \hline
    \end{tabular}
    }
    \label{tab:psrprop}
    \tablenotetext{a}{\citet{camilo06}}
    \tablenotetext{b}{The first braking index is the valued prefered by our modeling of the PWN (Table \ref{tab:model_pars}), the second is the value reported by \citealt{roy12}. }
    \tablenotetext{c}{\citet{livingstone11}.  Reported braking indices are the values measured before and after the observed change.}
    \tablenotetext{d}{\citet{ferdman15}}
    \tablenotetext{e}{\citet{marshall2016, kim19, wang20}.  Reported braking indices are the values measured before and after the observed change.}
    \tablenotetext{f}{\citet{gotthelf14}}
    \tablenotetext{g}{\citet{archibald16}}
\end{table*}

\added{As listed in Table \ref{tab:model_obs} and shown in Figure \ref{fig:model_sed}, qualitatively similar results are obtained when modeling this source by fixing the braking index of associated PSR J1833$-$1034 to the currently measured value ($p\equiv1.85690$; \citealt{roy12}) or treating it as a free parameter.  Models with $p \approx 3.1$ predict a higher flux densities at GHz frequencies and lower fluxes at X-ray energies, which improves the likelihood ${\mathcal L}$ (and correspondingly $\chi^2$) of the fits.  If this higher value of $p$ more accurately represents the time evolution of the rate energy is injected into the PWN by this pulsar, this suggests that its braking index of this pulsar may have changed over its lifetime.  Such behavior has been observed from other young pulsars, e.g. PSR J1846$-$0258 associated with SNR / PWN Kes 75 (e.g., \citealt{livingstone11}) and PSR B0540$-$69 (e.g., \citealt{kim19}).  In fact, the spin-down inferred surface dipole magnetic field strength and ages of both PSR J1833$-$1034 and PSR B0540$-$69 are quite similar (Table \ref{tab:psrprop}).  
However, the measured braking indices of both PSRs J1846$-$0258 and B0540$-$69 are $p<3$, suggesting that the observed spin-down is possibly the result of both magnetic dipole radiation and the particle outflow (e.g., \citealt{ou16} and references therein), while our modeling prefers that PSR J1833$-$1034 has $p > 3$ -- inconsistent with this physical model.}

\deleted{Furthermore, our modeling suggests that associated pulsar PSR J1833$-$1034 has a braking index $p \approx 3.127$. } \deleted{Again, extensive trials were conducted at $p \leq 3$, but they were resulted in significantly worse fits to the observed properties listed in Table \ref{tab:model_obs}.}  The first pulsar with a braking index $p > 3$ \deleted{(the canonical value for magnetic dipole radiation)} from a phase-connected timing solution is PSR J1640$-$4631 which has a measured braking index $p=3.15\pm0.03$ \citep{archibald16}.  As shown in Table \ref{tab:psrprop}, other than age, they are very few physical similarities between these two pulsars: PSR J1833$-$1034 has a period $P \sim 3\times$ smaller than PSR J1640$-$4631, a spin-down luminosity $\dot{E} \sim10\times$ larger, and a (spin-down inferred) surface dipole magnetic field strength $B_{\rm ns} \sim 4\times$ lower.

In addition, our modeling suggests the age of this system is less than the pulsar's spin-down timescale ($t_{\rm age} < \tau_{\rm sd}$; Tables \ref{tab:model_physprop} \& \ref{tab:model_pars}), as first suggested by \citet{camilo06}. As a result, the implied initial spin-down luminosity $\dot{E}_0$ (Equation \ref{eqn:e0dot}) and initial period $P_0$ (e.g., \citealt{pacini73, gaensler06} and references therein):
\begin{eqnarray}
\label{eqn:p0}
P_0 & = & P \left(1+\frac{t_{\rm age}}{\tau_{\rm sd}} \right)^{-\frac{1}{p-1}}
\end{eqnarray}
are quite close to their current values (Table \ref{tab:model_physprop}).  The derived initial spin period $P_0 \approx 50~{\rm ms}$ is slightly larger than expected for its surface magnetic field strength by models of fallback onto the proto-neutron star during the supernova (e.g., \citealt{watts02}).  Furthermore, the inferred initial spin-down luminosity $\dot{E}_0$ is somewhat lower than the $\dot{E}_0 \sim 10^{38}-10^{39}~\frac{\rm ergs}{\rm s}$ derived for other systems (e.g., \citealt{tanaka11, torres14, gelfand15}).

\deleted{In addition, the injected particle spectrum in PWN \pwn\ is different than that observed in other sources -- specifically $p_2 > p_1$, where for most PWNe, $p_1 < p_2$ (e.g., \citet{torres14, gelfand15}).} 
\added{The predicted injected particle spectrum in PWN \pwn\ is $p_1 \approx 2.9$ and $p_2 \approx 2.5$ (see Table \ref{tab:model_pars}). This relationship of $p_1 > p_2$ is different than that observed in other sources as for most PWNe $p_1 < p_2$ (e.g., \citet{torres14, gelfand15}).}
Extensive trials were conducted with $p_1 < p_2$, but were not able to reproduce the observed properties listed in Table \ref{tab:model_obs}. 
The low values of $p_1$ ($p_1 < 2$) inferred for other \replaced{PWN has}{PWNe have} been interpreted as magnetic reconnection dominating particle acceleration at low energies while Fermi acceleration dominating at higher energy (e.g., \citealt{sironi11}).  However, the required values of $p_1$ and $p_2$ \added{for \pwn} (Table \ref{tab:model_pars}) are both consistent with Fermi acceleration, and their different values possibly suggests particles are accelerated / injected at two sites within this PWN.  If correct, this could explain the spatial variations in $\Gamma$ observed near the center of this PWN (e.g., \citealt{guest19}).

Lastly, the results of our modeling can be used to interpret features in the observed SED of this PWN (Figure \ref{fig:model_sed}).  A particle of energy $E$ will generate synchrotron emission with a power $P_{\rm synch}$ (e.g., \citealt{pacholczyk70}):
\begin{eqnarray}
\label{eqn:psynch}
P_{\rm synch}(E) & = &  \frac{4e^4}{9m_e^4c^7}B^2E^2,
\end{eqnarray}
where $B$ is the strength of the nebular magnetic field, $e$ and $m_e$ are, respectively, the charge and mass of the electron while $c$ is the speed of light, and whose spectrum  will peak at a frequency $\nu_{\rm peak}$ (e.g., \citealt{pacholczyk70}):
\begin{eqnarray}
\label{eqn:nupeak}
\nu_{\rm peak}(E) & = & 0.29 \times \frac{3}{2} \left(\frac{E}{m_e c^2}\right)^2 \frac{e B}{m_e c}
\end{eqnarray}
For particles with a power-law energy distribution $\frac{dN}{dE} \propto E^{-p_{\rm par}}$, the synchrotron emission is also expected to have a power-law spectrum ($\frac{dN}{dE} \propto E^{-\Gamma}$) with:
\begin{eqnarray}
\label{eqn:alpha}
\alpha & = & \frac{1-p_{\rm par}}{2} \\
\label{eqn:gamma}
\Gamma & = & \frac{1+p_{\rm par}}{2}.
\end{eqnarray}
This synchrotron emission will cause a particle with energy $E$ to cool in time $t_{\rm cool}$:
\small
\begin{eqnarray}
t_{\rm cool} & \equiv & \frac{E}{P_{\rm synch}} = \frac{9m_e^4 c^7}{4e^4} B^{-2} E^{-1} \\
\label{eqn:tcool}
& \approx & 6.25 \left(\frac{B}{1~\mu{\rm G}}\right)^{-2} \left(\frac{E}{\rm 1~GeV}\right)^{-1} \times 10^{14}~{\rm years},
\end{eqnarray}
\normalsize
and a break in the electron spectrum will form at the energy $E_{\rm cool}$ whose synchrotron cooling time is equal to the age of the system:
\small
\begin{eqnarray}
E_{\rm cool}(B,t) & = &\frac{9m_e^4 c^7}{4e^4} B^{-2} t_{\rm age}^{-1} \\
& \approx & 1.26 \left(\frac{B}{1~\mu{\rm G}}\right)^{-2} \left(\frac{t_{\rm age}}{\rm 1~yr}\right)^{-1} \times 10^{19}~{\rm eV}
\end{eqnarray}
\normalsize
For the age $t_{\rm age}$ and current nebular magnetic field strength $B_{\rm pwn}$  predicted by our most likely set of model parameters (Table \ref{tab:model_physprop}), we have:
\begin{eqnarray}
E_{\rm cool}(B_{\rm pwn},t_{\rm age}) & \approx & 7.6~{\rm TeV}
\end{eqnarray}
and 
%\begin{eqnarray}
$\nu_{\rm peak}(E_{\rm min}) \approx  140~{\rm GHz}$, $\nu_{\rm peak}(E_{\rm break}) \approx 900~{\rm THz}$, $h\nu_{\rm peak}(E_{\rm cool})  \approx 0.2~{\rm keV}$, $h\nu_{\rm peak}(E_{\rm max}) \approx 0.1~{\rm MeV}$, 
%\end{eqnarray} 
where $h$ is Planck's constant.  As detailed below, we expect to see features in the observed SED at all of these frequencies.

At $\nu < \nu_{\rm peak}(E_{\rm min})$, the emission will be dominated by ``relic particles" injected into the PWN at earlier times and have since (primarily adiabatically) cooled to lower energies.  As a result, the ``flat" (spectral index $\alpha \approx 0$; flux density $S_\nu \propto \nu^\alpha$) observed at GeV frequencies does not necessarily reflect the spectrum of injected particles. Beginning at $\nu_{\rm peak}(E_{\rm min})$, the emitting particles will be a mix of freshly injected and ``relic" particles, and expect a change in $\alpha$ ($\Gamma$) and this point.  However, $t_{\rm cool} \gg t_{\rm age}$ at $\nu_{\rm peak}(E_{\rm min})$ and $\nu_{\rm peak}(E_{\rm break})$, so previously injected particles will dominate in this energy band and the emitted spectrum will be ``flatter" than that expected from the freshly injected particles:
\begin{eqnarray}
\label{eqn:alpha1}
\alpha_1 & = & \frac{1-p_1}{2} = -0.93 \\
\label{eqn:gamma1}
\Gamma_1 & = & \frac{1+p_1}{2} = 1.93.
\end{eqnarray}
At photon energy $h\nu \approx h\nu_{\rm peak}(E_{\rm cool})$, radiation from freshly injected particles should begin to dominate the observed emission.  This occurs well within the high-energy component of the injected broken power-law spectrum, and the observed synchrotron emission should have:
\begin{eqnarray}
\label{eqn:alpha2}
\alpha_2 & \approx & \frac{1-p_2}{2} = -0.76 \\
\label{eqn:gamma2}
\Gamma_2 & \approx & \frac{1+p_2}{2} = 1.76.
\end{eqnarray}
Indeed, the $\Gamma\sim 1.8 - 1.9$ measured between $0.8-3.0$~keV (where the emitting particles have $t_{\rm cool} \lesssim t_{\rm age}$) is similar to $\Gamma_2 = 1.76$.  At higher photon energies, the shorter cooling time $t_{\rm cool}$ results in a decrease in the average age, and therefore total number, of emitting particles, resulting in a softening (increase in $\Gamma$) of the spectrum.   However, due to the decreasing input of energy into the PWN by the pulsar, $\Delta \Gamma \neq 0.5$ as expected from standard synchrotron theory (e.g., \citealt{pacholczyk70}).  In fact, our simple model for the evolution of a PWN inside a SNR does a good job of reproducing the increasingly softening spectrum in the X-ray band (Figure \ref{fig:model_sed}, Table \ref{tab:model_obs}).  Lastly, we would expect little synchrotron emission at $h\nu_{\rm peak}(E_{\rm max}) \approx 0.1~{\mathrm MeV}$ -- suggesting that \pwn\ should not produce much MeV emission and therefore is not a promising target for proposed missions like {\it AMEGO}.

\subsection{SNR shell}
\label{sec:snrhydro}
The morphology of the SNR rim in \pwn\ suggests an interaction with dense material in the north. The shell is remarkably circular until an abrupt flattening that results in brightened X-ray emission and enhanced knot-like structures (Figure \ref{fig:g21_indiv_components}). Spectral investigations by \citet{guest19} suggest an ejecta-rich thermal component for which the density is $\sim 45 d_{4.6}^{-1/2}f^{-1/2}{\rm\ cm}^{-3}$, where $f$ is the filling factor of the X-ray gas. We note that this value is additionally uncertain due to the unknown composition of the ejecta. 

We have investigated a hydrodynamical model for the evolution of the SNR using the results from \S\ref{sec:pwnmodel} (summarized in Table \ref{tab:model_pars}) and assuming the presence of a dramatic density increase in regions north of the explosion center. The simulation was carried out with the grid-based hydrodynamics code VH1 (see \citealt{blondin2001, kolb2017}), which utilizes the PPMLR method \citep{colella1984} to resolve shock propagation. Here we have ignored the contributions from the pulsar since the PWN has no impact on the SNR morphology at this stage of evolution. We ran the simulation to an age of 1700~years (see \S\ref{sec:pwnmodel}), adjusting the position and magnitude of the density jump relative to the explosion center until the observed morphology reproduced that observed for \pwn.

\begin{figure}[t]
  \centering
    \includegraphics[width=0.48\textwidth]{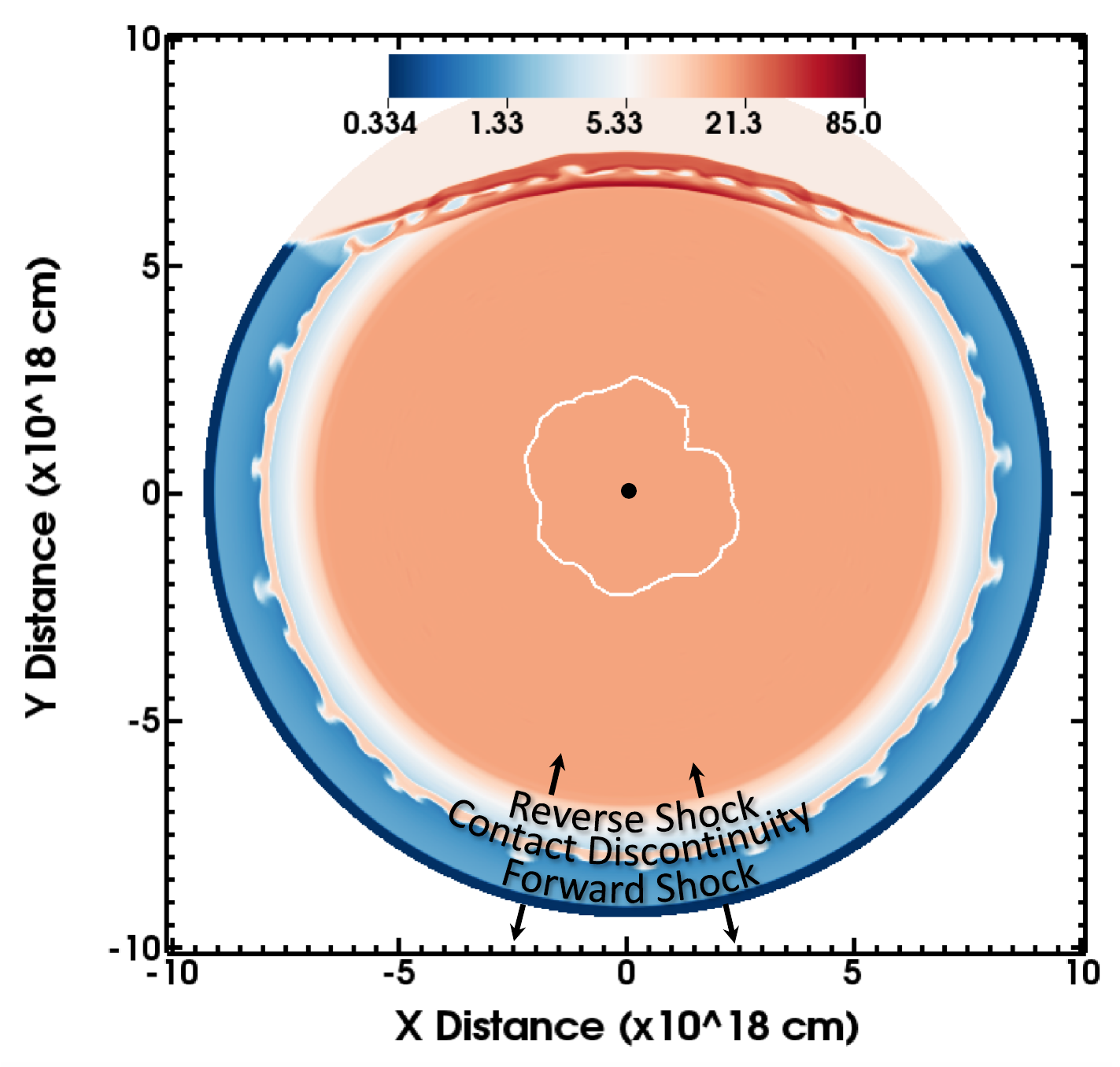}
   \caption{Hydrodynamical simulation of an SNR evolving in a medium with a density jump located to the north of the explosion site. The color bar indicates the density in units of $10^{-24}{\rm\ g\ cm}^{-3}$, and the positions of the Forward Shock, Reverse Shock, and Contact Discontinuity are indicated. The compression in the northern region is similar to that observed in \pwn. The white contour corresponds to the outer boundary of the PWN.}
\label{fig:hydro_simulation}
\end{figure}

We find that a reasonable representation of the SNR morphology can be obtained with a density jump by a factor of $\sim 20$
located $\sim 1.8$~pc north of the explosion center. The results are summarized in Figure \ref{fig:hydro_simulation} where we plot the density distribution from the simulation. The outermost boundary corresponds to the ambient density, and the position of the Forward Shock (FS), Reverse Shock (RS), and Contact Discontinuity (CD) are indicated. The peak density in the northern regions of the SNR is in the reverse-shocked ejecta, where $\rho_{\rm ej} \sim 3.1 \times 10^{-23}{\rm\ g\ cm}^{-3}$ ($n \sim 31 {\rm\ cm^{-3}}$), in reasonable agreement with the density estimate for the northern knot. While this solution is far from unique, it presents a reasonable interpretation of the basic conditions leading to the observed properties of the SNR. We note that, as expected, the RS (for which the outer contour is overlaid in white) is still far from the PWN boundary ($R_{\rm PWN} \sim 2.4 \times 10^{18}{\rm\ cm}$), consistent with our finding in \S\ref{sec:pwnmodel} that no RS/PWN interaction has occurred.

\section{Summary \& Conclusions}
\label{sec:summary}

We \added{have} reanalyzed archival IR (\textit{Herschel, Spitzer}; \S\ref{sec:ir}) and X-ray (\chandra, \nustar, \hitomi; \S\ref{sec:x-ray_analysis}) observations of PWN \pwn. The similar morphology observed in IR emission line and continuum maps of this source suggests surrounding dust and gas produce much of the observed radiation (\S\ref{dust}).  Our analysis of the X-ray observations shows that while there is an overall spectral softening within this band, discrepant power-law parameter values from different detectors indicate instrumental uncertainties should be taken into consideration when interpreting the values (\S\ref{sec:results}).

To quantify the degree and shape of the spectral softening in the X-ray band, we separately fit power laws over distinct energy bands (\textit{piecewise power law fits} \S\ref{sec:piecewise}), instead of fitting over the entire detector energy range with a single broken power law.  This shape is consistent with what is predicted by models for the evolution of a PWN inside a SNR, which find that the continuous injection of particles into, and changing magnetic field strength inside, the PWN does not result in a sharp break as required by broken power-law models.

We then used a one-zone model for the evolution of a PWN inside a SNR to reproduce the observed dynamical and broadband spectral properties of \pwn, taking into consideration that the IR emission is likely not dominated by synchrotron radiation from particles inside the PWN, and the increased uncertainty in the X-ray spectrum resulting from our comparison of different instruments (\S\ref{sec:pwnmodel}).  We found that this model can reproduce the properties of this source, but only if
%\begin{itemize}
%    \item 
the supernova ejecta had a low initial kinetic energy of $E_{\rm sn} \approx 1.2\times10^{50}~{\rm ergs}$ \added{and the}
\deleted{, associated pulsar PSR J1833$-$1034 has a braking index $p\approx 3.127 > 3$ (the canonical value for the braking index expected from magnetic dipole radiation) and a spin-down timescale $\tau_{\rm sd} \approx 2900~{\rm years} > t_{\rm age} \approx 1700~{\rm years}$, requiring that initial spin-period of this pulsar is close to its current value, and}
spectrum of particles injected into the PWN at the termination shock is softer at lower energies than at high energies $(p_1 \approx 2.9 > p_2 \approx 2.5)$ -- opposite of what is observed from most other PWNe.  Both values are consistent with what is expected from diffusive shock acceleration, suggesting that magnetic reconnection may not play an important role in accelerating particles in this PWN, and the different values may indicate two different acceleration sites.
%\end{itemize}
Furthermore,  we used a hydrodynamical model to determine the structure of the ambient medium needed to reproduce the morphology of the observed SNR shell (\S\ref{sec:snrhydro}). We are able to do so if there is a $\sim20\times$ increase in density ${\sim} 1.8$~pc north of the explosion center.

As a result, we have obtained an extensive picture of the supernova, neutron star, pulsar wind, and surrounding material of this source. The derived properties are useful for understanding how neutron stars are created in core-collapse supernovae and the different ways they energize their environment. \replaced{These techniques and tools are applicable to many other PWNe, which}{The techniques and tools presented in this study are applicable when analyzing many other PWNe, and their use} may provide a more comprehensive view of the different mechanisms by which neutron stars are formed and produce some of the highest energy particles in the Universe. 

\acknowledgments
\added{We would like to thank the anonymous referee for comments that improved the article.}  The contributions of JDG and SMS was supported by the National Aeronautics and Space Administration (NASA) under grant number NNX17AL74G issued through the NNH16ZDA001N Astrophysics Data Analysis Program (ADAP). JDG and SH are also supported by the NYU Abu Dhabi Research Enhancement Fund (REF) under grant RE022.  The research of JDG is also supported by NYU Abu Dhabi Grant AD022.  This research has made use of NASA's Astrophysics Data System Bibliographic Services.

\facility{CXO, NuSTAR, \added{Herschel}, Spitzer, Hitomi}
\software{HIPE \citep{ott2010}, XSpec \citep{xspec}, CIAO (v4.10; \citet{Fruscione2006}, HEASoft \citep{heasarc2014}, Sherpa \citep{freeman2001, doe2007}, SAOImage DS9 \citep{joye2003, sao2000}, Matplotlib \citep{Hunter:2007}}, NumPy \citep{oliphant2006guide, van2011numpy}, SciPy \citep{virtanen2020}

\clearpage

% Acknowledge facilities using \facility{facility ID} https://journals.aas.org/facility-keywords/ 
% Software: \software{CASA (McMullin et al. 2007), XSPEC (Arnaud 1996),astropy (The Astropy Collaboration 2013, 2018), ... }

%\begin{thebibliography}{}

%\bibitem[Astropy Collaboration et al.(2013)]{astropy2013} Astropy Collaboration, Robitaille, T.~P., Tollerud, E.~J., et al.\ 2013, \aap, 558, A33 

\clearpage
\appendix
\section{Systematic Errors in \nustar\ Spectrum}
% \subsection{\nustar\ Independent Fit results} 
\label{sec:nustar_independent_fits}

\begin{figure*}
    \centering
        \includegraphics[width=0.48\textwidth]{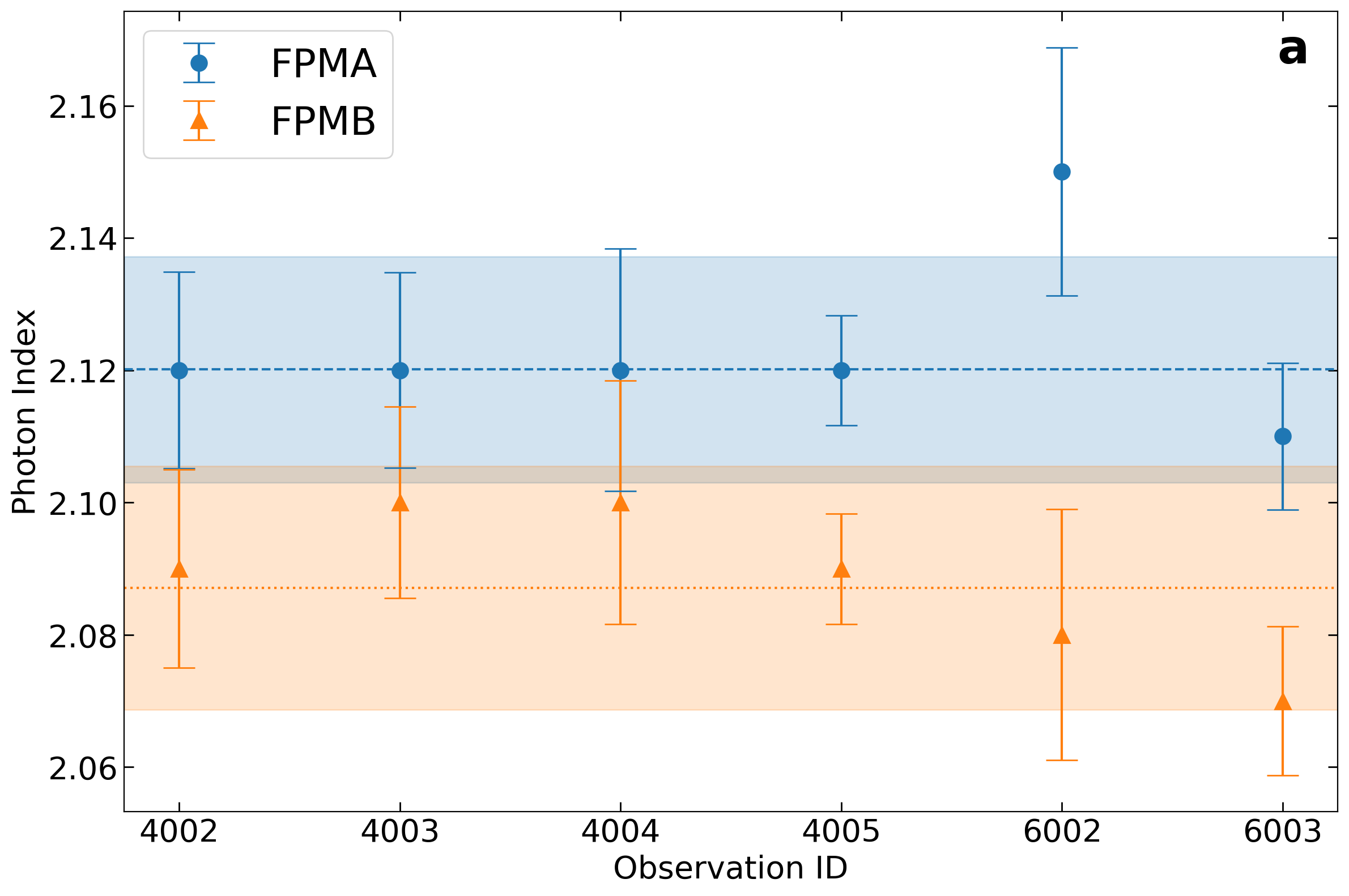}
        \includegraphics[width=0.48\textwidth]{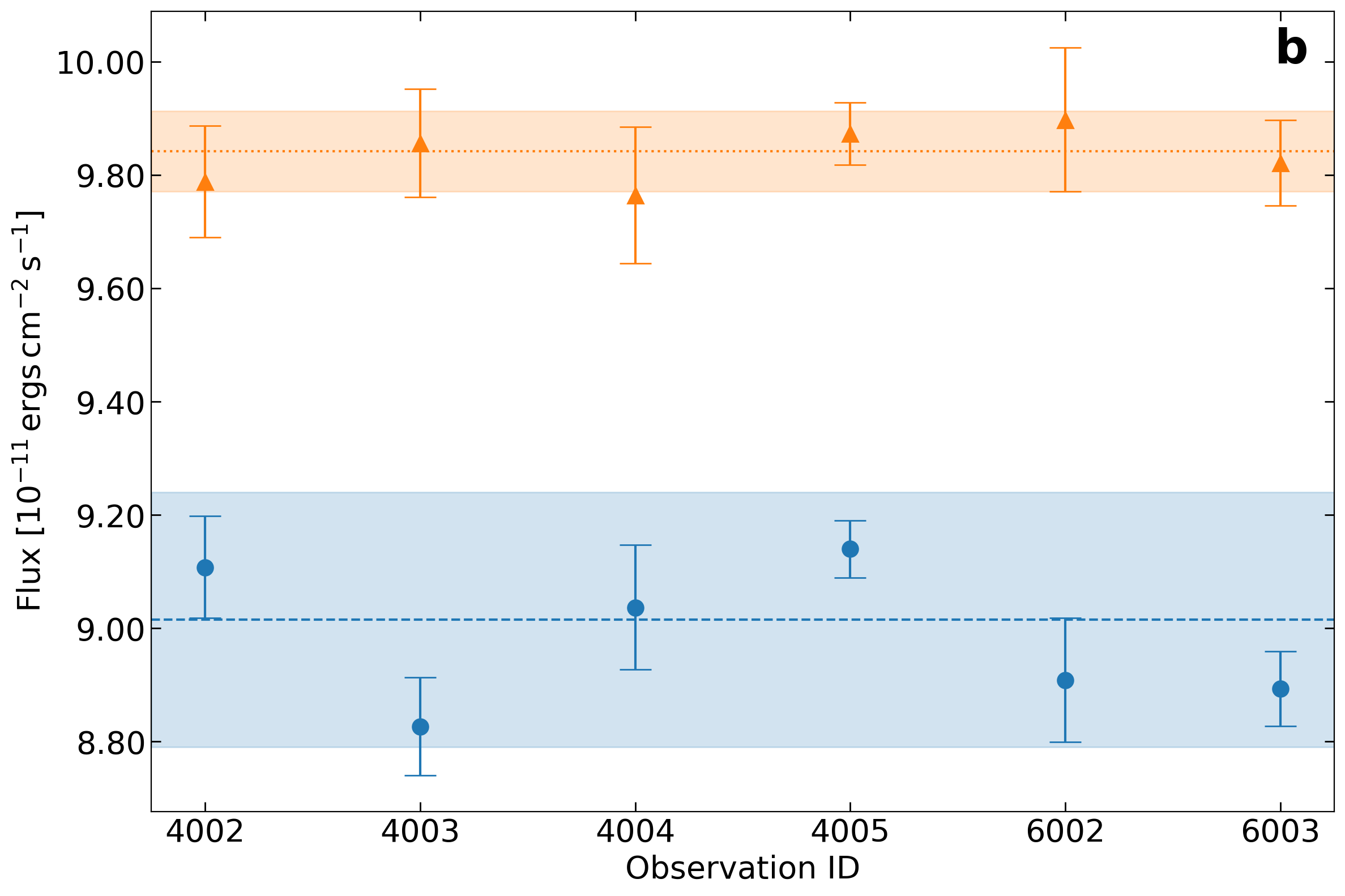}\\%
        \includegraphics[width=0.48\textwidth]{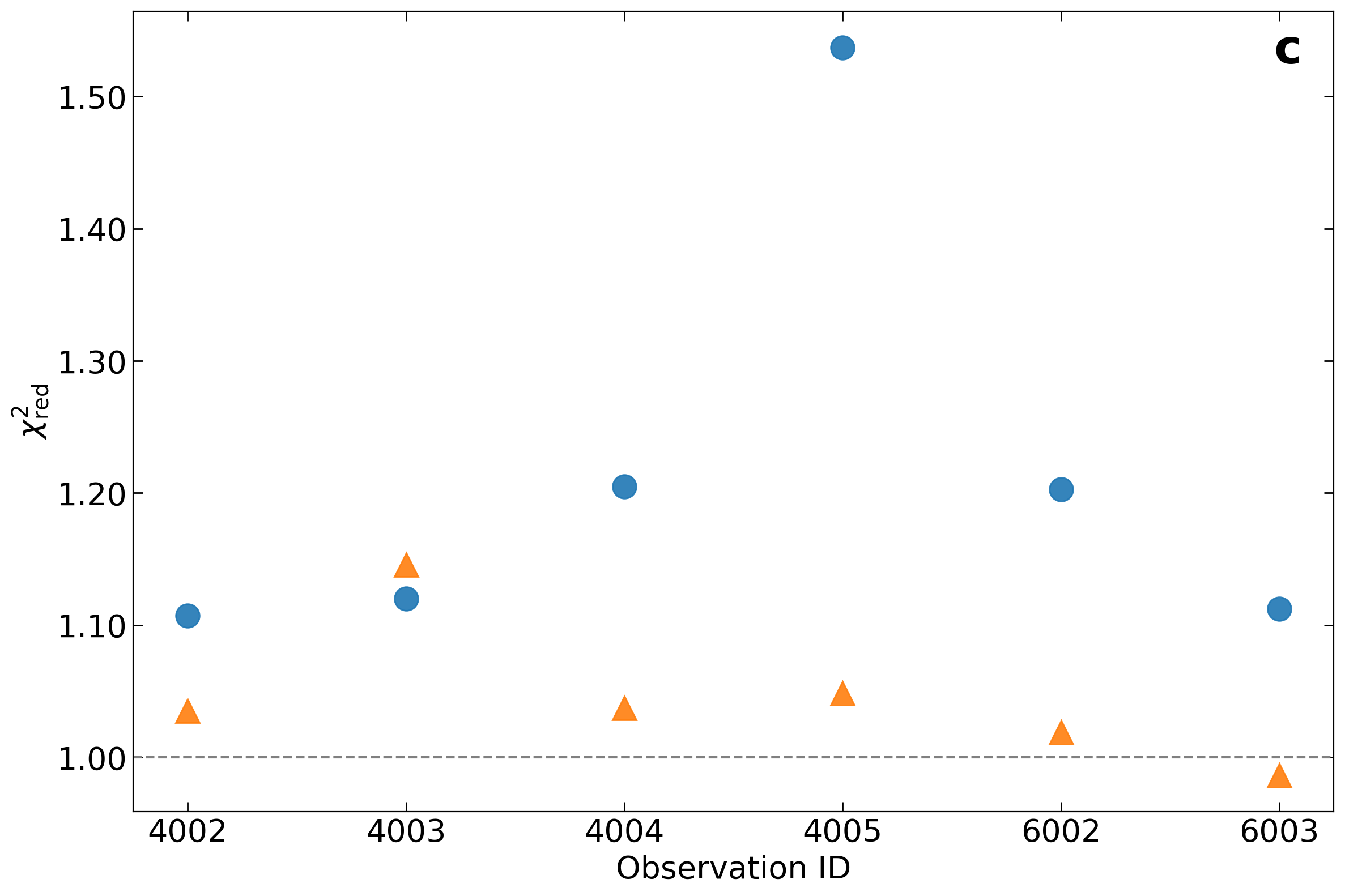}
    \caption{Photon index (a), normalization (b), and reduced $\chi^2$ (c) for each observation over the entire energy range (3--45 keV). The fitted model is {\it with} the black-body component. Shaded regions indicate 90\% confidence intervals.} 
    \label{fig:nustar_indiv_3-45_with_BBody}
\end{figure*}

\begin{figure*}
    \centering
        \includegraphics[width=0.32\textwidth]{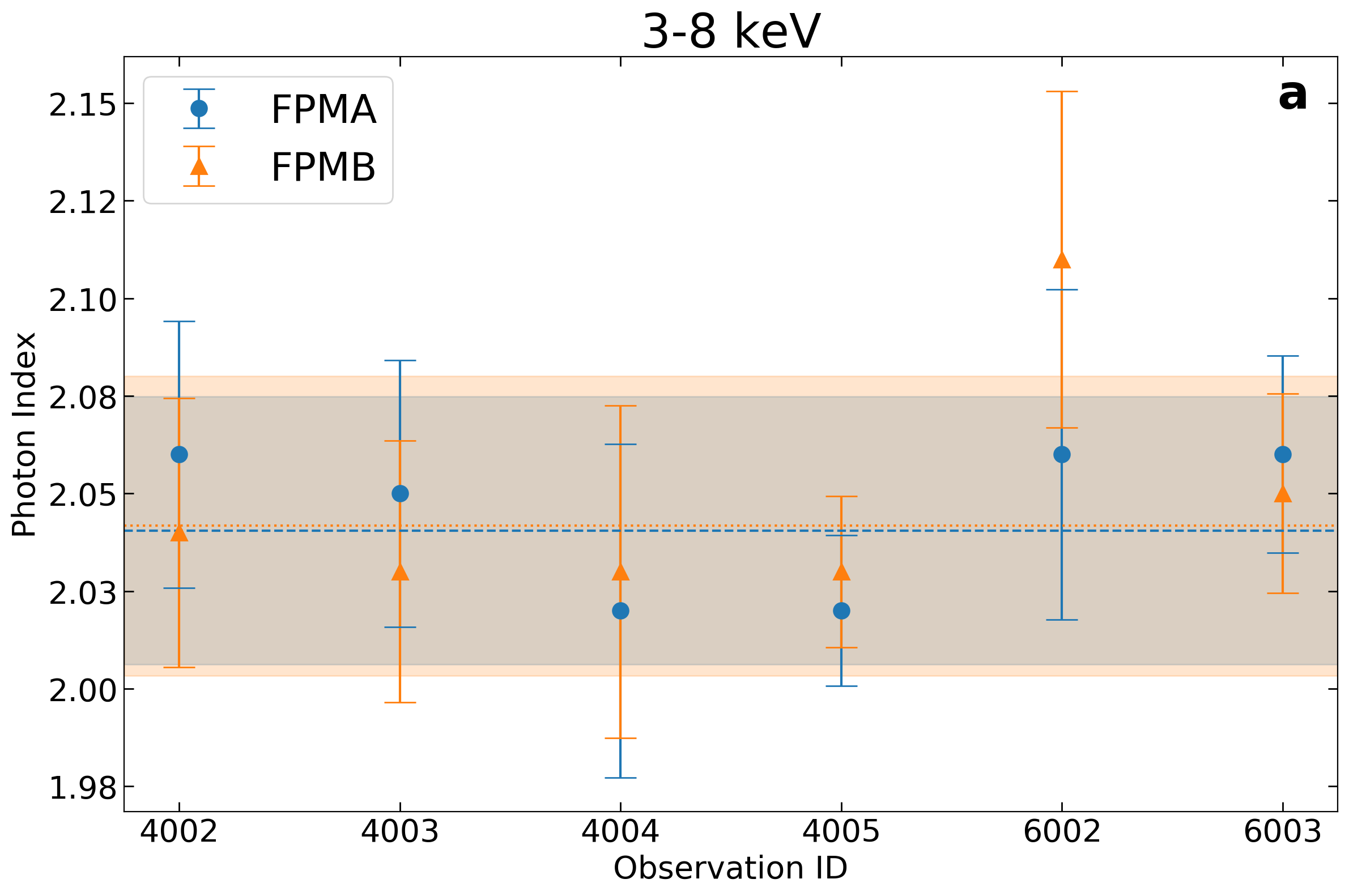}
        \includegraphics[width=0.32\textwidth]{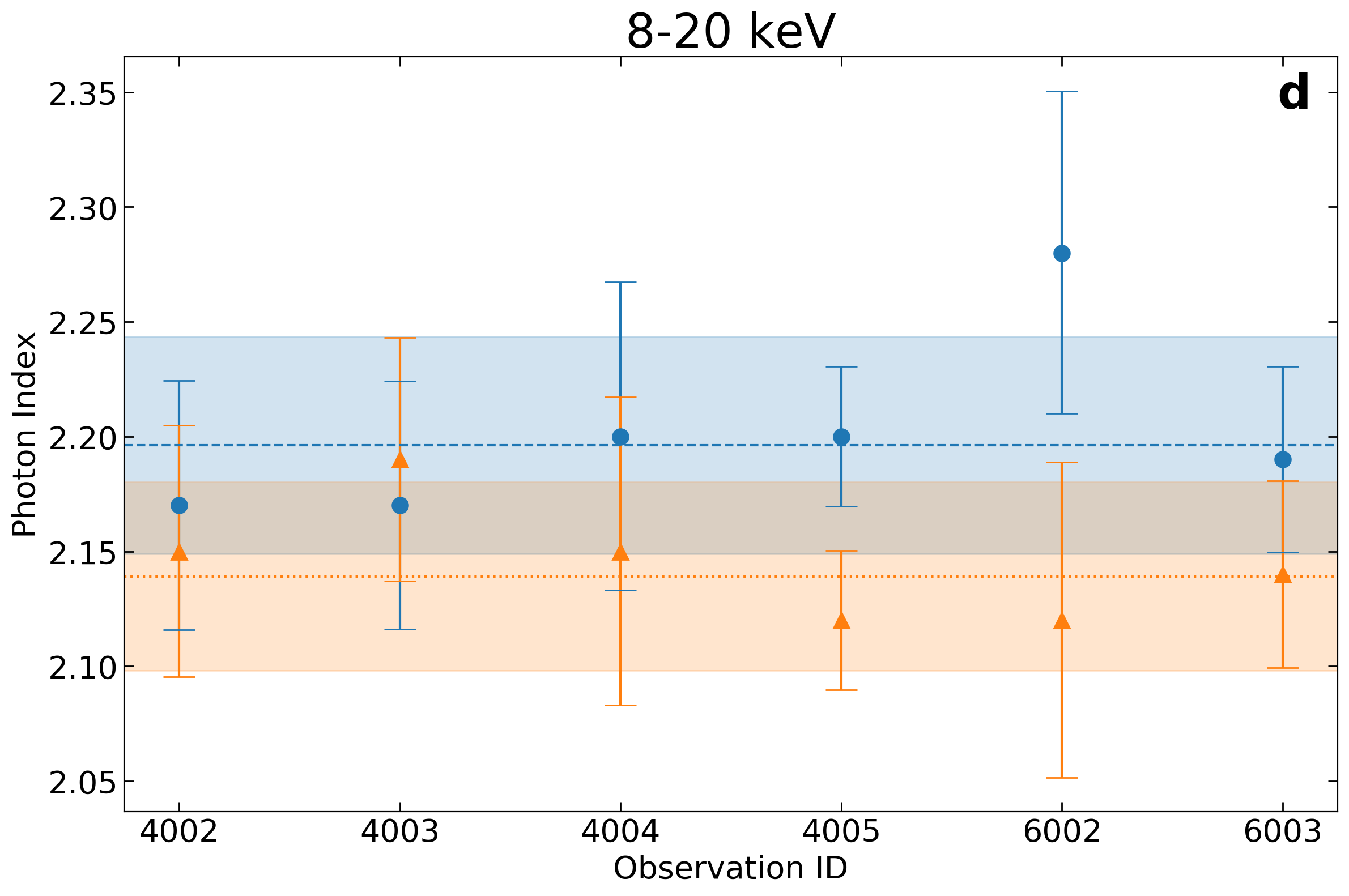}
        \includegraphics[width=0.32\textwidth]{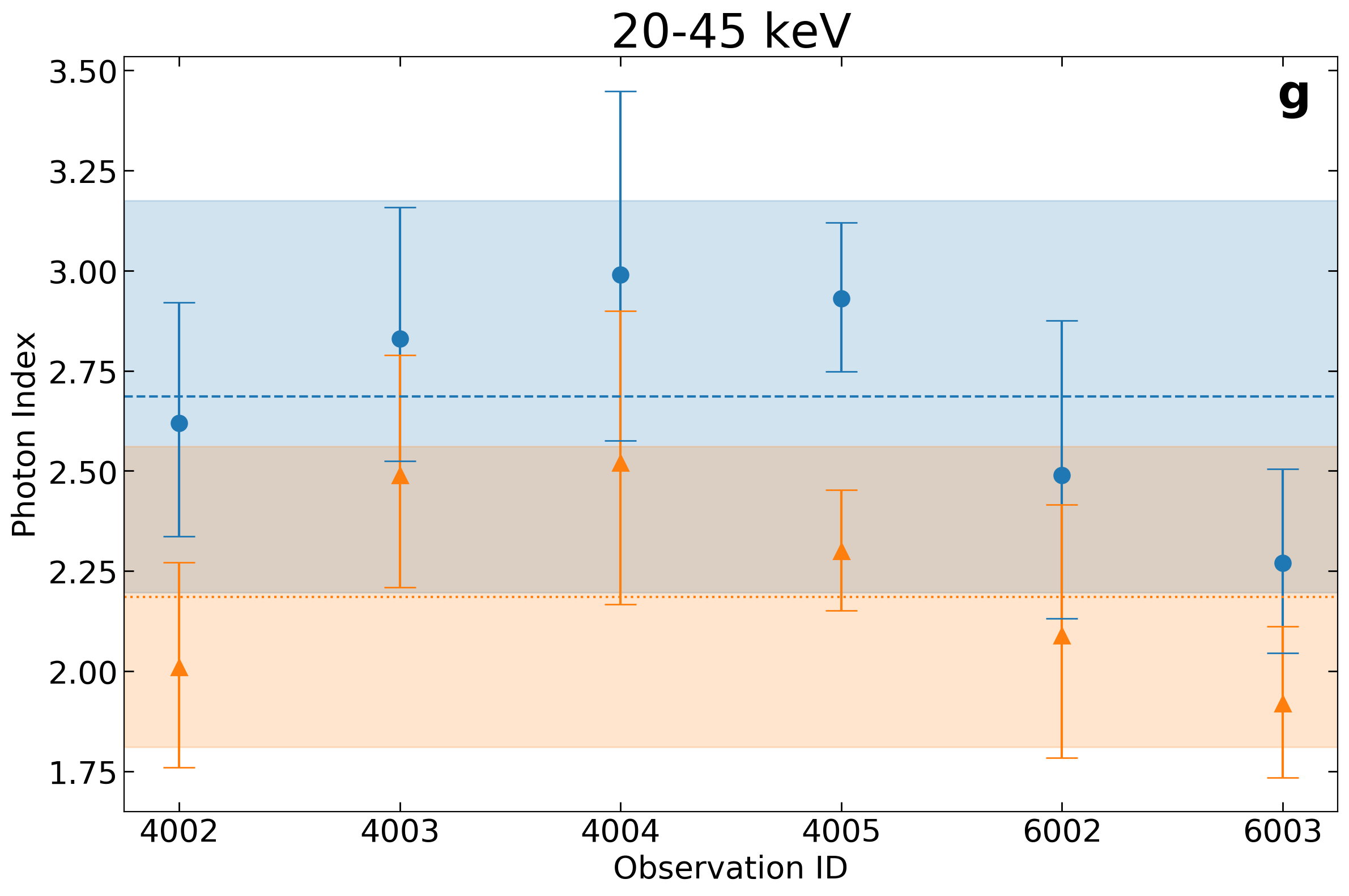} \\ 
        \includegraphics[width=0.32\textwidth]{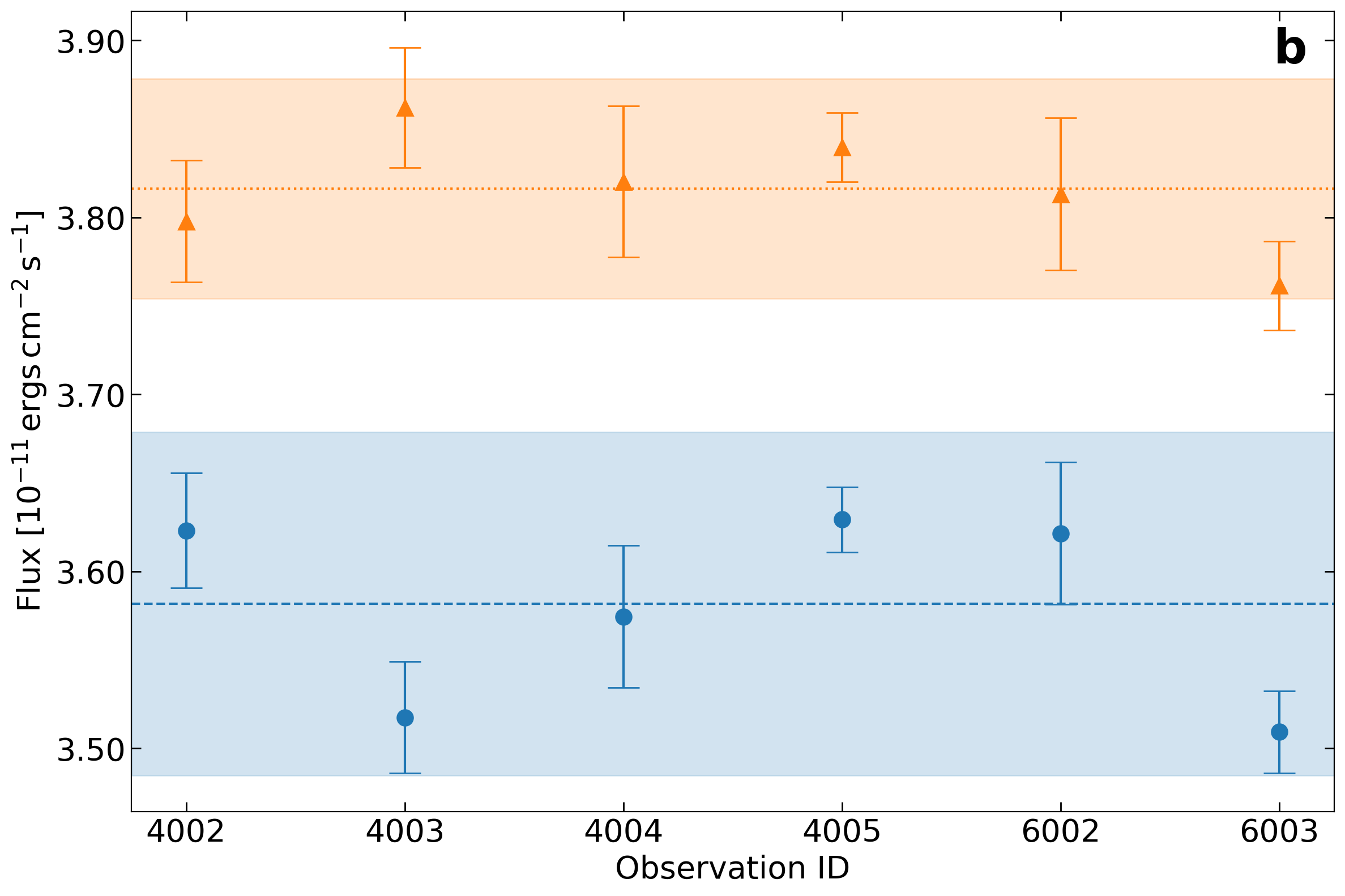}
        \includegraphics[width=0.32\textwidth]{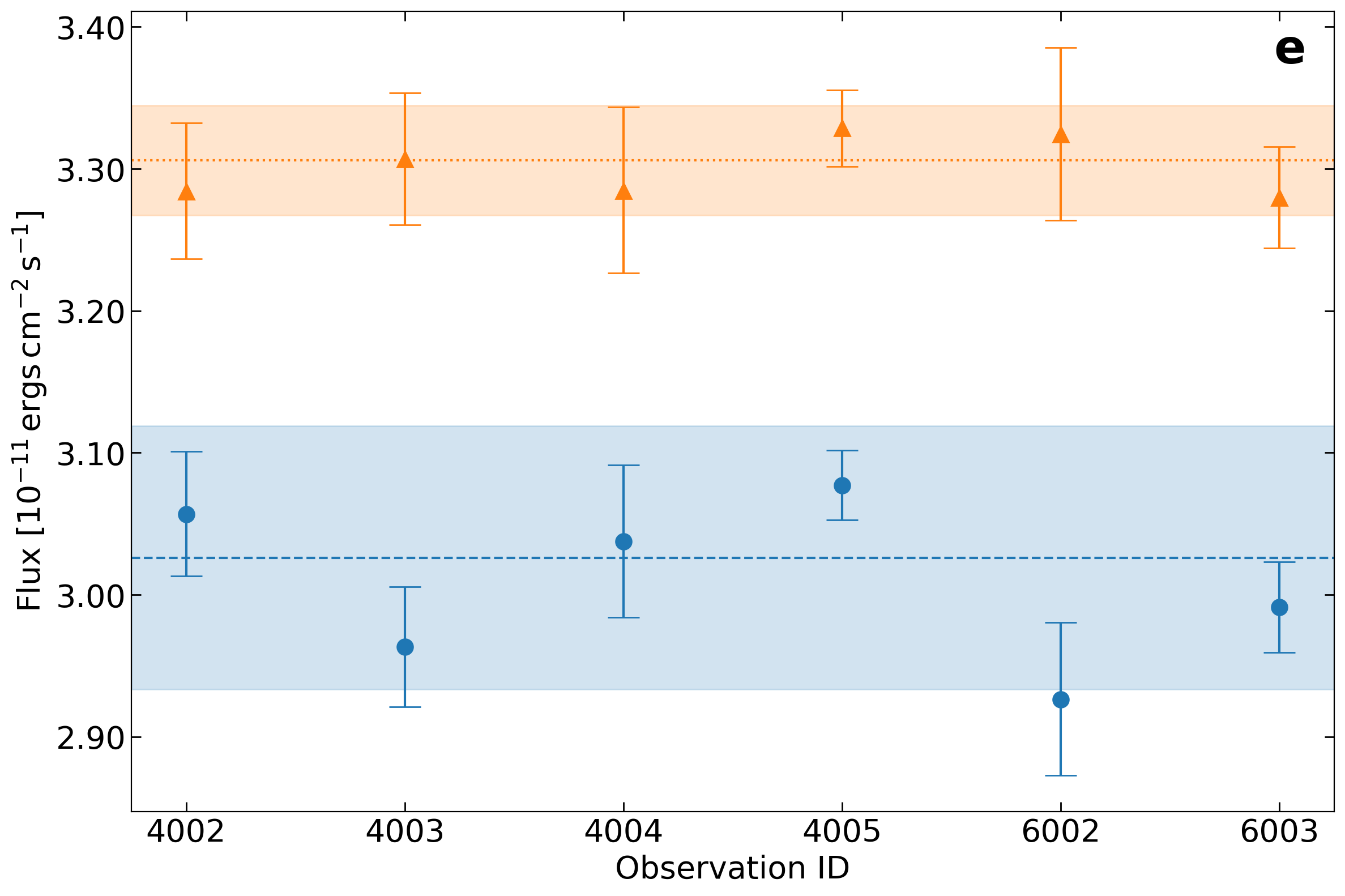}
        \includegraphics[width=0.32\textwidth]{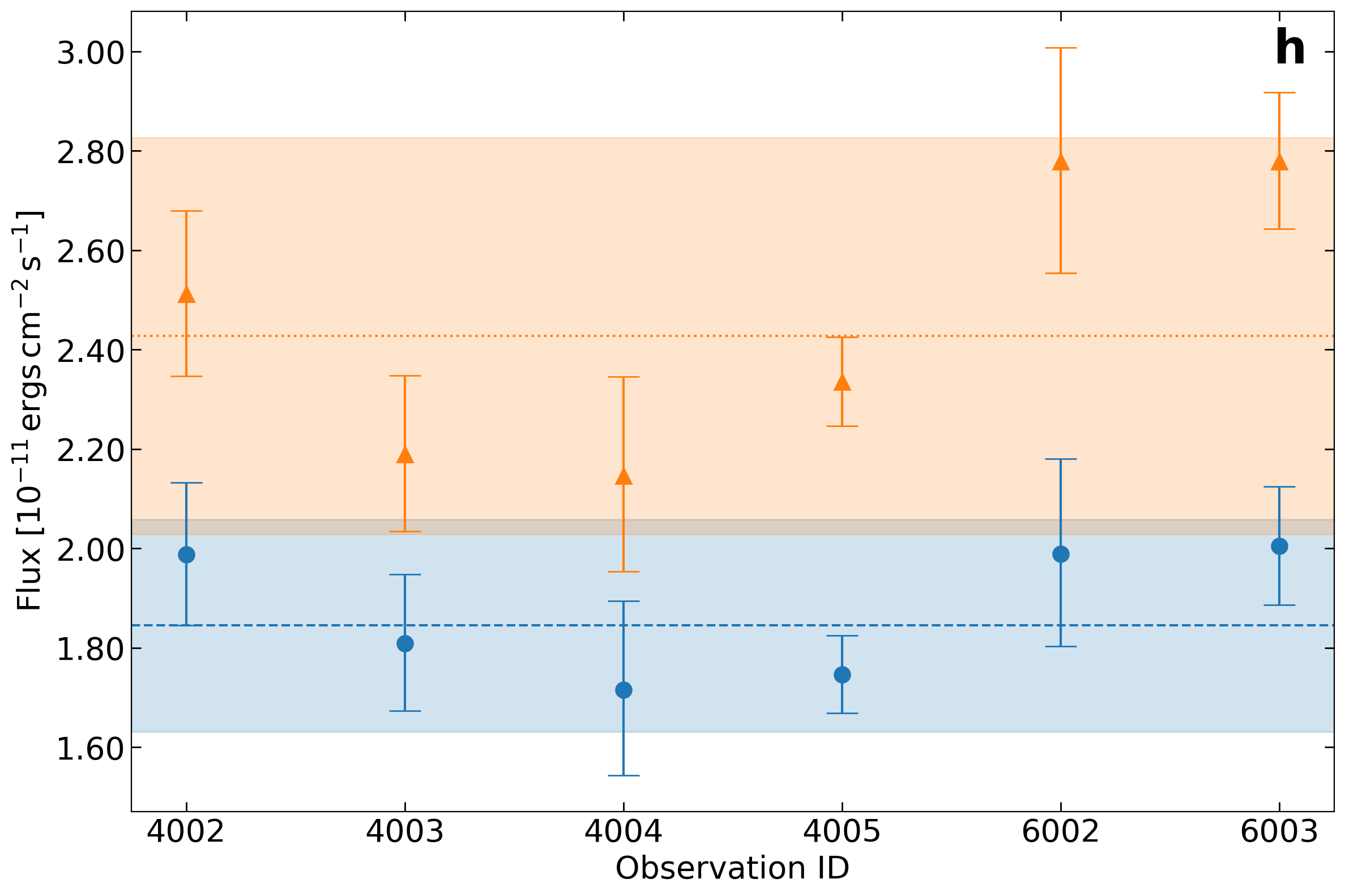} \\ 
        \includegraphics[width=0.32\textwidth]{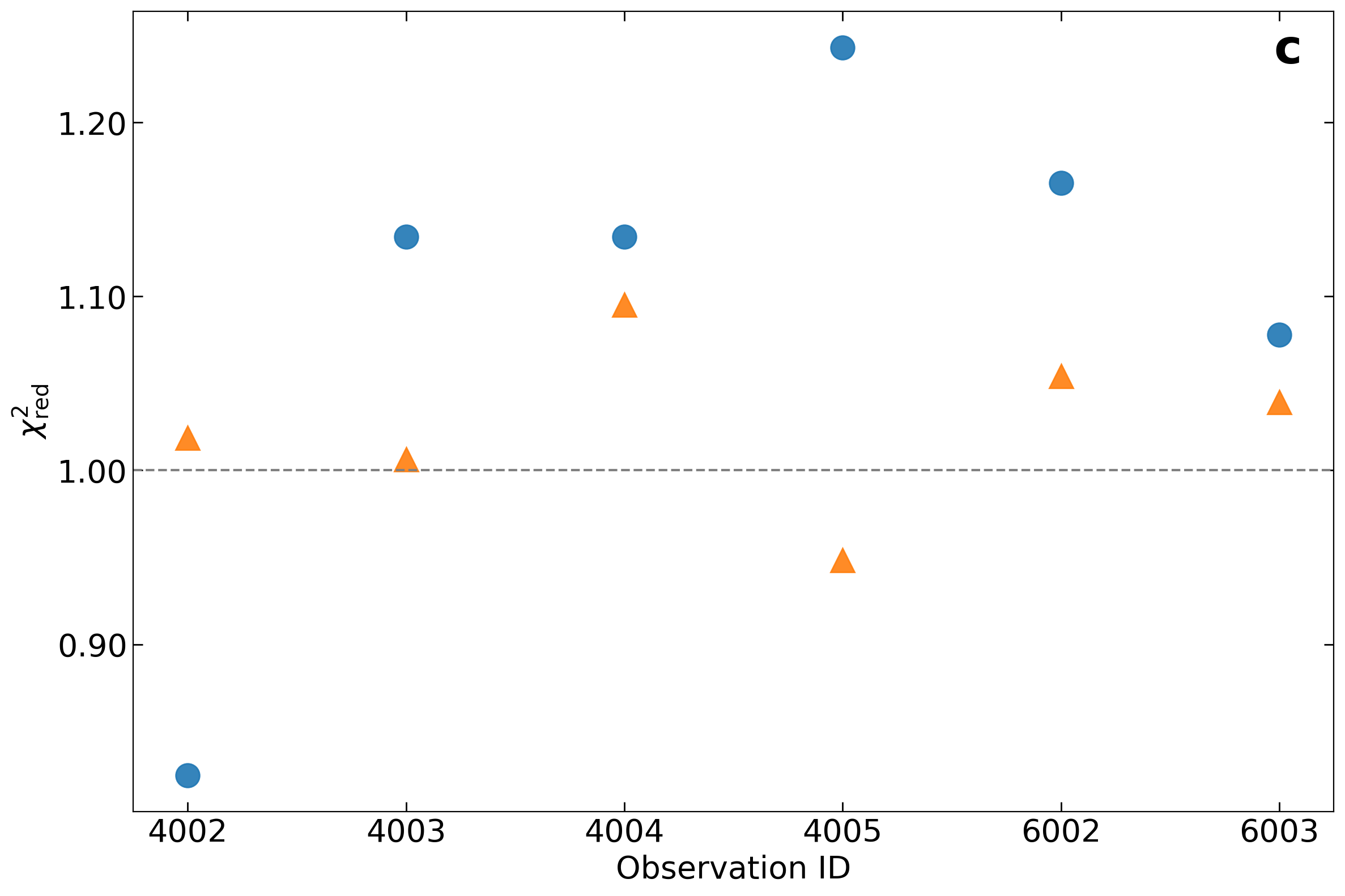}
        \includegraphics[width=0.32\textwidth]{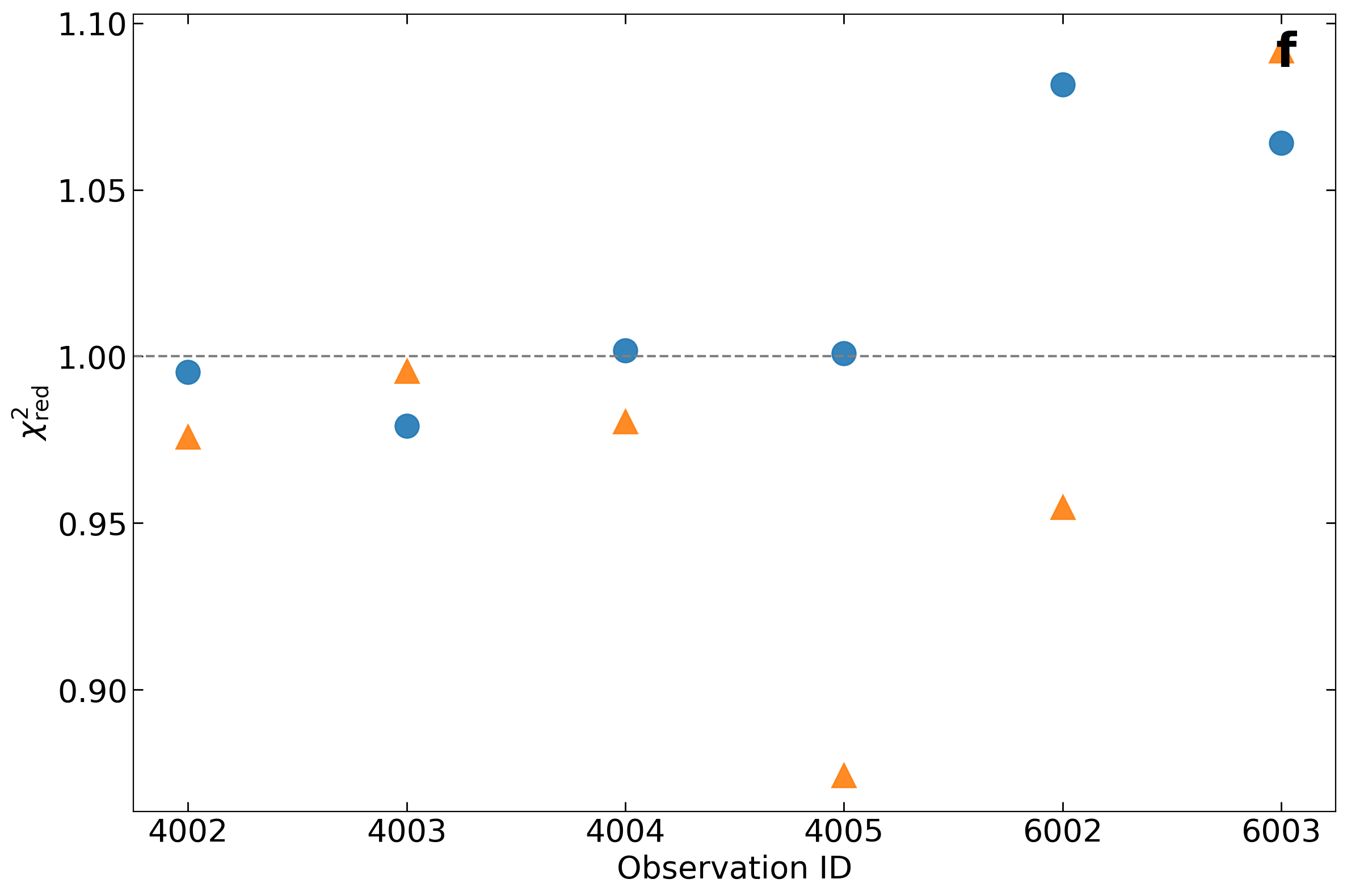}
        \includegraphics[width=0.32\textwidth]{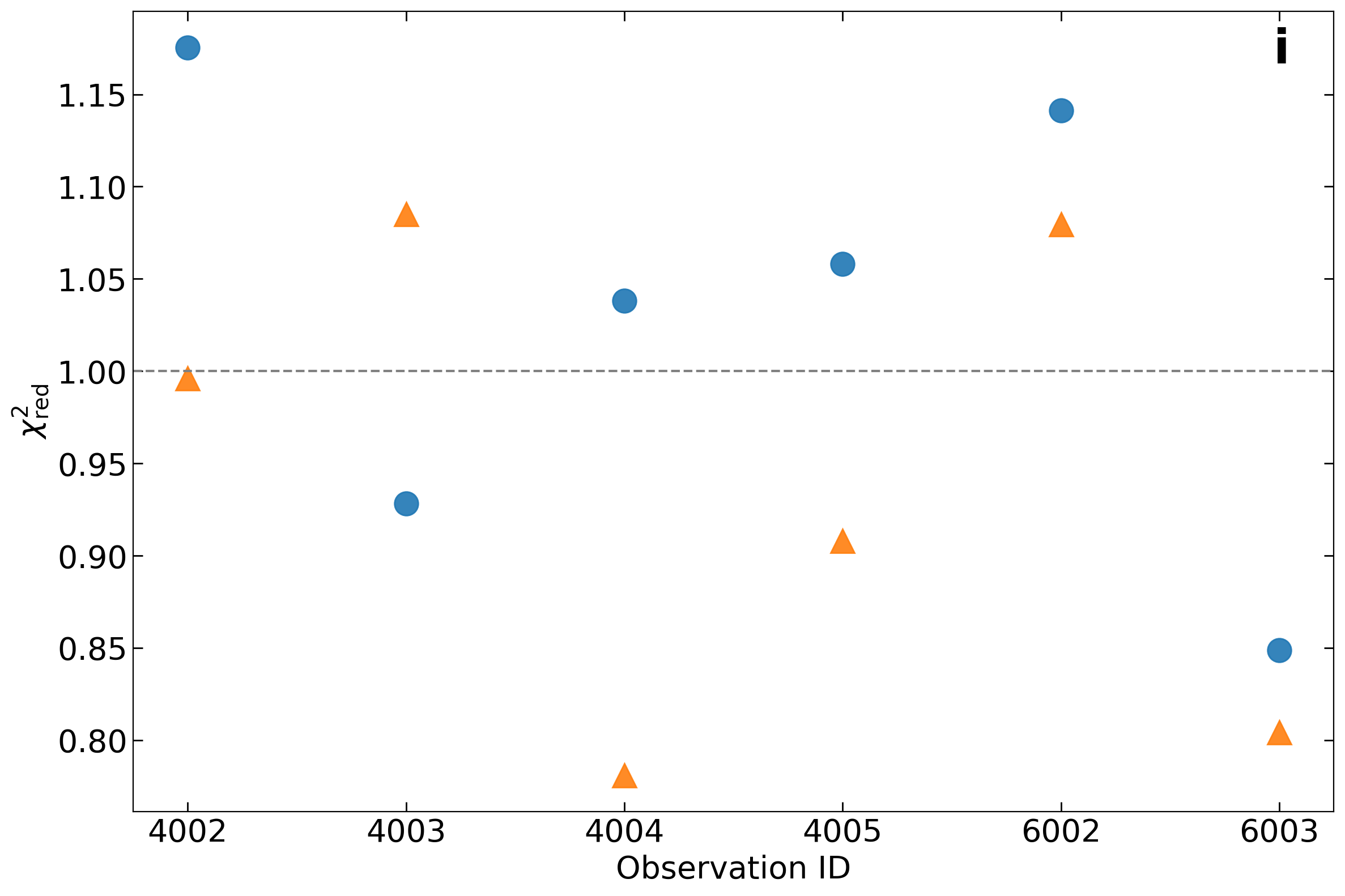}
    \caption{Photon index, normalization, and reduced $\chi^2$ for each observation in the energy bands: 3--8 keV (a, b, c), 8--20 keV (d, e, f), 20--45 keV (g, h, i). The fitted model is {\it with} the black-body component. Shaded regions indicate 90\% confidence intervals.} 
    \label{fig:nustar_indiv_each_eband_with_BBody}
\end{figure*}

Here we discuss the systematic differences between FPMA and FPMB spectra. We initially fit all twelve \nustar\ spectra independently over the entire 3--45 keV range without dividing the energy ranges (Figure \ref{fig:nustar_indiv_3-45_with_BBody}) using the model with the pulsar black-body component (explained in \S\ref{sec:x-ray_analysis}). While we also fit the model without the pulsar black-body component and obtained similar results, here we only report on the results of fitting the model with the pulsar black-body component as we are simply trying to highlight the differences between FPMA and FPMB spectra. 

We found that the photon index $\Gamma$ was consistently higher for FPMA spectra compared to FPMB spectra, indicating spectra from FPMA was softer (i.e., a lower fraction of higher energy X-ray photons). The weighted average (inverse variance weighting), across observations, of the photon index for spectra from FPMA was $\Gamma_{\mathrm{A}} = 2.12 \pm 0.01$ and the weighted average of of the photon index for spectra from FPMB was $\Gamma_{\mathrm{B}} = 2.09 \pm 0.01$. The uncertainties reported here are the $1\sigma$ weighted sample standard deviations calculated with the formula $\sigma = \sqrt{\frac{N}{N-1}\frac{\Sigma_i w_i (x_i - \bar{x})^2}{\Sigma_i w_i}}$ where $w_i \equiv 1/\sigma_i^2$, $\bar{x}$ is the weighted average, and $N=6$ (the number of observations) in our case. 
The standard deviation of $\pm 0.01$ for each photon index is within what is mentioned as the approximate repeatability error of the spectral slope ($\pm 0.01$) in the {\it NuSTAR} calibration paper \citep{nustar_calibration}, indicating that the discrepancy across different observations from each FPM is within the calibration uncertainty.
While \citet{nustar_calibration} report offsets of $\Delta \Gamma \approx 0.1$ between $\Gamma_\mathrm{A}$ and $\Gamma_\mathrm{B}$ for the source 3C273 during certain cross-calibration campaign observations, they do not address $\Gamma_{\mathrm{A}}$ being consistently higher than $\Gamma_{\mathrm{B}}$, which is what we observe for \pwn. They do note that if the signal to noise ratio is high enough, which could be the case for a bright source such as \pwn, the inter-instrumental slope differences between FPMA and FPMB could be significant.

In addition to the discrepancy in the photon indices between FPMA and FPMB spectra, the unabsorbed flux values in the 3--45\,keV range are also different. In units of $10^{-11}$\ \flux, we obtain $F_{\mathrm{A (3-45)}} = 9.02 \pm 0.14$\ and $F_{\mathrm{B (3-45)}} = 9.84 \pm 0.04$. The two weighted average flux values differ by $\sim 8\%$, which is slightly larger than the potential $5\%$ flux difference mentioned in the \nustar\ calibration paper \citep{nustar_calibration}. We find that the unabsorbed flux for spectra from FPMB is consistently higher than that of spectra from FPMA. As with the photon index, this consistent offset may be due to the brightness of \pwn.

We then repeated the above analysis over each energy band; 3--8\,keV, 8--20\,keV, and 20--45\,keV (Figure \ref{fig:nustar_indiv_each_eband_with_BBody}). The photon indices $\Gamma_\mathrm{A}, \Gamma_\mathrm{B}$ agree in the 3--8 keV band and there is no consistent offset. However, while the 90\% confidence intervals overlap for the 8--20 keV and 20--45 keV bands, we do see that in most cases $\Gamma_\mathrm{A}$ is higher than $\Gamma_\mathrm{B}$. For the unabsorbed flux we see that the value is consistently higher for FPMB spectra compared to FPMA spectra in all energy bands. The 90\% confidence intervals for the unabsorbed flux do not overlap in the 3--8 keV and 8--20 keV bands, and slightly overlap in the 20--45 keV band due to the large spread of values for FPMB spectra. 

While there exist discrepancies in the PWN photon index and unabsorbed flux values between the spectra from FPMA and FPMB, the fit results between spectra from the same FPM across different observations are within the calibration uncertainty. As such, we believe the appropriate approach is to do joint fits of spectra from each FPM separately.

%\end{thebibliography}
\bibliography{ms}
\bibliographystyle{aasjournal}
\end{document}